\documentclass[sn-mathphys-num]{sn-jnl}

\usepackage{graphicx}%
\usepackage{multirow}%
\usepackage{amsmath,amssymb,amsfonts}%
\usepackage{amsthm}%
\usepackage{mathrsfs}%
\usepackage{subfig}
\usepackage{placeins}
\newcommand{\gfx}[1]{\frame{\includegraphics[width=\dimexpr\linewidth-2\fboxrule\relax]{#1}}}
\usepackage[inline]{enumitem}

\begin{document}

	\title[Article Title]{Segre Classification as a Framework for Generating Exact Solutions in Static Cylindrical Spacetimes with Levi-Civita Matching.}
	\author[1]{\fnm{Wardah} \sur{Aroosh Afzal}}\email{arooshw@gmail.com}
	
	\author*[2]{\fnm{Tooba} \sur{Feroze}}\email{tferoze@sns.nust.edu.pk}

	\affil[1,2]{\orgdiv{School of Natural Sciences}, \orgname{National University of Sciences and Technology}, \orgaddress{\street{H-12}, \city{Islamabad}, \country{Pakistan}}}
	
	\abstract{\noindent The main objective of this article is to analyze cylindrically symmetric static spacetime based on the Segre classification, where each Segre type corresponds to a set of canonical forms of Ricci spinors represented by Newman-Penrose scalars. The surviving Segre types of trace-free Ricci tensor in the cylindrical geometry are $[1,(111)]$, $[1,1(11)]$, $[(1,1)11]$, $[1,111]$, $[(1,11)1]$, $[(1,1)(11)]$ and $[(1,111)]$. Different combinations of eigenvalues and eigenvectors formed via Segre classification aid in understanding the distribution of matter and energy within that particular kind. Solutions are presented for the isotropic type $[1,(111)]$, for the anisotropic type $[1,1(11)]$, where a new metric function is introduced, and for type $[1,111]$, where all three pressures differ and the Krori-Barua functions are used. They satisfy the physical and stability criteria, namely the Zeldovich condition, causality condition, energy conditions and adiabatic index. The types $[(1,1)11]$ and $[(1,11)1]$ have $\rho=-P_r$ and give dark energy like configurations in which the strong energy condition is violated in the core, the latter having the tachyonic fluid form of the energy momentum tensor. For type $[(1,1)(11)]$, a charged solution in the presence of a non-null electromagnetic field is obtained and joined to an electrovac exterior. Matching with the Levi-Civita exterior shows that the vanishing of the radial pressure at the boundary is equivalent to the Darmois conditions. The C-energy and the curvature invariants are evaluated for each solution, and the results are compared with earlier works.}
	

	\maketitle

	\section{Introduction}
	\noindent In General Relativity (GR), the spacetime geometry is described by the Einstein equation. At present, finding the analytical solution remains a significant challenge due to the essential feature of the non-linearity of the field equation. Several approaches are taken to understand it deeply, the most important of which, still being worked out, is finding the exact solutions to the Einstein field equations. A substantial amount of work is done, considering spacetimes with specific symmetries, and many theoretical models obeying the physical situations are found along with the solutions that provide a great understanding of various gravitational phenomena. One of the vastly studied spacetime geometries is spherical symmetry; several authors have worked on and obtained models validating the physical criteria and complying with the astrophysical observations in GR, a few of which are \cite{0, 1, 2, 3, 4, 5}. Moreover, due to the limitations in GR, the studies are extended to different theories of gravity like $f(R)$, $f(R, T)$, $f(G)$, $f(T)$, $f(Q)$ etc. An extensive amount of literature is present on the solutions obtained while considering the different theories of gravity with authors considering different ansatz and equations of state \cite{6,7,8,9, n1, n3}. In addition, over the period, GR and gravitational theories have looked at dark energy and the expanding universe. The evidence obtained from measurements of numerous sources, such as supernova type Ia, large-scale structure, the Wilkinson microwave anisotropy probe, and cosmic microwave background radiation, has confirmed the present acceleration of the Universe \cite{10,11}. Acceleration of the universe can be explained by Dark Energy (DE) present in the universe, which is about $69\%$ of the total mass energy. Dark energy, which is nothing more than matter with negative pressure properties, is now thought to be the primary cause of the Universe's acceleration. Several authors have proposed various approaches and models to understand the nature of dark energy \cite{12, 13, 14, 15}.
	
	\noindent  Apart from spherical symmetry, a less encountered class of solutions is that with cylindrical symmetry, as it does not retain the same simplifications as spherical symmetry. These solutions are relevant when exploring rotating systems or infinitely long cylinders.  The general static vacuum solutions, including translational invariance, were studied by Levi-Civita \cite{15a} in 1919. Later, \cite{16, 17} obtained its extension considering the stationary spacetime. Exact solutions were analyzed and interpreted, representing the propagation of cylindrical gravitational waves \cite{18}. Interestingly, in the pivotal work on the non-linearity of gravitational waves, Einstein and Rosen \cite{19} rediscovered a similar class of solutions; besides this, a rigidly rotating infinitely long cylinder of dust was analyzed \cite{20}. Since then, cylindrically symmetric spacetime has been explored with diverse motives by many researchers, namely Bonnor, Weber and Wheeler \cite{21, 22}. Static perfect-fluid cylinders have been extensively studied in \cite{Evans,Bronnikov79,Kramer88,Davidson,BonnorDavidson,Philbin,Haggag,Bicak}. These studies demonstrate that the solutions admit an arbitrary generating function \cite{Kramer88,Philbin} and can be matched to a Levi-Civita exterior for $0\leq\sigma<1/2$ \cite{BonnorDavidson,Philbin}.

	\noindent A comprehensive review of cylindrical geometry has been done, focusing in detail on the solutions of field equations, where some key features like wormholes, gravitational Faraday rotation, polarization of gravitational waves, and gravitational collapse are examined \cite{23}. Using 1+3 formalism, a comprehensive analysis of the self-gravitating fluids in cylindrical spacetime is carried out through structure scalars, and all solutions to Einstein's equation can be written in these scalars for static cases \cite{25}. The non-static solution in cylindrical geometry having a charged perfect fluid is discussed, where the gravitational collapse occurs in the presence of an electromagnetic field \cite{26} and the end state of collapse is a conical singularity. 
	
	\noindent The cylindrically symmetric spacetime is meticulously studied in the modified theories of gravity. In $f(R)$ gravity \cite{32}, three families of exact solutions were explored, out of which two satisfied the physical criteria (energy density, pressure, etc.). In the $f(R,\phi, X)$ gravitational theory, six cases of exact solution are discussed where the violation of null energy conditions occurred, signifying the existence of cylindrical wormholes \cite{33}. Furthermore, \cite{n37} investigated cylindrically symmetric solutions in $f(R,G)$ gravity and discovered six families of exact solutions with a null energy condition analysis that revealed potential wormhole regions. Analyzing anisotropic perfect fluid solutions, \cite {n38} investigated static, cylindrically symmetric spacetimes under the framework of $f(Q)$ gravity. It reveals that the Levi-Civita solution violates energy conditions in the power law function of $f(Q)$ but satisfies them in limited regions for the exponential function, whereas cosmic strings violate energy criteria in both functions. Additionally, the cylindrical geometry has been taken into consideration in other modified theories by several researchers \cite{27, 28, 29, 30, 31}. More recently, general exact solutions for cylindrically symmetric stiff-fluid sources were obtained in \cite{HLM}. Anisotropic cylindrical configurations have also been investigated in the context of $f(R,G)$ and $f(R,L_m,T)$ gravity, including the complexity of the matter distribution \cite{RYJ,RYRa} and the effect of an electromagnetic field \cite{RYR}. In addition, the Segre type of the Ricci and energy-momentum tensors has been used to construct exact solutions in a recently proposed theory of gravity \cite{SN}. The present work builds on these studies by classifying static interior configurations according to the algebraic type of their matter source. We further examine their matching to the Levi-Civita vacuum, which allows the integration constants appearing in the solutions to be interpreted in terms of physically meaningful quantities.
	
	\noindent This article aims to analyze cylindrically symmetric static spacetime based on the Segre classification. As this is the classification of rank two tensors, it provides a clear understanding of the underlying structure of equations.  Moreover, it is an effective tool that helps in obtaining the exact solutions of field equations by guiding appropriate coordinate choice, reducing variables, and improving the numerical and analytical methods. The earlier classification of spherically symmetric spacetime is studied in GR and $f(R)$ theory of gravity by \cite{34} and \cite{35}. Here, we will look into the cylindrical geometry and find the exact solutions of the corresponding Segre types. In addition, the physical and stability conditions like the behavior of density, pressure, speed of sound, etc., will be analyzed to determine whether they are admissible with this geometry, and a detailed graphical analysis will be carried out for this purpose. We further match each interior solution to the Levi-Civita vacuum exterior and calculate its C-energy, thereby assigning a clear physical interpretation to the constants appearing in each model. 
	
	\noindent The breakdown of the paper is as follows: In Section \ref{S2}, the corresponding Segre types for cylindrically symmetric static metric are obtained by defining the NP scalars. All the Segre types are thoroughly analyzed, and solutions are proposed. Firstly, corresponding to Segre type $[1,(111)]$, the isotropic fluid distribution is considered, and a solution is obtained. For Segre type $[1,1(11)]$, an anisotropic fluid is considered, and a class of solutions is produced by introducing a new metric function. The presence of dark energy is noticed owing to negative pressure when looking for a solution to type $[(1,1)11]$. In addition, a class of solutions for type $[1,111]$ is produced by considering principal stresses in the radial, axial, and azimuthal directions. For type $[(1,11)1]$, we obtain a regular solution whose energy-momentum tensor corresponds to a tachyonic fluid. In contrast, type $[(1,1)(11)]$ leads to a charged anisotropic solution coupled to an electromagnetic field.  Furthermore, for Segre type $[(1,111)]$ a brief analysis is provided, highlighting previous work. Section \ref{S8} is devoted to the matching conditions with the Levi-Civita exterior, the calculation of C-energy, and a comparative analysis of the resulting solution types. Final remarks are presented in Section \ref{S7}.
	
	\section{Segre Classification of Spacetime Geometry with Cylindrical Symmetry} \label{S2}
	
	\noindent The general spacetime metric that describes the cylindrical symmetric static geometry is given as
	\begin{equation}\label{1}
		ds^2=-e^{\nu(r)}dt^2+e^{\lambda(r)}dr^2+e^{\xi(r)}d\theta^2+e^{\mu(r)}dz^2,
	\end{equation} 
	where $\nu(r)$, $\lambda(r)$, $\xi(r)$ and $\mu(r)$ are the unknown functions. In correspondence to the standard interior metric having spherically symmetric spacetime geometry, the coordinate $r$ is defined such that the coefficient of $d\theta^2$ becomes equivalent to $r^2$. Thus the metric in \eqref{1} takes the form
	\begin{equation}\label{2}
		ds^2=-e^{\nu(r)}dt^2+e^{\lambda(r)}dr^2+r^2d\theta^2+e^{\mu(r)}dz^2.
	\end{equation}
	The Einstein field equation is defined as 
	\begin{equation}\label{3}
		G_{ab}= \mathcal{R}_{ab}-\frac{1}{2}g_{ab}\mathcal{R}=\kappa T_{ab},
	\end{equation}
	where $\kappa$ is the gravitational constant (set to unity in the expressions that follow), $\mathcal{R}_{ab}$ is the Ricci tensor and its nonzero components for the metric \eqref{2} are given as
	\begin{align}
		\mathcal{R}_{00}&=\frac{e^{\nu-\lambda}}{4}\left(2\nu''+\nu'^2-\nu'\lambda'+\nu'\mu'+\frac{2\nu'}{r}\right)\label{4},
		\\
		\mathcal{R}_{11}&=\frac{1}{4}\left(-2\mu''-2\nu''-\mu'^2-\nu'^2+\mu'\lambda'+\nu'\lambda'+\frac{2\lambda'}{r}\right)\label{5},
		\\
		\mathcal{R}_{22}&=\frac{re^{-\lambda}}{2}\left(\lambda'-\nu'-\mu'\right)\label{6},
		\\
		\mathcal{R}_{33}&=\frac{e^{\mu-\lambda}}{4}\left(-2\mu''-\mu'^2+\mu'\lambda'-\mu'\nu'-\frac{2\mu'}{r}\right)\label{7},
	\end{align}	
	and the Ricci scalar becomes  
	\begin{equation}\label{8}
		\mathcal{R}=\frac{-e^{-\lambda}}{2}\left(2\mu''+2\nu''+\mu'^2+\nu'^2+\mu'\nu'-\lambda'(\nu'+\mu')+\frac{2}{r}\left(\nu'+\mu'-\lambda'\right)\right).
	\end{equation}
	By equations \eqref{4}-\eqref{8} the non zero components of the Einstein tensor on the left side of field equation \eqref{3} are expressed as
	\begin{align}
		G_{00}&=\frac{e^{\nu-\lambda}}{4}\left(-2\mu''-\mu'^2+\mu'\lambda'+\frac{2}{r}(\lambda'-\mu')\right),\label{9}
		\\
		G_{11}&=\frac{1}{4}\left(\mu'\nu'+\frac{2}{r}(\mu'+\nu')\right)\label{10},
		\\
		G_{22}&=\frac{r^2e^{-\lambda}}{4}\left(2\mu''+2\nu''+\mu'^2+\nu'^2+\mu'\nu'-\mu'\lambda'-\nu'\lambda'\right)\label{11},
		\\
		G_{33}&=\frac{e^{\mu-\lambda}}{4}\left(2\nu''+\nu'^2-\nu'\lambda'+\frac{2}{r}(\nu'-\lambda')\right)\label{12}.
	\end{align}	
	\noindent The right side of field equation \eqref{3} describes the distribution of matter and energy in spacetime, which is an essential part that determines the spacetime curvature and, moreover, leads to different solutions of field equations. As the form of energy momentum tensor (EMT) is a crucial aspect to be considered, the most reasonable and systematic approach to determine the distribution of matter and energy in a specific spacetime is the Segre classification scheme. In this scheme, each Segre type is assigned to the symmetric tensor that provides the algebraic structure of the tensor field. Several authors in GR earlier discussed the algebraic properties of trace-free Ricci tensors based on these Segre types and categorized them into different classes. McIntosh \textit{et al} \cite{36} gave the classification of Ricci tensor where for each Segre type they gave a set of canonical forms of Ricci spinors using the well-known notation for the Ricci spinor by Newman and Penrose, i.e., Newman-Penrose (NP) scalars $\phi_{ab}$.

	\noindent For cylindrically symmetric static spacetime geometry given by \eqref{2} the non-zero components of the NP scalars are 
	\begin{align}
		\phi_{00}&= \phi_{22}=\frac{e^{-\lambda}}{16}\left(-2\mu''-\mu'^2+\mu'(\nu'+\lambda')+\frac{2}{r}\left(\nu'+\lambda'\right)\right)\label{13}
		\\
		\phi_{02}&=\phi_{20}=\frac{e^{-\lambda}}{16}\left(2\mu''+\mu'^2+\mu'(\nu'-\lambda')+\frac{2}{r}\left(\lambda'-\nu'\right)\right)\label{14}
		\\
		\phi_{11}&=\frac{e^{-\lambda}}{16}\left(	2\nu''+\nu'^2-\nu'\lambda'-\frac{2\mu'}{r}\right)\label{15}
	\end{align}
	
	\noindent The aforementioned non-zero NP scalars result in various Segre types, each associated with specific canonical forms. These forms impose constraints on cylindrical geometry by standardizing the Ricci tensor, thereby determining the possible shapes and physical properties of cylindrical spacetimes. The obtained Segre types corresponding to the non-zero NP scalars for cylindrically symmetric static spacetime metric \eqref{2} are given in Table \ref{T1}.
	\begin{center}
		\begin{table}[ht]
			\centering
			\setlength{\tabcolsep}{8pt}
			\renewcommand{\arraystretch}{1.5}
			\begin{tabular}{|c|c|c|}
				\hline
				Sr No. &Segre Characteristic & Non-zero $\phi_{\alpha\beta}$
				\\  \hline
				I& $[1,(111)]$ & $\phi_{00}=\phi_{22}=2\phi_{11}$
				\\ \hline
				II& $[1,1(11)]$ & $\phi_{00}=\phi_{22}, \phi_{11}$    
				\\ \hline
				III& $	[(1,1)11]$  & $\phi_{11}, \phi_{02}=\phi_{20}$ 
				\\ \hline
				IV& $	[1,111]$ & $\phi_{00}=\phi_{22}, \phi_{11}, \phi_{02}=\phi_{20}$  
				\\ \hline
				V& $[(1,11)1]$ & $\phi_{02}=\phi_{20}=2\phi_{11}$
				\\ \hline
				VI& $[(1,1)(11)]$	& $ \phi_{11}$    
				\\ \hline
				VII& $[(1,111)]$ &    $-$ 
				\\ \hline
			\end{tabular}
			\caption{Frame of Segre types of trace-free Ricci tensor for metric \eqref{2}}\label{T1}
		\end{table}
	\end{center}
	\noindent By analyzing the non-zero NP scalars and their corresponding Segre types and canonical forms, we can understand the limitations and characteristics of cylindrical spacetime geometries. Now in the framework of GR, we will explore solutions considering an appropriate EMT corresponding to the Segre types. Moreover, the physical and stability criteria will be analyzed in the subsequent sections, whether the solutions obey the physically admissible conditions, stated as
	\begin{itemize}
		\item The metric potentials must be singularity-free.
		
		\item Energy density and pressures must be positive and decreasing towards the boundary $(r=R)$, whereas pressure in the presence of dark energy must be negative. Throughout, $r$ is measured in units of the boundary radius, so $R=1$ in the constants that come out of the boundary condition.
		
		\item The radial pressure must inevitably vanish at $(r=R)$, and the pressures in radial and azimuthal direction must be equal at origin i.e. $(P_r=P_\theta)|_{r=0}$
		
		\item The density and pressure gradients must necessarily be negative. 
		
		\item The Zeldovich \cite{38} criterion needs to be satisfied, i.e. $\left(\frac{P_\alpha}{\rho}\right)|_{r=0}\le1$.
		
		\item The profile of surface and gravitational redshift must increase and decrease respectively, wherever they are defined.
		
		\item For the stability of the system, the causality condition, energy conditions like Null Energy Condition (NEC), Weak Energy Conditions (WEC), Strong Energy Conditions (SEC), Dominant Energy Conditions (DEC) and adiabatic index need to be satisfied essentially.
	\end{itemize}
	
	\section*{I:\ \ Segre Type $[1,(111)]$}\label{S3}
	
	\noindent The Segre type $[1,(111)]$ has a unique eigenvalue corresponding to a timelike eigenvector, whereas the eigenvectors in spatial dimension have similar eigenvalues. Thus, this reflects isotropy, indicating the fluid's uniformity in spatial directions, so the considered EMT in this case is that of a perfect fluid, in general given as
	\begin{equation}
		T^{\ \ (pf)}_{ab}=(\rho+P)u_au_b+Pg_{ab},\label{I1}
	\end{equation}
	here $\rho$ is the energy density and $P$ is the pressure.
	Thus the field equations \eqref{9}-\eqref{12} by taking into account the isotropic distribution of matter become \begin{align}
		\rho&=\frac{e^{-\lambda}}{4}\left(-2\mu''-\mu'^2+\mu'\lambda'+\frac{2}{r}(\lambda'-\mu')\right)\label{I2},
		\\
		P&=\frac{e^{-\lambda}}{4}\left(\mu'\nu'+\frac{2}{r}(\mu'+\nu')\right),\label{I3}
		\\
		P&=\frac{e^{-\lambda}}{4}\left(2\mu''+2\nu''+\mu'^2+\nu'^2+\mu'\nu'-\mu'\lambda'-\nu'\lambda'\right)\label{I4},
		\\
		P&=\frac{e^{-\lambda}}{4}\left(2\nu''+\nu'^2-\nu'\lambda'+\frac{2}{r}(\nu'-\lambda')\right)\label{I5}.
	\end{align}
	
	\noindent The corresponding non-zero NP scalars for this Segre type are 
	\begin{equation}\label{I6}
		\phi_{00}=2\phi_{11}=\phi_{22},
	\end{equation} 
	which leads to the constraint equations that the solution needs to satisfy, given as 
	\begin{align}
		-2\mu''-\mu'^2+\mu'(\lambda'+\nu')+\frac{2}{r}\left(\nu'+\lambda'\right)&\ne0\label{I7},
		\\
		2\nu''+\nu'^2-\nu'\lambda'-\frac{2\mu'}{r}&\ne0\label{I8}.
	\end{align}
	
	\noindent Moreover, in this case the two equations obtained by $\phi_{02}=0$ and $\phi_{00}-2\phi_{11}=0$ are given as 
	\begin{align}
		2\mu''+\mu'^2+\mu'(\nu'-\lambda')+\frac{2}{r}\left(\lambda'-\nu'\right)&=0\label{I9},
		\\
		-2\mu''-\mu'^2-4\nu''-2\nu'^2+\mu'\nu'+\mu'\lambda'+2\nu'\lambda'+\frac{2}{r}\left(\nu'+\lambda'+2\mu'\right)&=0\label{I10}.
	\end{align}
	
	\noindent The equations \eqref{I9} and \eqref{I10} are also clearly obtained by simplifying the equations \eqref{I3}-\eqref{I5}. For finding the solution we have four linearly independent equations and five unknowns $(\nu,\lambda,\mu,\rho, P)$, so one ansatz is considered
	\begin{equation}\label{I11}
		\mu=\ln(1+a_1r^2).
	\end{equation}
	Substituting $\mu$ in equations \eqref{I9} and \eqref{I10} and further solving the differential equation yields the other two metric functions in this case, given as 
	\begin{align}
		\nu&=\frac{b_1r^2}{2}+c_0\label{I12},
	\end{align}
	\begin{align}
		\lambda&=\frac{b_1r^2}{2}-\ln(1+a_1r^2)+c_1\label{I13}.
	\end{align}

	\noindent The obtained metric potentials  are singularity free, substituting them in equations \eqref{I2}-\eqref{I5} the expression of energy density and pressure become
	\begin{align}
		\rho&=\frac{\exp\left(\frac{-b_1r^2-2c_1}{2}\right)\left(-6a_1+b_1+ 2a_1b_1r^2\right)}{2},\label{I14}
		\\
		P&=\frac{\exp\left(\frac{-b_1r^2-2c_1}{2}\right)\left(2a_1+b_1+ 2a_1b_1r^2\right)}{2}.\label{I15}
	\end{align}
	Since pressure must vanish at the boundary i.e. $P|_{(r=R)}=0$, it gives the expression of the constant parameter $b_1$ as
	\begin{equation}
		b_1=\frac{-2a_1}{1+2a_1}\label{I16}.
	\end{equation}
	\begin{figure}
		\centering
		\begin{minipage}[t]{0.31\textwidth}\subfloat[]	
			{\gfx{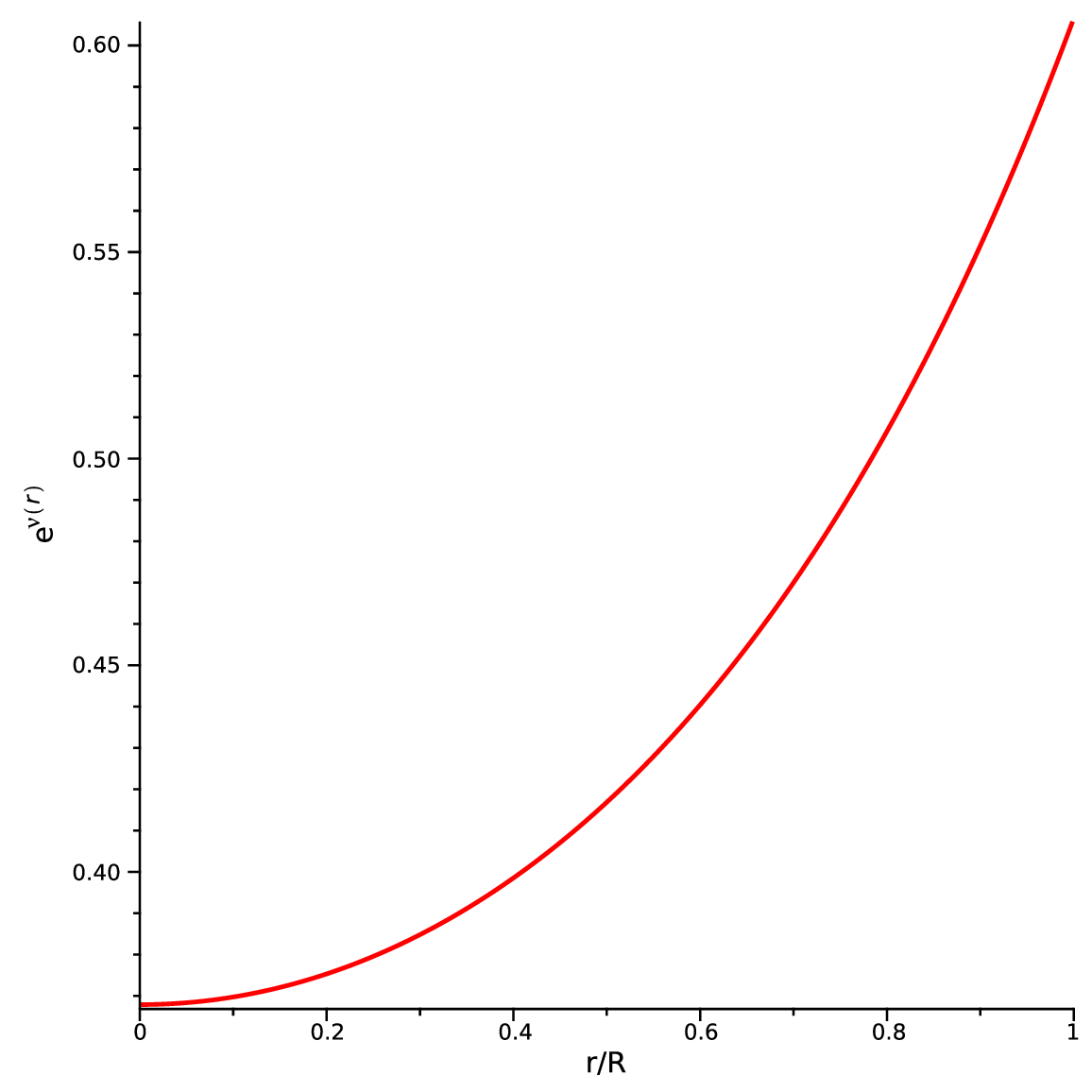}}
		\end{minipage}
		\hfill
		\begin{minipage}[t]{0.31\textwidth}\subfloat[]	
			{\gfx{IMp2.eps}}
		\end{minipage}
		\hfill
		\begin{minipage}[t]{0.31\textwidth}\subfloat[]	
			{\gfx{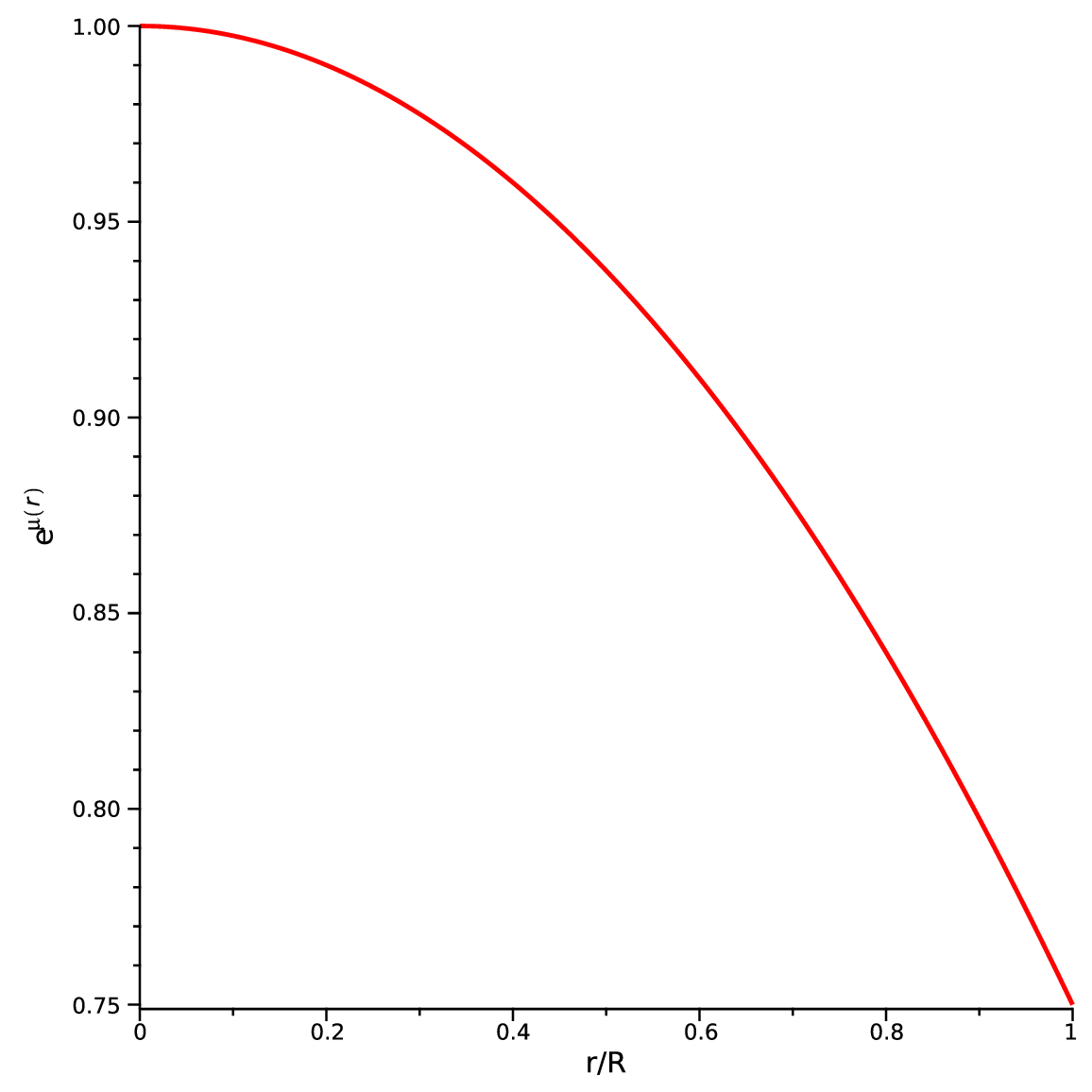}}
		\end{minipage}
		
		\caption{Graph of metric potentials for $a_1=-0.25$, $b_1=1$ and $c_0=-1$ }\label{fig.I1}
	\end{figure}
	
	\begin{figure}
		\centering
		\begin{minipage}[t]{0.31\textwidth}
			\centering
			{\gfx{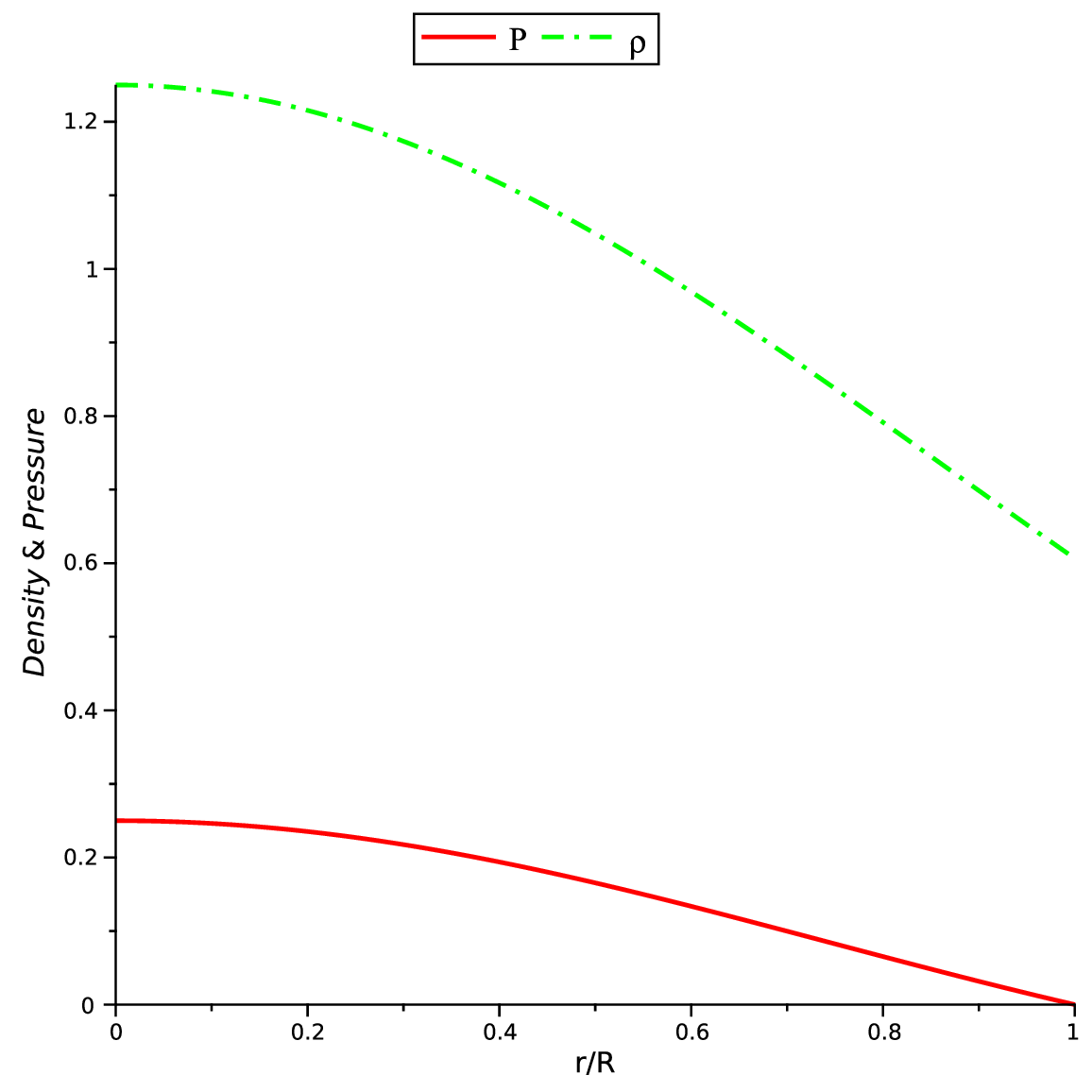}}
			\caption{Graph of energy density and pressure for $a_1=-0.25$, and $b_1=1$}
			\label{fig.I2}
		\end{minipage}
		\hfill
		\begin{minipage}[t]{0.31\textwidth}
			\centering
			{\gfx{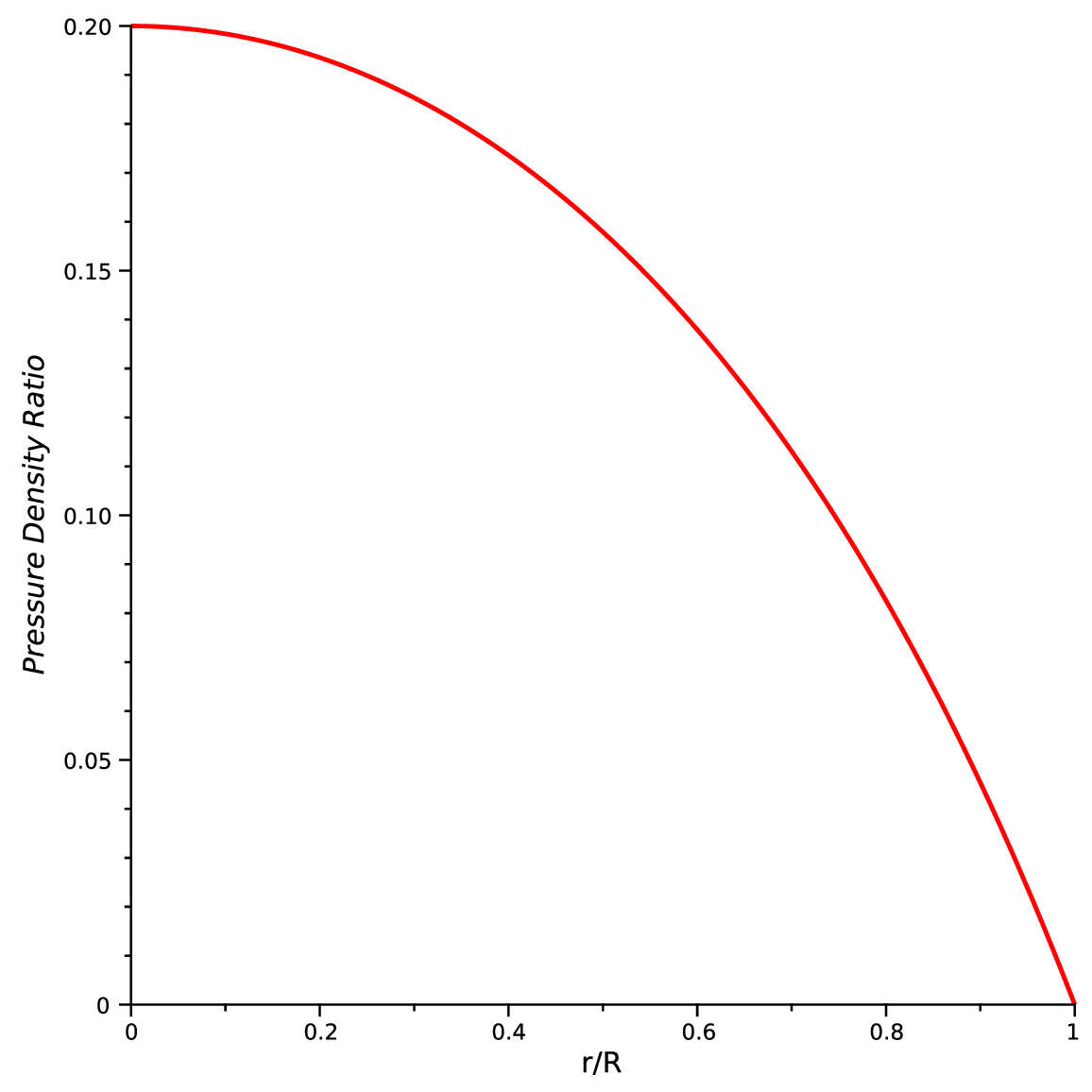}}
			\caption{Graph of pressure density ratio for $a_1=-0.25$ and $b_1=1$}
			\label{fig.I4}
		\end{minipage}
		\hfill
		\begin{minipage}[t]{0.31\textwidth}
			\centering
			{\gfx{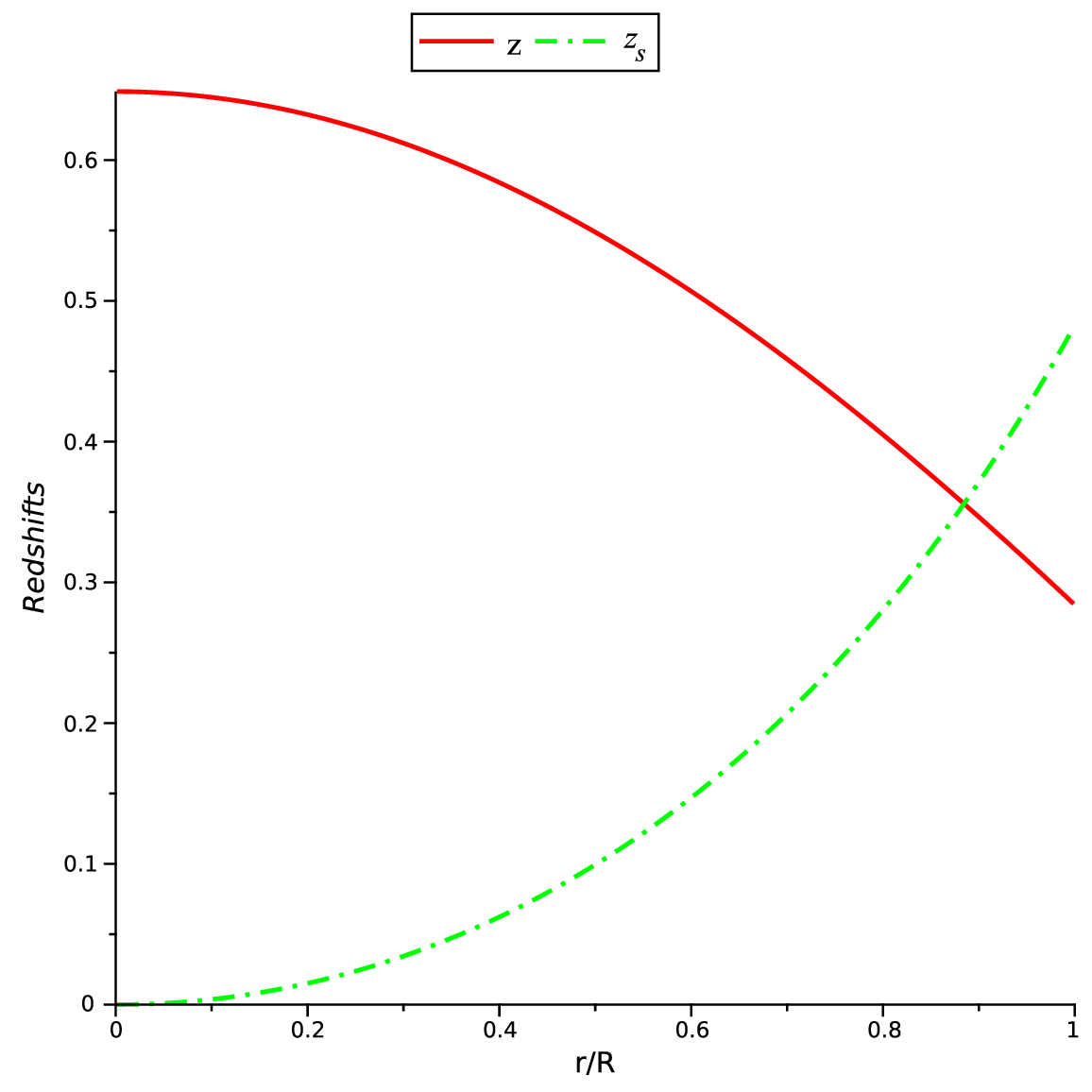}}
			\caption{Graph of redshifts for $a_1=-0.25$ and $b_1=1$}
			\label{fig.I5}
		\end{minipage}
	\end{figure}
	
	\noindent As the metric function $\left(e^{\lambda}\right)|_{r=0}=1$ implies the constant of integration $c_1$ to be zero, also $e^\nu$ at the center must be equal to some positive constant, so consider $a_1=-0.25$, which yields $b_1=1$ by equation \eqref{I16}. As a consistency check, letting $a_1\to0$ in equation \eqref{I16} gives $b_1\to0$. Then $\rho\to0$, $P\to0$, and the metric reduces to flat spacetime. The profiles of metric potentials for these constants are shown in Fig. \ref{fig.I1} adhering to the regularity condition. 
	
	\noindent Moreover, the behavior of density and pressure is illustrated in Fig. \ref{fig.I2}, both decrease when moving towards the boundary, and $P$ vanishes.

	\noindent The decreasing nature of density and pressure can be verified by taking the first derivative of equations \eqref{I14} and \eqref{I15} which yields the expression of gradients as
	\begin{align}
		\frac{d\rho}{dr}&=-rb_1\exp\left(\frac{-b_1r^2-2c_1}{2}\right)\left(\frac{-10a_1+b_1+ 2a_1b_1r^2}{2}\right),
		\\
		\frac{dP}{dr}&=-rb_1\exp\left(\frac{-b_1r^2-2c_1}{2}\right)\left(\frac{-2a_1+b_1+ 2a_1b_1r^2}{2}\right).
	\end{align}

	\noindent The negative sign in expressions ensures the decreasing behavior. To validate the regularity condition, the Zeldovich \cite{38} criterion must be fulfilled, that is, the pressure density ratio must be less than $1$, positive and continuous throughout as illustrated in Fig. \ref{fig.I4}. 
	
\begin{figure}
		\centering
		\begin{minipage}[t]{0.46\textwidth}
			\centering
			\subfloat[]
			{\gfx{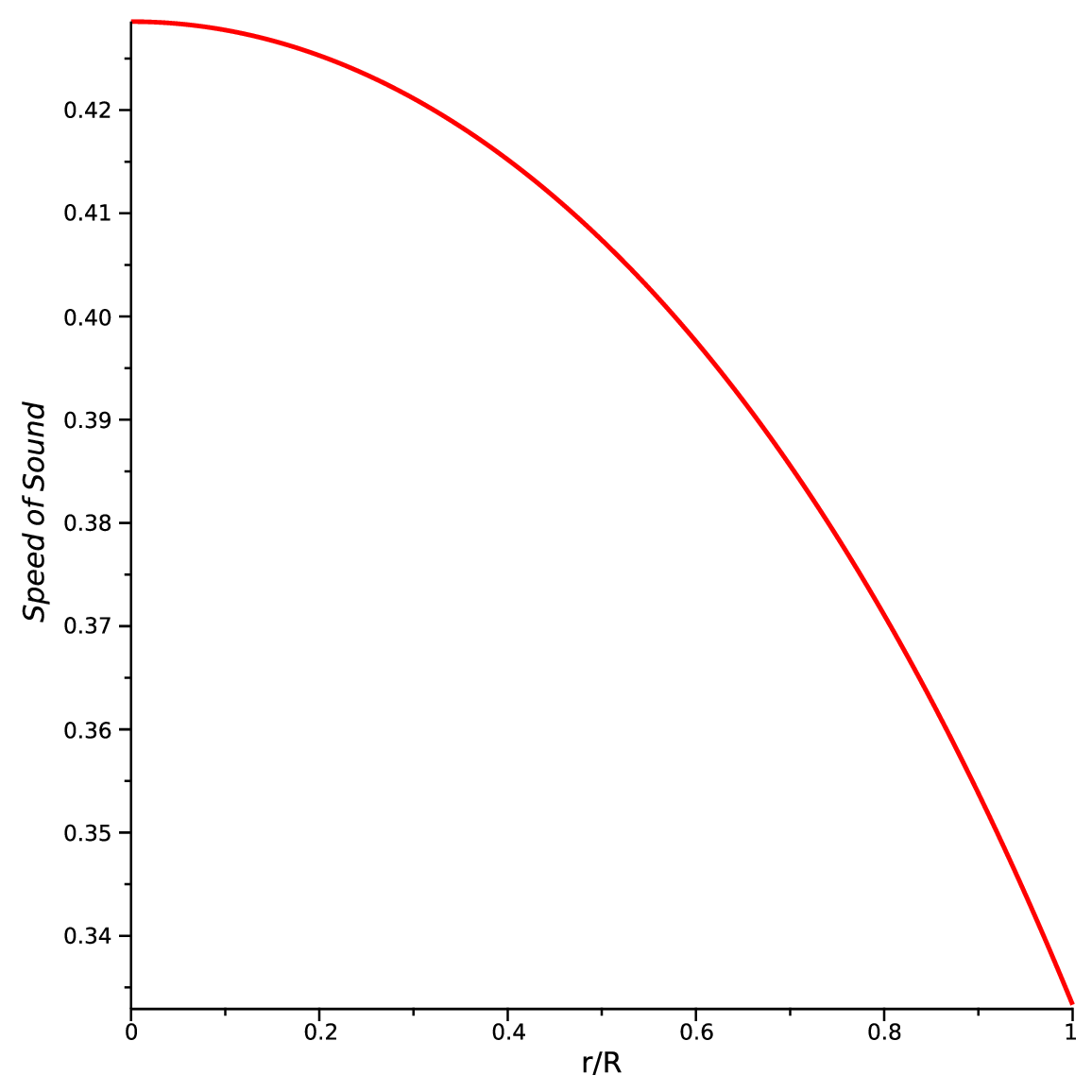}}
		\end{minipage}
		\hfill
		\begin{minipage}[t]{0.46\textwidth}
			\centering
			\subfloat[]
			{\gfx{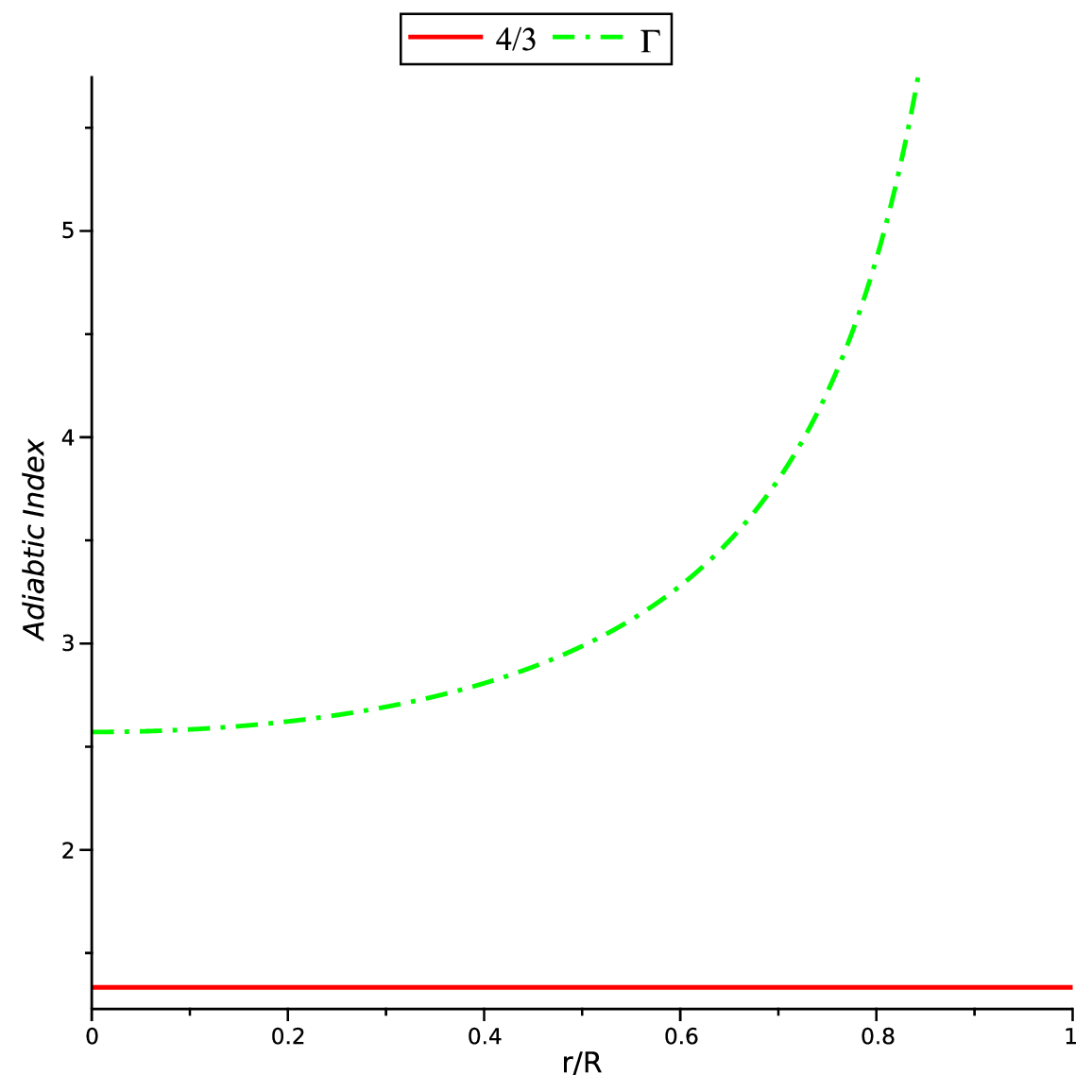}}
		\end{minipage}
		
		\caption{Graphs of stability conditions for $a_1=-0.25$ and $b_1=1$} \label{fig.I6}
	\end{figure}

	\noindent Moving onto the important phenomenon of redshifts, where the overall mass and radius of a star affect the surface redshift, and gravitational redshift is directly tied to the strength of the gravitational field. The gravitational and surface redshifts are defined as
	\begin{align}
		z&=e^{\frac{-\nu}{2}}-1,
		\\
		z_s&=e^{\frac{\lambda}{2}}-1.
	\end{align}  
	For this type, the redshifts turn out to be
	\begin{align}
		z&=\exp\left(\frac{-b_1r^2}{4}-\frac{c_0}{2}\right)-1,
		\\
		z_s&=\frac{\exp\left(\frac{b_1r^2}{4}+\frac{{c_1}}{2}\right)-\sqrt{1+a_1r^2}}{\sqrt{1+a_1r^2}},
	\end{align}  
	\noindent and the profiles are shown in Fig. \ref{fig.I5}.
	\noindent Now for the stability conditions, we first check the causality condition, i.e. the speed of sound is less than the speed of light, where in general the speed of sound is defined as 
	\begin{equation}
		v_\alpha^2=\frac{dP_\alpha}{d\rho}.
	\end{equation}
	For this Segre type, the speed of sound becomes 
	\begin{equation}
		v^2=\frac{dP}{d\rho}=\frac{2a_1b_1r^2-2a_1+b_1}{2a_1b_1r^2-10a_1+b_1},
	\end{equation}
	the graph of which is given in Fig. \ref{fig.I6}$(a)$ fulfilling the essential requirement. 
	
	\noindent For a physically reasonable solution, the energy conditions must be satisfied; without them solutions could enable superluminal energy flows, negative densities, or some non-physical attributes. So for a system to be stable and physically realistic the following energy conditions \cite{HawkingEllis,Visser97} must be satisfied in the isotropic distribution of matter 
	\begin{itemize}
		\item NEC:\ $\rho + P\ge 0$,
		\item WEC:\ $\rho \ge 0$ and $\rho + P\ge 0$,
		\item SEC: \ \ $\rho + 3P \ge 0$,
		\item DEC:\ $\rho \ge 0$ and $\rho \ge|P|$.
	\end{itemize}
	\noindent Because $\rho$ and $P$ stay positive across the interior (Fig. \ref{fig.I2}), NEC, WEC and SEC hold directly. DEC follows from the pressure-density ratio staying below $1$ (Fig. \ref{fig.I4}). So the essential requirements for all energy conditions are fulfilled and the mathematical solution is physically reasonable and stable in cylindrical geometry. For a stable configuration, the adiabatic index is another important parameter defined as the ratio of two specific heats \cite{39}  
	\begin{equation}
		\Gamma= \left(\frac{\rho+P}{P}\right)\left(\frac{dP}{d\rho}\right).
	\end{equation}
	\noindent The stability is achieved when $\Gamma>4/3$, in this case $\Gamma$ is given as
	\begin{equation}
		\Gamma=\frac{2\left(-2a_1+b_1+2a_1b_1r^2\right)^2}{\left(2a_1b_1r^2-10a_1+b_1\right)\left(2a_1b_1r^2+2a_1+b_1\right)}.
	\end{equation}
	Fig. \ref{fig.I6}$(b)$ depicts that the condition is satisfied. It is worth noting that $\Gamma>4/3$ is the usual stability criterion for spherical configurations. In the case of an infinitely long cylinder, however, the Newtonian critical value for radial perturbations is $\Gamma=1$ \cite{CF}. Thus, the condition $\Gamma>4/3$, adopted here and throughout the following analysis, represents a more restrictive stability requirement. The isotropic model constructed under Segre type $[1,(111)]$ satisfies all the required physical acceptability and stability conditions. In comparison with earlier studies of perfect-fluid cylinders \cite{Evans,Kramer88,BonnorDavidson,Philbin,Haggag}, the present solution corresponds to the particular choice of the generating function given in Eq.~\eqref{I11}. The resulting configuration does not satisfy a linear equation of state of the form $P=w\rho+c$. Indeed, Eqs.~\eqref{I14} and \eqref{I15} give
	\[
	\rho-P=-4a_1\exp\left(-\frac{b_1r^2+2c_1}{2}\right).
	\]
	Thus, although $\rho$ and $P$ share the same Gaussian factor, their polynomial prefactors in $r^2$ differ. Consequently, a linear equation of state can be satisfied only in the special case $a_1=0$.

	\section*{II:\ \ Segre Type $[1,1(11)]$} \label{S4}
	
	\noindent We will derive the solution using the EMT corresponding to the Segre type $[1,1(11)]$. In this arrangement, a similar eigenvalue is shared by two eigenvectors in the spatial directions and the third eigenvector has a different eigenvalue along with a timelike eigenvector. This configuration between the timelike and spatial components highlights the inimitable behavior and relation for the EMT i.e. corresponding EMT here is anisotropic fluid given as 
	\begin{equation}\label{II1}
		T^{\ \ (apf)}_{ab
		}=(\rho+P_\theta)u_au_b+P_\theta g_{ab}+(P_r-P_\theta)v_av_b.
	\end{equation}
	Here $P_r$ and $P_\theta$ are radial and azimuthal pressures respectively. Thus the non-zero components of EMT are 
	\begin{equation}\label{II2}
		T^a_{b}= \mathrm{diag}(-\rho,P_r,P_\theta,P_\theta).
	\end{equation}
	\begin{figure}
		\centering
		\begin{minipage}[t]{0.31\textwidth}\subfloat[]	
			{\gfx{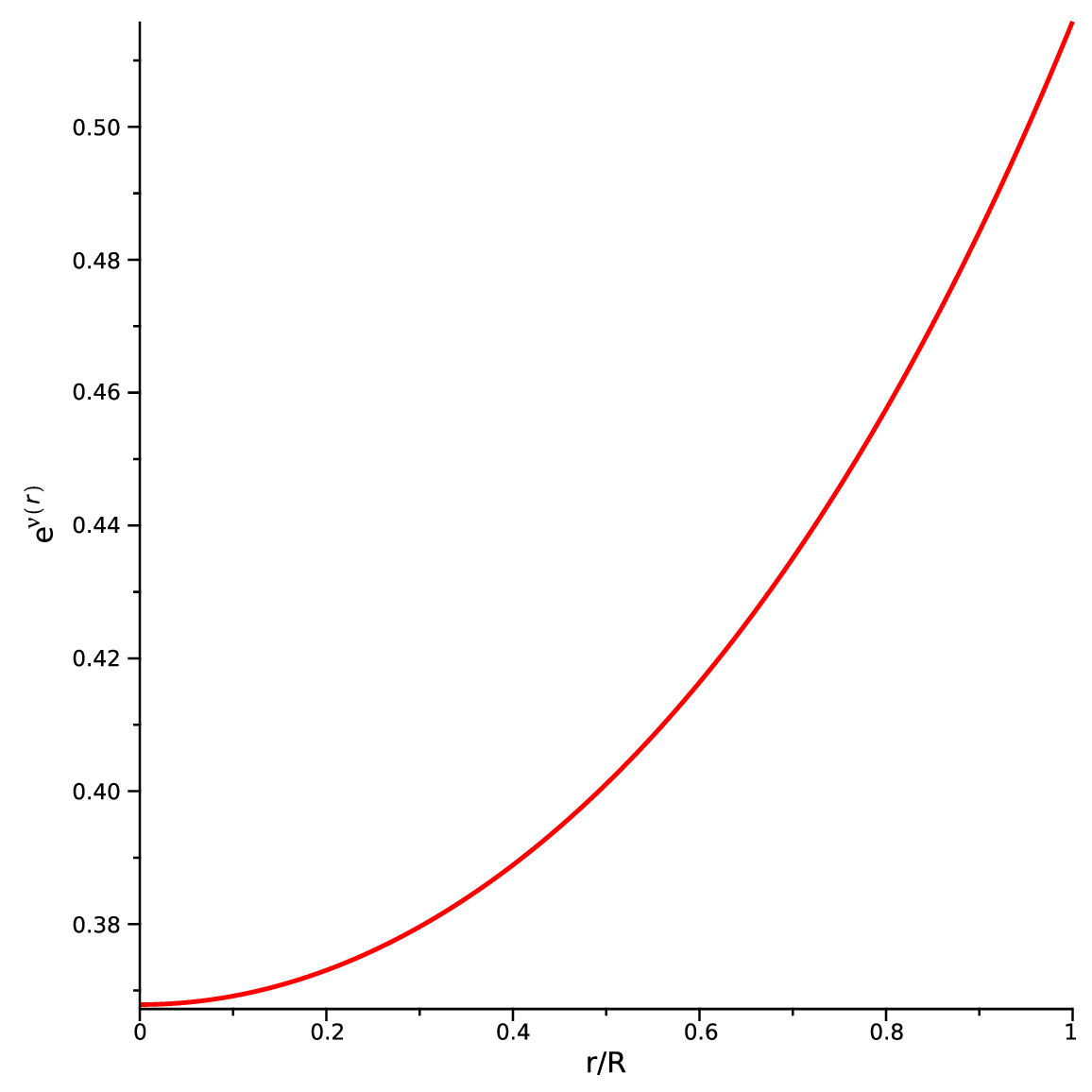}}
		\end{minipage}
		\hfill
		\begin{minipage}[t]{0.31\textwidth}\subfloat[]	
			{\gfx{IIMp2.eps}}
		\end{minipage}
		\hfill
		\begin{minipage}[t]{0.31\textwidth}\subfloat[]	
			{\gfx{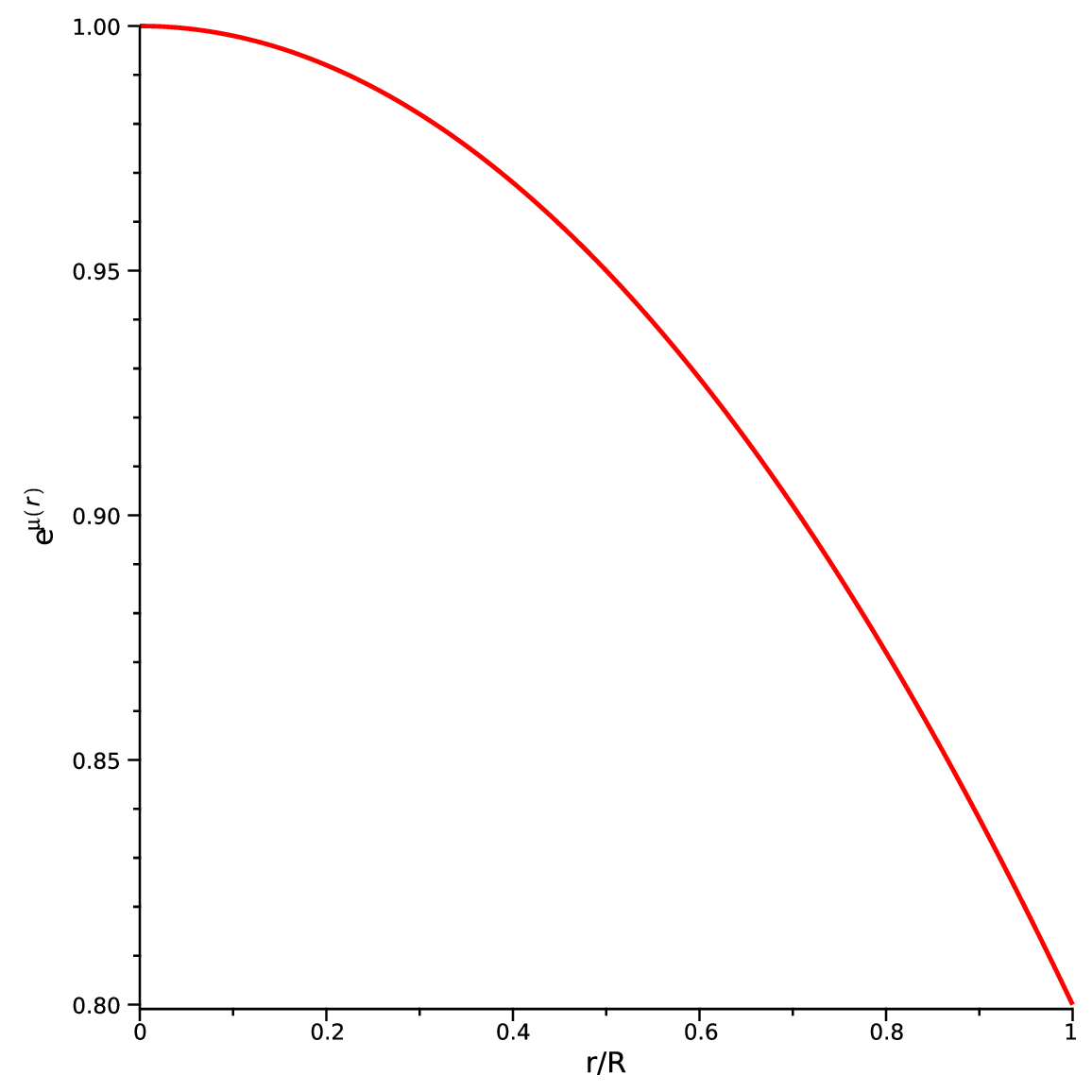}}
		\end{minipage}
		
		\caption{Graph of metric potentials for $a_2=0.5501411418$, $b_2=-0.2$ and $c_2=-1$}
		\label{fig.II1}
	\end{figure}\noindent The field equations \eqref{9}-\eqref{12} by taking into account the anisotropic distribution become 
	\begin{align}
		\rho&=\frac{e^{-\lambda}}{4}\left(-2\mu''-\mu'^2+\mu'\lambda'+\frac{2}{r}(\lambda'-\mu')\right)\label{II3},
		\\
		P_r&=\frac{e^{-\lambda}}{4}\left(\mu'\nu'+\frac{2}{r}(\mu'+\nu')\right)\label{II4},
		\\
		P_\theta&=\frac{e^{-\lambda}}{4}\left(2\mu''+2\nu''+\mu'^2+\nu'^2+\mu'\nu'-\mu'\lambda'-\nu'\lambda'\right)\label{II5},
		\\
		P_\theta&=\frac{e^{-\lambda}}{4}\left(2\nu''+\nu'^2-\nu'\lambda'+\frac{2}{r}(\nu'-\lambda')\right)\label{II6}.
	\end{align}
	
	\noindent The corresponding non-zero NP scalars for this Segre type are 
	\begin{equation}\label{II7}
		\phi_{00}=\phi_{22}\ne0, \hspace{1cm}\phi_{11}\ne0,
	\end{equation} 
	which leads to the constraint equations \eqref{I7} and \eqref{I8}. Here, $\phi_{02}=0$ as given by \eqref{I9} which is also obtained by simplifying equations \eqref{II5} and \eqref{II6}. For finding the solution here we have four linearly independent equations and six unknowns $(\nu,\lambda,\mu,\rho, P_r, P_\theta)$. There are two arbitrary choices present in this case, so we introduce a new form of function that would yield a new metric potential. Moreover, this function is required to fulfill the criteria mentioned in Section \ref{S2}, so that the metric potential adheres to the Segre type and gives a reasonable solution. New function $\lambda$ along with $\mu$ is defined as 
	
	\begin{align}
		\lambda&= a_2r^2\cosh(b_2r^2),\label{II8}
		\\
		\mu&=\ln(1+b_2r^2).\label{II9}
	\end{align} 
	
	\noindent Substituting these potentials in \eqref{I9} and solving the differential equation leads to 
	\begin{equation}
		\nu= \ln(1+b_2r^2)+a_2r^2\cosh(b_2r^2)+c_2. \label{II10}
	\end{equation}
	Using these metric potentials in equations \eqref{II3}-\eqref{II6}, we get the expressions of density, radial and tangential pressure given as
	\begin{align} \label{II11}
		\rho&=\left(\frac{2\exp\left(-a_2r^2\cosh\left(b_2r^2\right)\right)}{(1+b_2r^2)^2}\right)\left(a_2\left(b_2r^2+\frac{1}{2}\right)\left(1+b_2r^2\right)\cosh(b_2r^2)\right.\nonumber
		\\
		&\left.\hspace{1.75cm}+b_2\left(-\frac{b_2r^2}{2}+a_2\left(b_2r^2+\frac{1}{2}\right)\left(1+b_2r^2\right)r^2\sinh\left(b_2r^2\right)-1\right)\right),
		\\
		P_r&=\left(\frac{2\exp\left(-a_2r^2\cosh(b_2r^2)\right)}{\left(1+b_2r^2\right)^2}\right)\left(a_2\cosh(b_2r^2)(1+b_2r^2)\left(b_2r^2+\frac{1}{2}\right)\right.\nonumber
		\\
		&\left. \hspace{2.2cm}+b_2\left(\frac{3b_2r^2}{2}+1+ a_2r^2\sinh(b_2r^2)(1+b_2r^2)\left(b_2r^2+\frac{1}{2}\right)\right)\right),		
		\\
		P_\theta&=\left(\frac{2\exp\left(-{a_2}r^2\cosh\left(b_2r^2\right)\right)}{\left(1+b_2r^2\right)^2}\right)\left(a_2(1+b_2r^2)\cosh\left(b_2r^2\right)\left(b_2^3r^6+b_2^2r^4\right.\right.\nonumber
		\\ 
		&\left.\left. \hspace{0.5cm}+b_2r^2+\frac{1}{2}\right)+3b_2\left(a_2r^2(1+b_2r^2)\sinh(b_2r^2)\left(b_2r^2+\frac{5}{6}\right)+\frac{b_2r^2}{6}+\frac{1}{3}\right)\right).
	\end{align} \begin{figure}
		\centering
		\begin{minipage}[t]{0.31\textwidth}
			\centering
			{\gfx{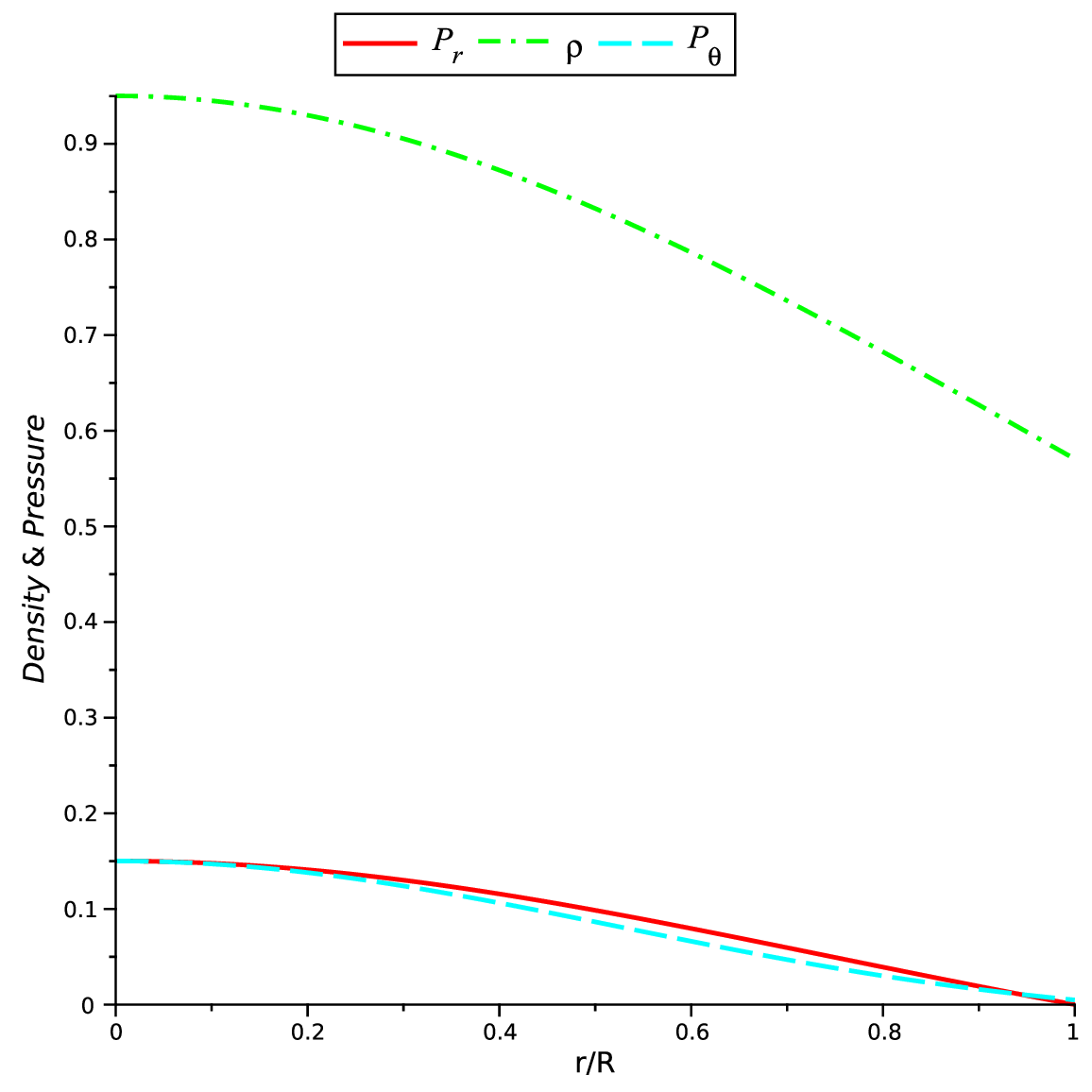}}
			\caption{Graph of density, radial and tangential pressures for $a_2=0.5501411418$ and $b_2=-0.2$}
			\label{fig.II2}
		\end{minipage}
		\hfill
		\begin{minipage}[t]{0.31\textwidth}
			\centering
			{\gfx{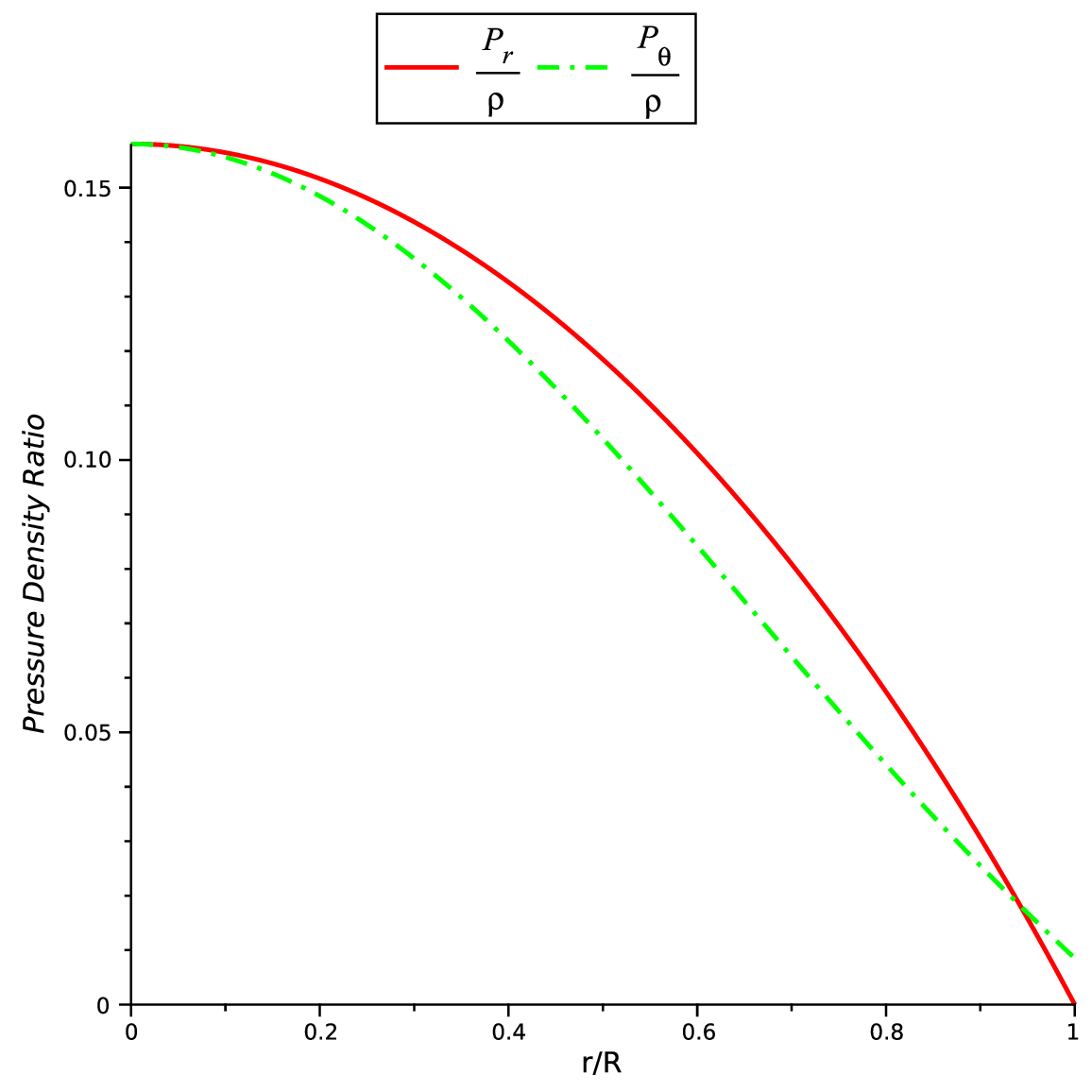}}
			\caption{Graph of pressure density ratio for $a_2=0.5501411418$ and $b_2=-0.2$}
			\label{fig.II4}
		\end{minipage}
		\hfill
		\begin{minipage}[t]{0.31\textwidth}
			\centering
			{\gfx{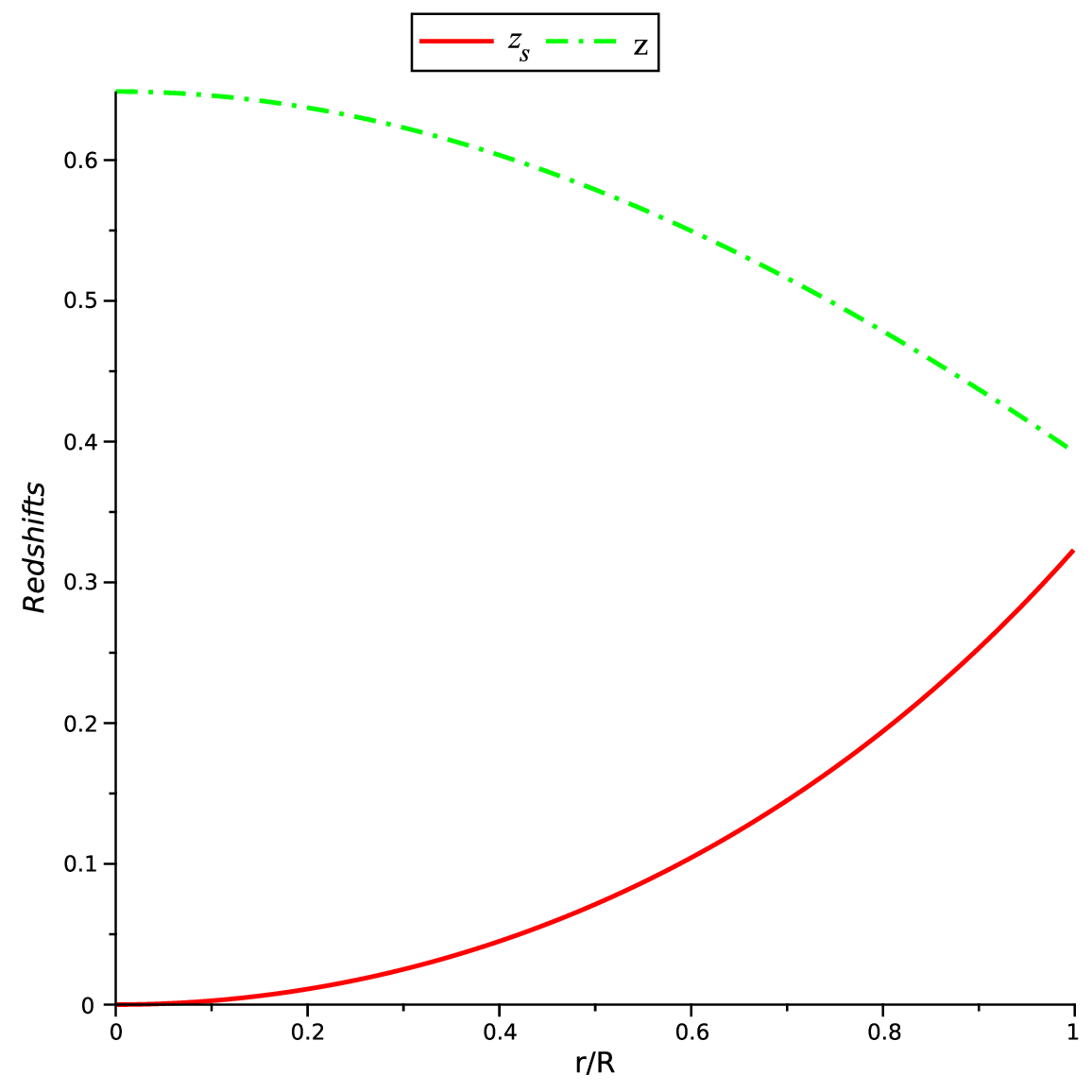}}
			\caption{Graph of redshifts for $a_2=0.5501411418$, $b_2=-0.2$ and  $c_2=-1$ }
			\label{fig.II5}
		\end{minipage}
	\end{figure} \noindent Since radial and tangential pressure equations must be equal at the center, so we get 
	\begin{equation}\label{II12}
		P_\theta=P_r|_{r=0}=a_2+2b_2,
	\end{equation}
	and radial pressure must vanish at the boundary, $P_r|_{(r=R)}=0$, it gives the expression for the constant parameter $a_2$ as 
	\begin{equation}\label{II13}
		a_2=\frac{-b_2(2+3b_2)}{(1+b_2)(1+2b_2)(\cosh(b_2)+b_2\sinh(b_2))}.
	\end{equation}
	\noindent Considering $b_2=-0.2$, equation \eqref{II13} yields the value of constant $a_2=0.5501411418$. The graphs of metric functions for these constants are shown in Fig. \ref{fig.II1}, satisfying the requirement of being singularity-free. $\left(e^{\lambda}\right)|_{r=0}$ is $1$ and $e^\nu$ at the center is equal to the positive constant i.e. $0.3678794412$. 
	
	\noindent Moreover, the behavior of density and pressure is illustrated in Fig. \ref{fig.II2}, both are decreasing when moving towards the boundary and $P_r$ vanishes. This again can be verified by taking the derivative of density and pressure which remain negative throughout the interior validating the decreasing nature of density and pressure. The Zeldovich criteria are obeyed as the ratio of pressure in the radial and azimuthal direction with density is less than one, as depicted in Fig. \ref{fig.II4}.
	
	\noindent In this case, the expressions for surface and gravitational redshifts become
	\begin{align}
		z_s&=\exp\left(\frac{a_2r^2\cosh(b_2r^2)}{2}\right)-1,
		\\
		z&=\frac{\exp\left(\frac{-a_2{r^2}\cosh(b_2r^2)-c_2}{2}\right)-\sqrt{1+b_2r^2}}{\sqrt{1+b_2r^2}}.
	\end{align}  
	\noindent Fig. \ref{fig.II5} illustrates the increasing and decreasing nature of redshifts respectively. 
	
\begin{figure}
		\centering
		\begin{minipage}[t]{0.46\textwidth}
			\centering
			\subfloat[]
			{\gfx{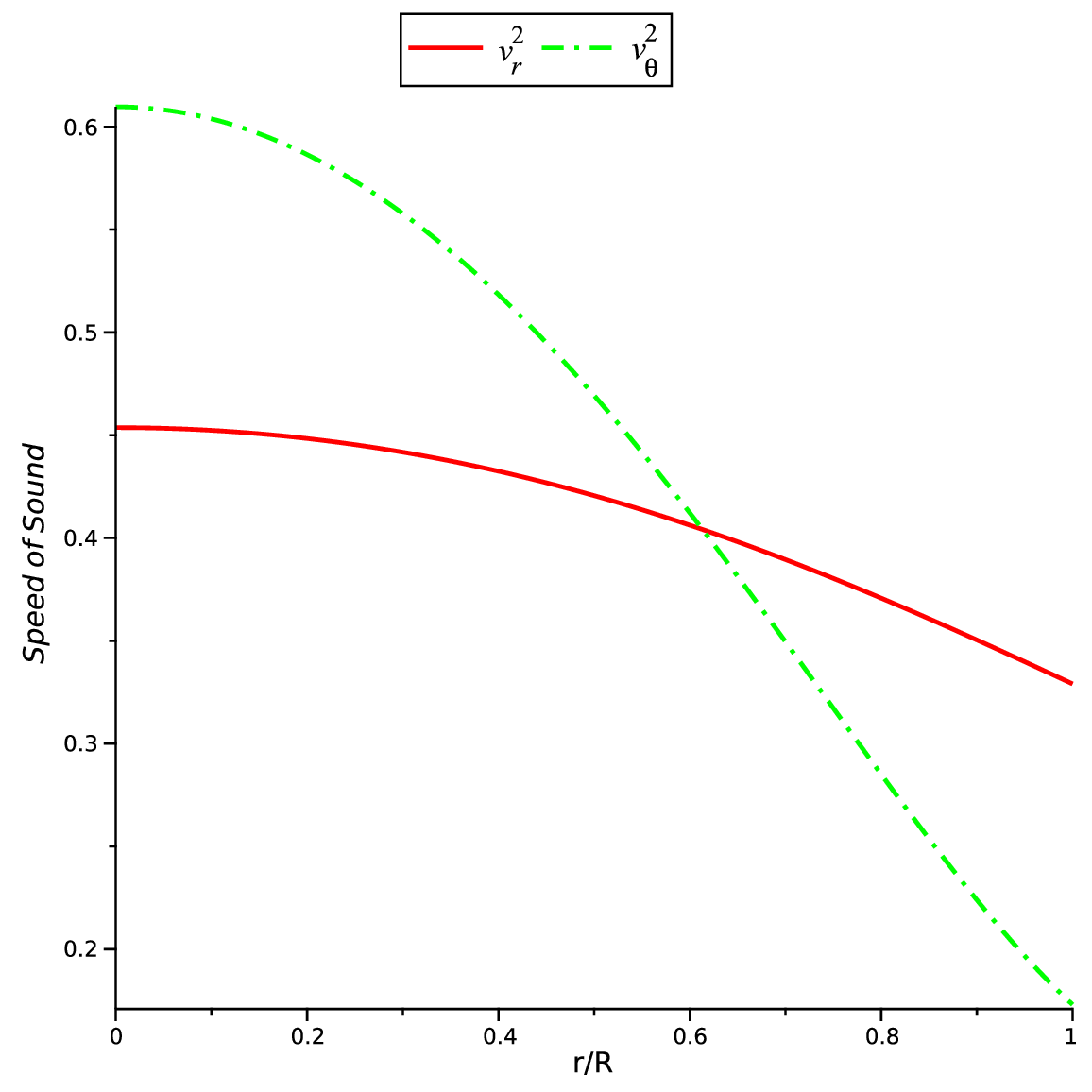}}
		\end{minipage}
		\hfill
		\begin{minipage}[t]{0.46\textwidth}
			\centering
			\subfloat[]
			{\gfx{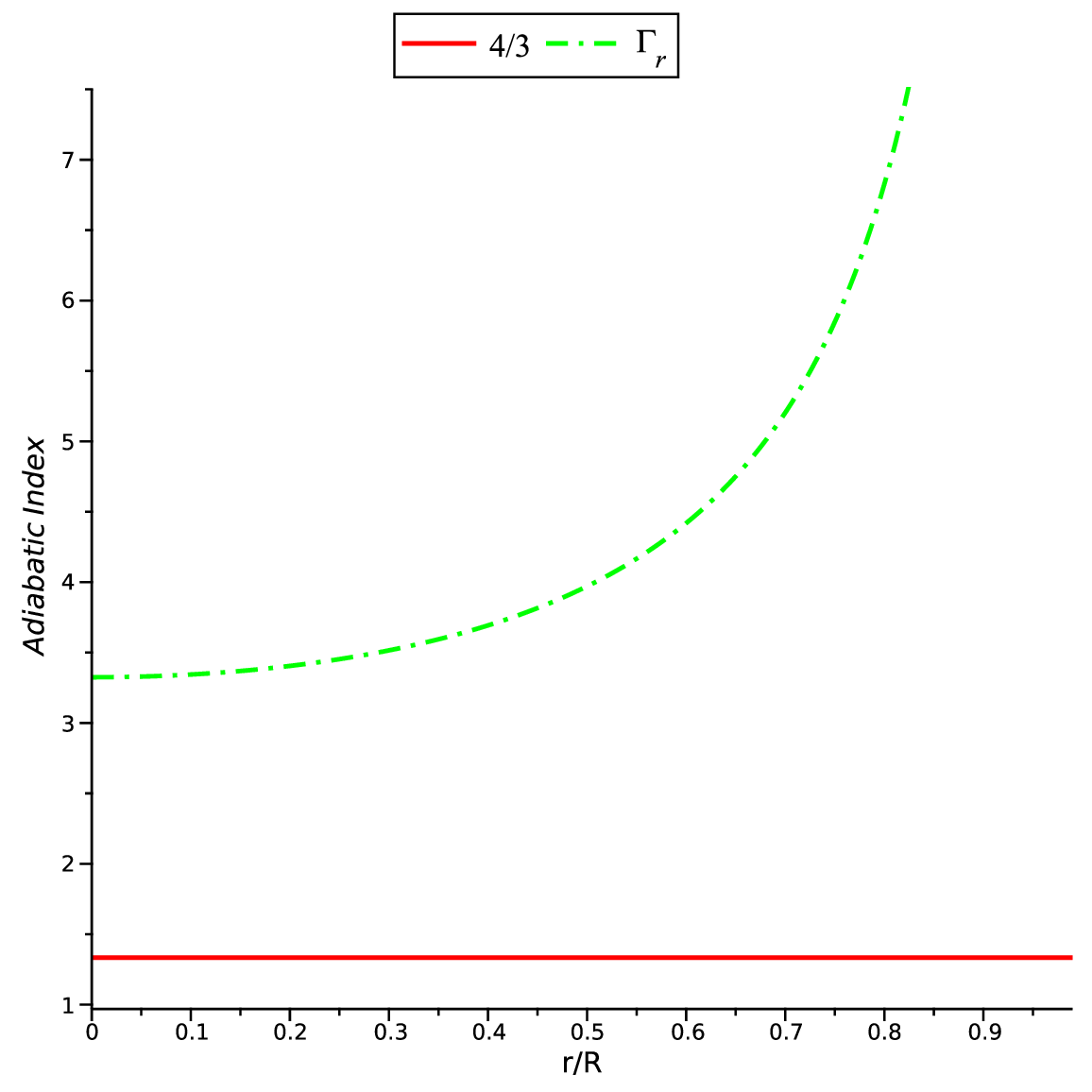}}
		\end{minipage}
		
		\caption{Graphs of stability conditions for $a_2=0.5501411418$ and $b_2=-0.2$} \label{fig.II6}
	\end{figure}

	\noindent For the stable configuration, we look for the causality condition in this case, Fig. \ref{fig.II6}$(a)$ elucidates the behavior of the speed of sound in radial and azimuthal directions that is less than the speed of light. The energy conditions in anisotropic fluid distribution are defined as 
	\begin{itemize}
		\item NEC:\ \ $\rho + P_n\ge 0$ for $n= 1,2$,
		\item WEC:\ $\rho \ge 0$ and $\rho + P_n\ge 0$ for $n= 1,2$,
		\item SEC: \ \ $\rho + P_1 + 2P_2 \ge 0$ and $\rho + P_n\ge 0$ for $n= 1,2$,
		\item DEC:\ $\rho \ge 0$ and $\rho \ge|P_n|$.
	\end{itemize}

	\noindent As shown in Fig. \ref{fig.II2}, the energy density and both pressure components remain positive throughout the configuration. This ensures that the NEC, WEC, and SEC are satisfied. The DEC is also fulfilled, since the ratios of both pressures to the energy density remain below unity, as illustrated in Fig. \ref{fig.II4}. Furthermore, Fig. \ref{fig.II6}$(b)$ indicates that the adiabatic index defined in this case as $\Gamma_r= \left(\frac{\rho+P_r}{P_r}\right)\left(\frac{dP_r}{d\rho}\right)$ is greater than 4/3 in the anisotropic distribution. Thus the solution obtained for Segre type $[1,1(11)]$ corresponding to an anisotropic fluid distribution is physically reasonable and stable. In addition, the new ansatz presented is precise and acceptable being singularity-free and satisfying the remaining necessities. Earlier studies by \cite{HLMS} established the field equations, regularity conditions, and matching requirements for static anisotropic cylinders, with particular attention to conformally flat sources. It was shown that such sources cannot be matched to a Levi-Civita exterior. The solution considered here differs in this respect: it is not conformally flat and arises from the generating function specified in Eq.~\eqref{II8}.
	
	\section*{III:\ \ Segre Type $[(1,1)11]$} \label{S5}
	
	\noindent Now we will explore the Segre type $[(1,1)11]$. The eigenvalue degeneracy refers to the symmetry in a timelike and one spatial direction, and the two spacelike eigenvectors have different eigenvalues. As a result, the fluid distribution in this Segre type is considered as 
	\begin{equation}
		T_{ab}=(\rho+P_\theta)u_au_b+P_\theta g_{ab}+(P_r-P_\theta)\eta_a\eta_b+(P_z-P_\theta)\chi_a\chi_b, \label{III4}
	\end{equation}
	
	\noindent Here, $P_r, P_\theta, P_z$ are the principal stresses in radial, azimuthal and axial directions, $u^a$ is timelike unit four velocity, $\eta^a$ and $\chi^a$ are spacelike four vectors where $u^au_a=-1$, $\eta^a\eta_a=1=\chi^a\chi_a$ and $u^a\eta_a=u^a\chi_a=\chi^a\eta_a=0$. Thus, the non-zero EMT components are given as
	\begin{equation}
		T^a_b=\mathrm{diag}(-\rho,P_r,P_\theta,P_z)\label{III5}.
	\end{equation}
	\noindent Using equations \eqref{9}-\eqref{12} and the aforementioned EMT the field equations are transformed as
	\begin{align}
		\rho&=\frac{e^{-\lambda}}{4}\left(-2\mu''-\mu'^2+\mu'\lambda'+\frac{2}{r}(\lambda'-\mu')\right) \label{III6}
		\\
		P_r&=\frac{e^{-\lambda}}{4}\left(\mu'\nu'+\frac{2}{r}(\mu'+\nu')\right) \label{III7}
		\\
		P_\theta&=\frac{e^{-\lambda}}{4}\left(2\mu''+2\nu''+\mu'^2+\nu'^2+\mu'\nu'-\mu'\lambda'-\nu'\lambda'\right) \label{III8}
		\\
		P_z&=\frac{e^{-\lambda}}{4}\left(2\nu''+\nu'^2-\nu'\lambda'+\frac{2}{r}(\nu'-\lambda')\right) \label{III9}
	\end{align}
	
	\noindent The corresponding non-zero NP scalars for this Segre type are
	\begin{equation}\label{III1}
		\phi_{02}=\phi_{20}\ne0, \hspace{1cm}\phi_{11}\ne0
	\end{equation}
	The constraints for this type are given by equation \eqref{I8} and
	\begin{equation}
		2\mu''+\mu'^2+\mu'(\nu'-\lambda')+\frac{2}{r}\left(\lambda'-\nu'\right)\ne0 \label{III2},
	\end{equation}
	this implies that $\phi_{00}=\phi_{22}=0$, given as 
	\begin{equation}
		-2\mu''-\mu'^2+\mu'(\lambda'+\nu')+\frac{2}{r}\left(\nu'+\lambda'\right)=0 \label{III3}.
	\end{equation}
	
	\begin{figure}
		\centering
		\begin{minipage}[t]{0.31\textwidth}\subfloat[]	
			{\gfx{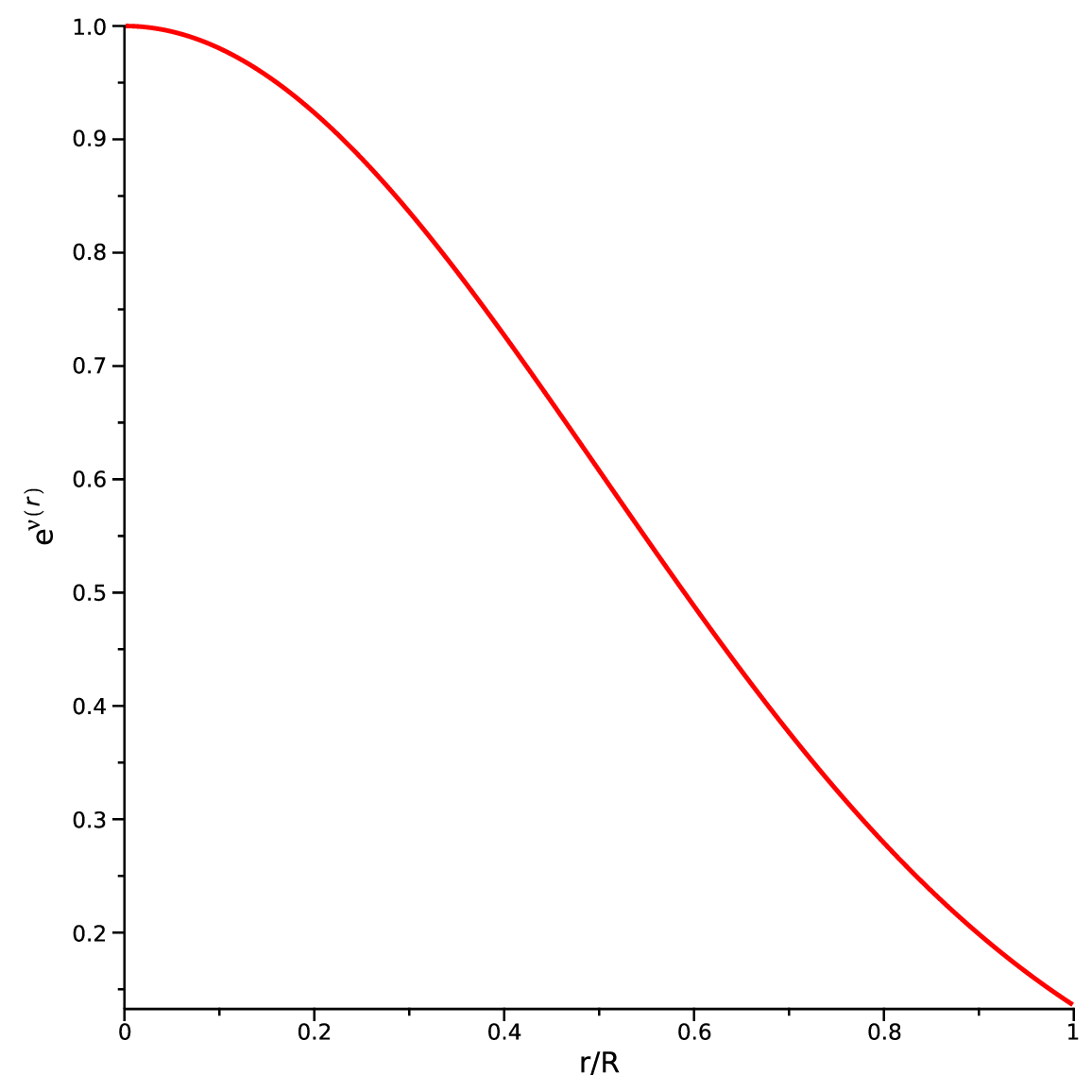}}
		\end{minipage}
		\hfill
		\begin{minipage}[t]{0.31\textwidth}\subfloat[]	
			{\gfx{IIIMp2.eps}}
		\end{minipage}
		\hfill
		\begin{minipage}[t]{0.31\textwidth}\subfloat[]	
			{\gfx{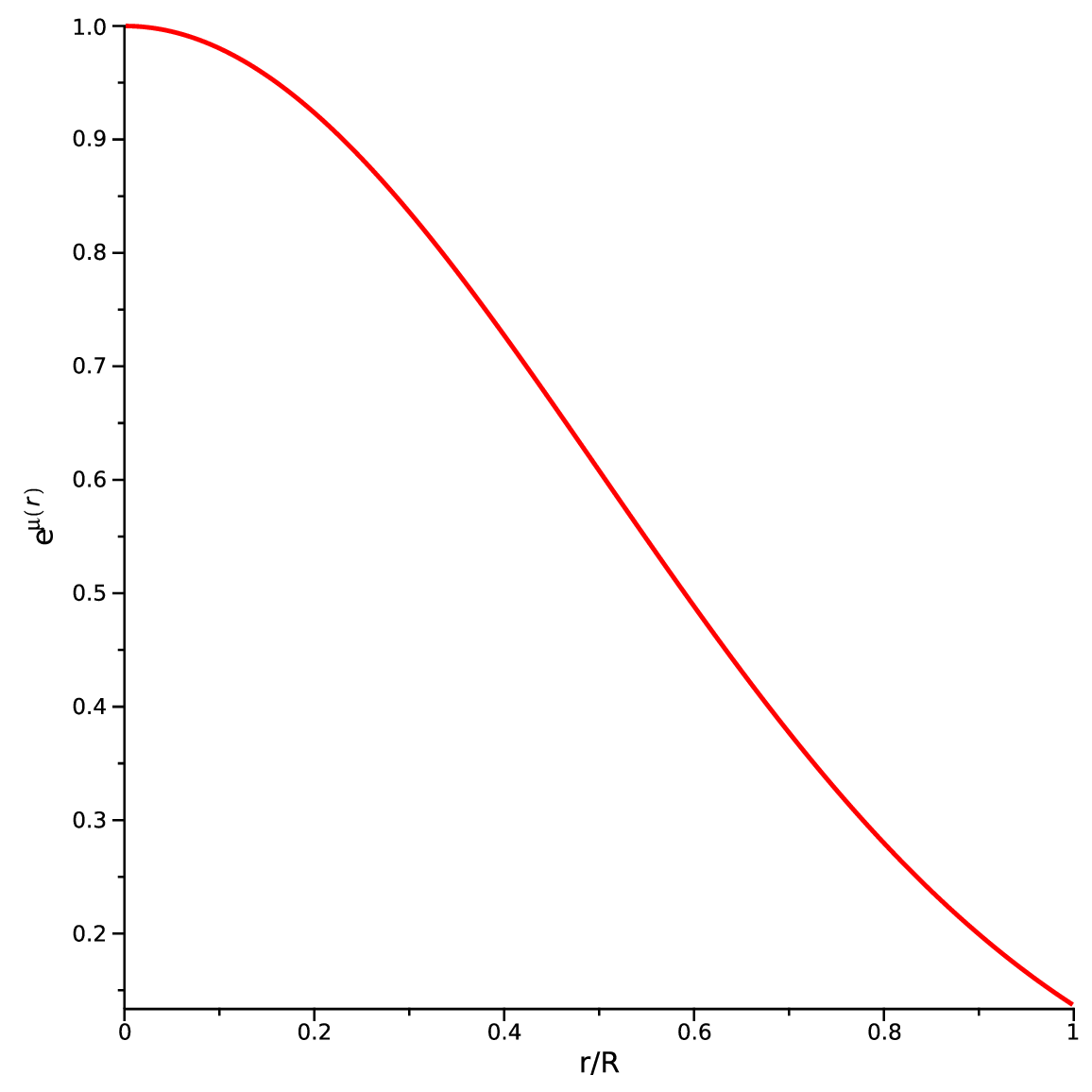}}
		\end{minipage}
		
		\caption{Graph of metric potentials for $a_3=0.001849379103$ and $b_3=-1.99$}
		\label{fig.III1}
	\end{figure}
	\noindent The symmetry in timelike and spatial eigenvectors implies that density and radial pressure must be equal in magnitude. By adding equations \eqref{III6},\eqref{III7} and using equation \eqref{III3} in it yields the relation 
	\begin{equation}
		\rho=-P_r \label{III10}
	\end{equation}

	\noindent As we know the relation in equation \eqref{III10} depicts one of the main features of the dark energy case i.e. the pressure is negative. Dark energy is one of the most enigmatic forms of energy, the negative pressure drives the repulsive gravitational effect which leads to the accelerated expansion of the universe. Here in this Segre type, the number of unknowns is seven $(\nu,\lambda,\mu,\rho, P_r, P_\theta, P_z)$ and we have five equations, this leaves us with two arbitrary choices to solve the system. 
	
	\noindent Consider the new metric function ($\lambda$), the same as introduced in II, and $\mu$, given as  
	\begin{align}
		\lambda&=a_3r^2\cosh(b_3r^2), \label{III11}
		\\
		\mu&=b_3r^2, \label{III12}
	\end{align}
	substituting these in equation \eqref{III3} yields
	\begin{equation}
		\nu= b_3r^2-a_3r^2\cosh(b_3r^2)+c_3. \label{III13}
	\end{equation}	
	\noindent Taking the constant of integration $c_3$ to be zero, putting equations \eqref{III11}-\eqref{III13} in equations \eqref{III6}-\eqref{III9}, we get the expressions of energy density and pressures given as 
	
	\begin{align}\label{III14}
		\rho&= e^{\left(-a_3r^2\cosh(b_3r^2)\right)}\left(a_3(b_3r^2+1)\cosh(b_3r^2)\right.\nonumber
		\\
		&\left. \hspace{3cm}+a_3b_3r^2(b_3r^2+1)\sinh(b_3r^2)-b_3^2r^2-2b_3\right)=-P_r,
		\\
		{P_\theta}&{=\frac{e^{-\lambda}}{4}\left(8b_3+12b_3^2r^2-2\lambda''-10b_3r\lambda'+2\lambda'^2\right),}
		\\
		{P_z}&{=\frac{e^{-\lambda}}{4}\left(8b_3+4b_3^2r^2-2\lambda''-6b_3r\lambda'+2\lambda'^2-\frac{4\lambda'}{r}\right),}
	\end{align}
	where, from equation \eqref{III11}, $\lambda'=2a_3r\left(\cosh(b_3r^2)+b_3r^2\sinh(b_3r^2)\right)$ and $\lambda''=2a_3\left(1+2b_3^2r^4\right)\cosh(b_3r^2)+10a_3b_3r^2\sinh(b_3r^2)$.
	\begin{figure}
		\centering
		\begin{minipage}[t]{0.46\textwidth}
			{\gfx{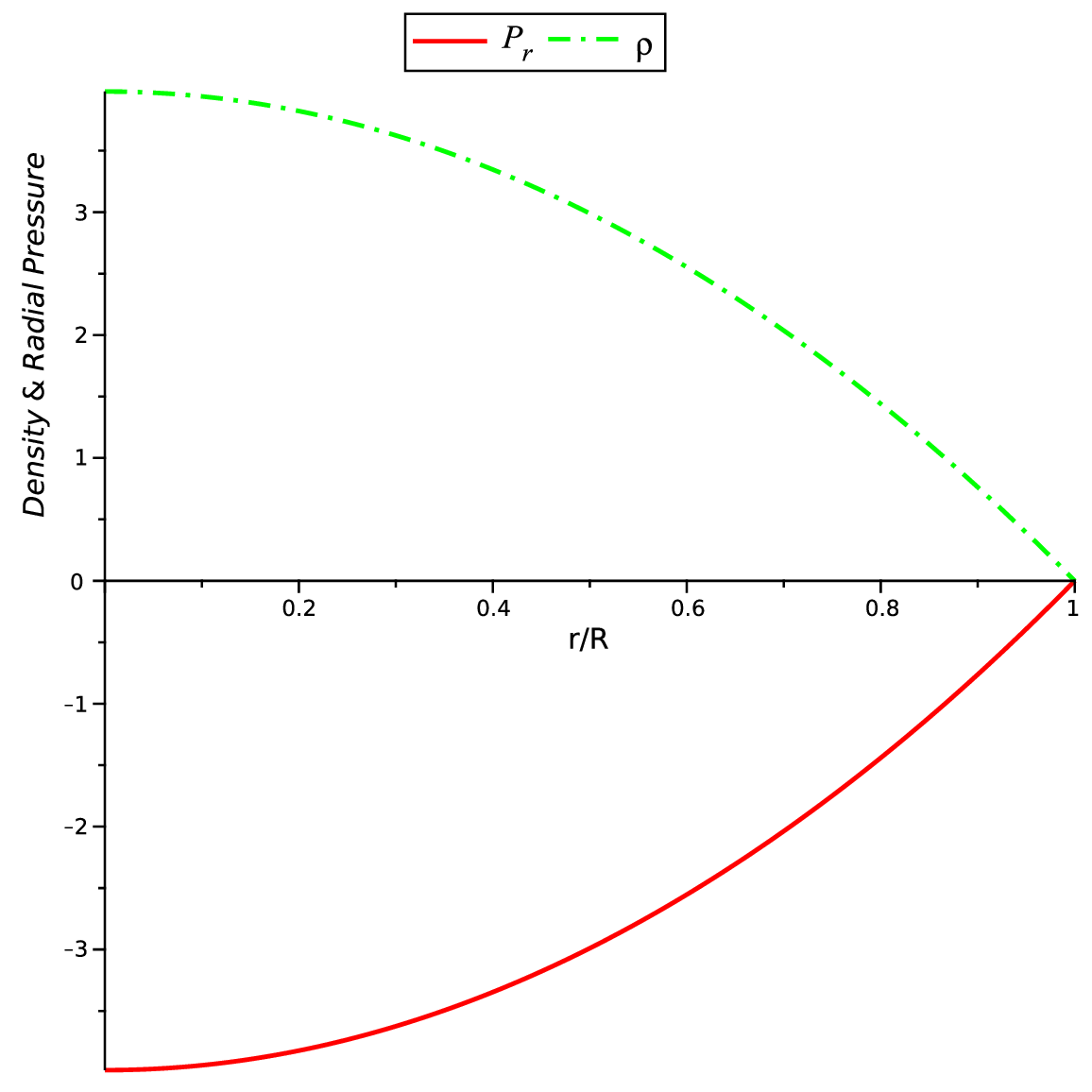}}
			\caption{Graph of energy density and radial pressure for $a_3=0.001849379103$ and $b_3=-1.99$}
			\label{fig.III2}
		\end{minipage}
		\hfill
		\begin{minipage}[t]{0.46\textwidth}
			{\gfx{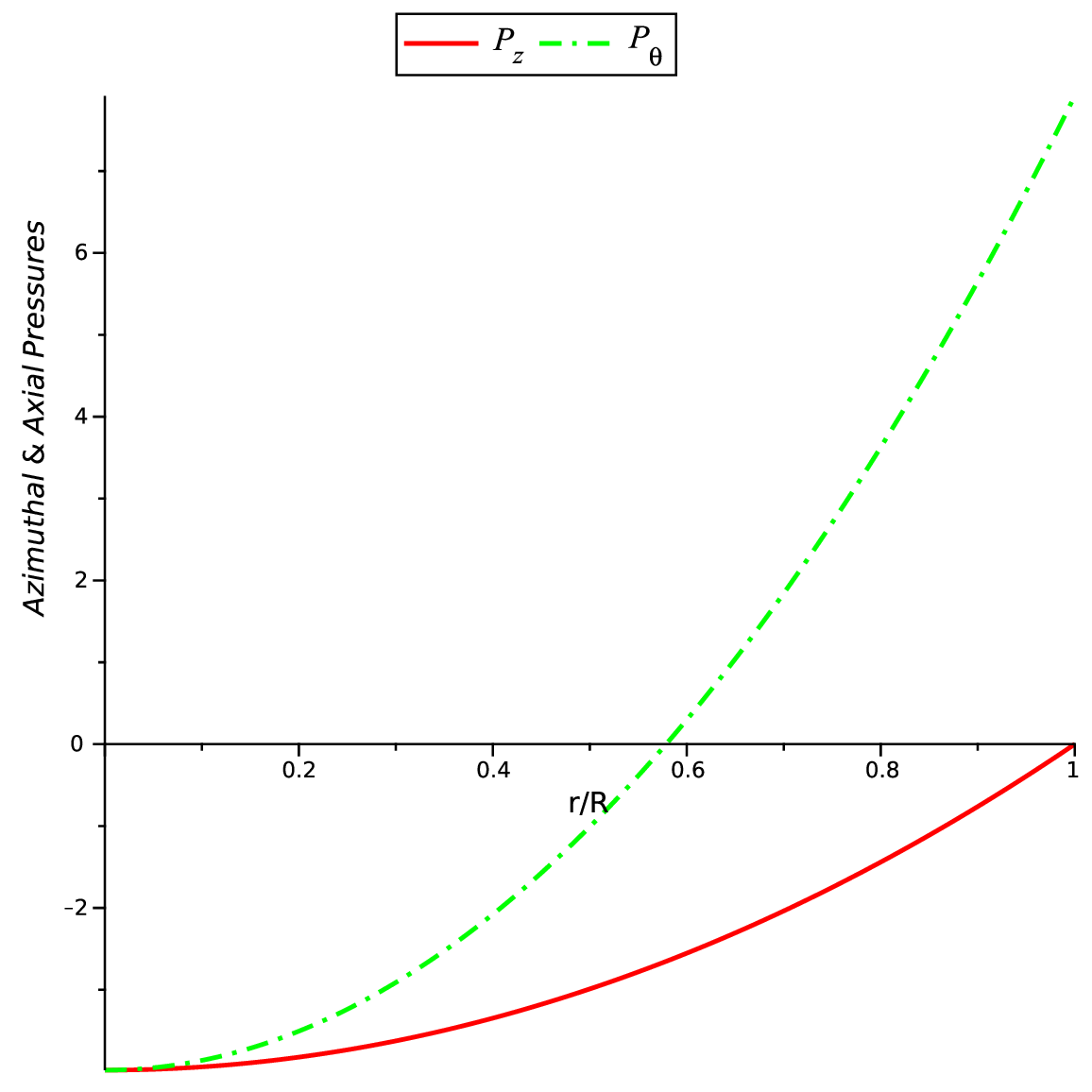}}
			\caption{Graph of azimuthal and axial pressures for $a_3=0.001849379103$ and $b_3=-1.99$}
			\label{fig.III3}
		\end{minipage}
	\end{figure}
	
	\noindent Since radial pressure must vanish at the boundary, $P_r|_{(r=R)}=0$, it gives the expression for the constant parameter $a_3$ as 
	\begin{equation}\label{III15}
		a_3=\frac{b_3(2+b_3)}{(1+b_3)(\cosh(b_3)+b_3\sinh(b_3))}.
	\end{equation}
	\noindent Considering $b_3=-1.99$ gives $a_3=0.001849379103$, the profiles of metric potentials for these constants are shown in Fig. \ref{fig.III1}, all are regular at origin implying the singularity-free nature. 
	
	\noindent Moreover the profiles of energy density and pressures in radial, axial and azimuthal directions are illustrated in Fig.s \ref{fig.III2} and \ref{fig.III3}. The radial and axial pressures stay negative across the whole interior referring to the presence of dark energy, while the azimuthal pressure is negative in the core, changes sign at $r=0.579R$, and turns positive toward the boundary; in addition the radial and azimuthal pressure are equal at center i.e. 
	\begin{equation}\label{III16}
		(P_r=P_\theta)|_{r=0}=-a_3+2b_3,
	\end{equation}
	
	\noindent Dark energy does not behave as ordinary matter and violation of SEC is a defining characteristic of dark energy \cite{Visser97} due to the significance of negative pressure outweighing the energy density. Furthermore, depending on the specific properties of the energy-momentum tensor or metric potential the presence of dark energy may contradict the NEC and WEC. For this Segre type, the energy conditions are defined as   
	\begin{itemize}
		\item NEC:\ \ $\rho + P_n\ge 0$ for $n= 1,2,3$,
		\item WEC:\ $\rho \ge 0$ and $\rho + P_n\ge 0$ for $n= 1,2,3$,
		\item SEC: \ \ $\rho + \sum_{n=1}^{3} P_n \ge 0$ and $\rho + P_n\ge 0$ for $n= 1,2,3$,
		\item DEC:\ \ $\rho \ge 0$ and $\rho \ge|P_n|$.
	\end{itemize}
	\begin{figure}
		\centering
		\begin{minipage}[t]{0.46\textwidth}
			{\gfx{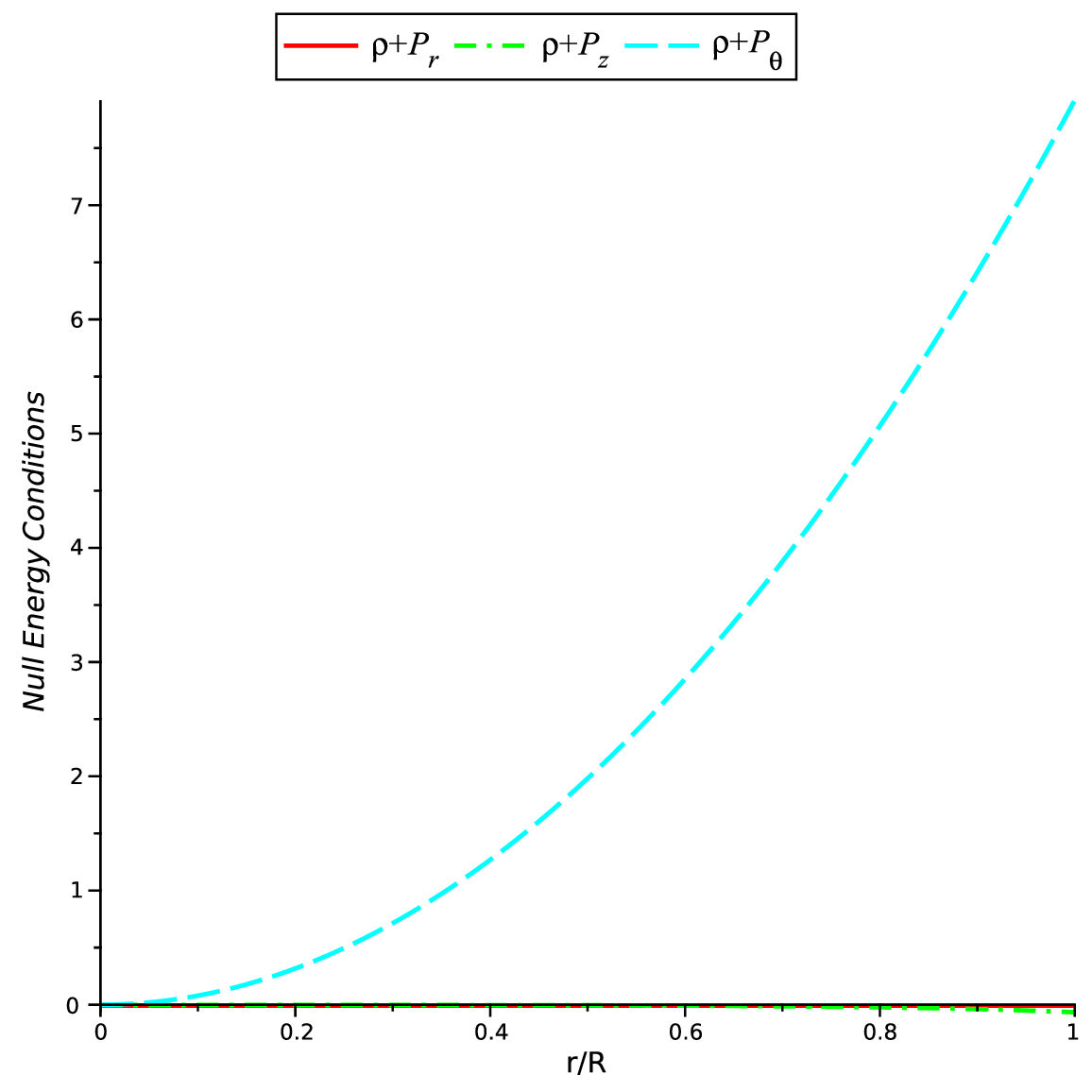}}
			\caption{Graph of NEC for $a_3=0.001849379103$ and $b_3=-1.99$}
			\label{fig.III4}
		\end{minipage}
		\hfill
		\begin{minipage}[t]{0.46\textwidth}
			{\gfx{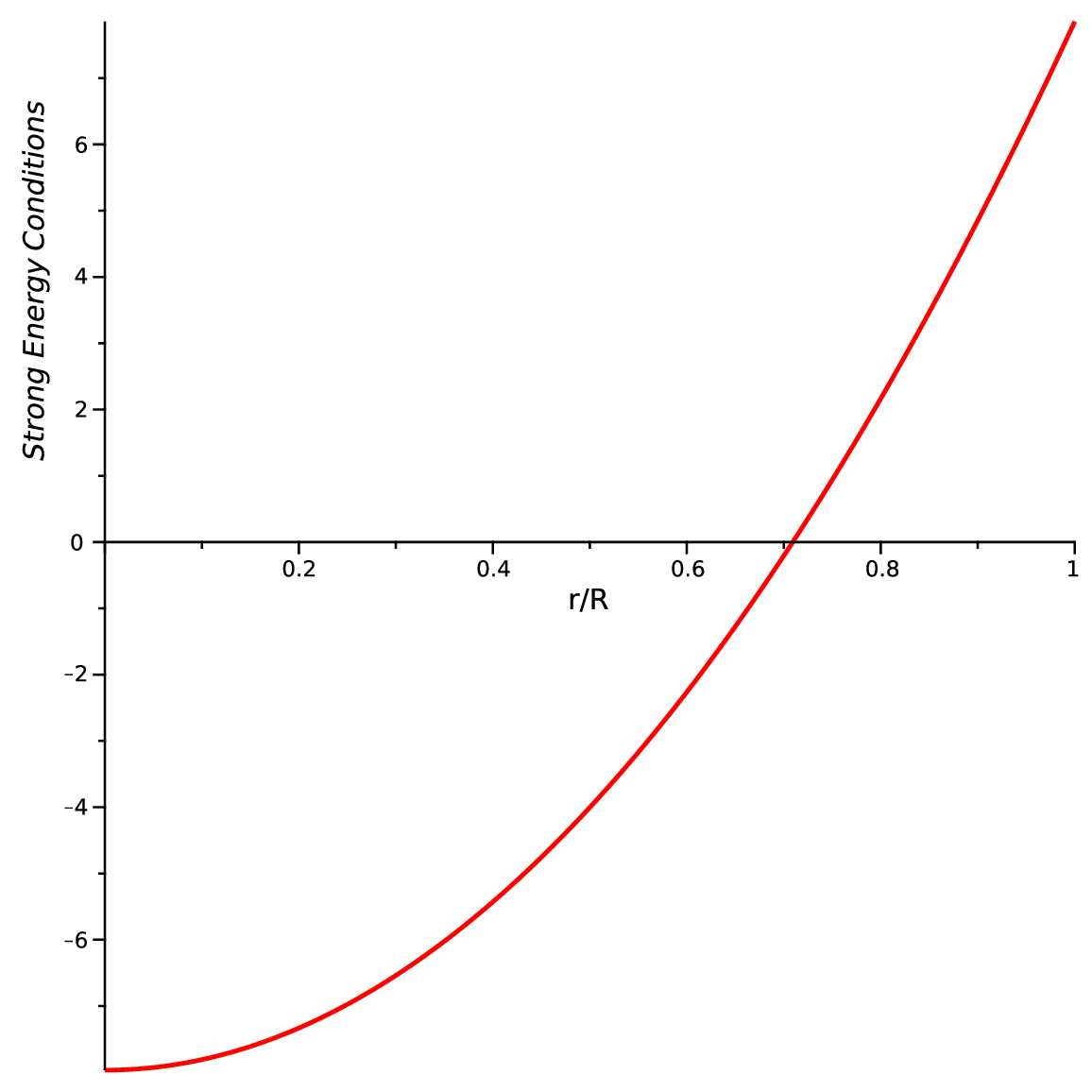}}
			\caption{Graph of SEC for $a_3=0.001849379103$ and $b_3=-1.99$}
			\label{fig.III5}
		\end{minipage}
	\end{figure}
	
	\noindent Fig.s \ref{fig.III4} and \ref{fig.III5} illustrate the violation of energy conditions, mainly the violation of SEC is what allows accelerated expansion in the presence of dark energy.  In particular, $\rho+P_r=0$ and $\rho+P_\theta\geq0$ hold throughout the configuration, whereas $\rho+P_z$ remains slightly negative, with $-0.066\leq\rho+P_z\leq-0.0035$. Thus, the NEC and WEC are violated only marginally, due entirely to the axial stress. The SEC trace, $\rho+P_r+P_\theta+P_z$, is negative for $r<0.709R$ and becomes positive beyond this point, indicating that the dark-energy behaviour is confined to the core. The DEC is marginally violated through $P_z$ and, for $r>0.709R$, through $P_\theta$, which remains positive as $\rho$ approaches zero at the boundary. Since $P_r/\rho=dP_r/d\rho=-1$ identically, the Zeldovich criterion, sound speed, and adiabatic index are not applicable, and the analysis instead relies on the energy conditions.
	
	\noindent In relation to earlier studies, $\rho=-P_r$ represents vacuum-like behaviour in the radial direction \cite{Dym}, while $P_z\approx-\rho$ corresponds to the equation of state of a cosmic string oriented along the symmetry axis \cite{Vil,Gott,His}. The present solution provides a regular, bounded, and non-conformally-flat interior of this kind, with distinct azimuthal and axial stresses. Thus the solution obtained for Segre type $[(1,1)11]$ with the matter distribution of density and pressure along radial, azimuthal, and axial direction obeys the symmetry condition of the similarity in the timelike and spacelike eigenvectors. This specifies that in cylindrical geometry the respective Segre type results in the existence of dark energy.

	\section*{IV:\ \ Segre Type $[1,111]$} \label{S6}
	
	\noindent Segre type $[1,111]$ reveals the presence of all four unique eigenvalues reflecting distinct timelike and spacelike eigenvectors. This type represents the fluid distribution in the same way as in III i.e. denoted as $\mathrm{diag}(-\rho,P_r,P_\theta,P_z)$, additional analysis and a solution would be obtained by employing this. The constraints corresponding to this Segre type are given in equations \eqref{I7},\eqref{I8}, and \eqref{III2}. Here, there are no vanishing NP scalars so we have only four equations as given by equations \eqref{III6}-\eqref{III9} and seven unknowns $(\nu,\lambda,\mu,\rho, P_r, P_\theta, P_z)$. This leads to three free functions; we consider the Krori and Barua \cite{40} forms of functions
	\begin{align}
		\nu&=a_4r^2+b_4, \label{IV1}
		\\
		\lambda&=c_4r^2, \label{IV2}
	\end{align}
	in addition to that the third free function is taken as 
	\begin{equation}
		\mu=k_4r^2. \label{IV3}
	\end{equation}
	\begin{figure}
		\centering
		\begin{minipage}[t]{0.31\textwidth}\subfloat[]
			{\gfx{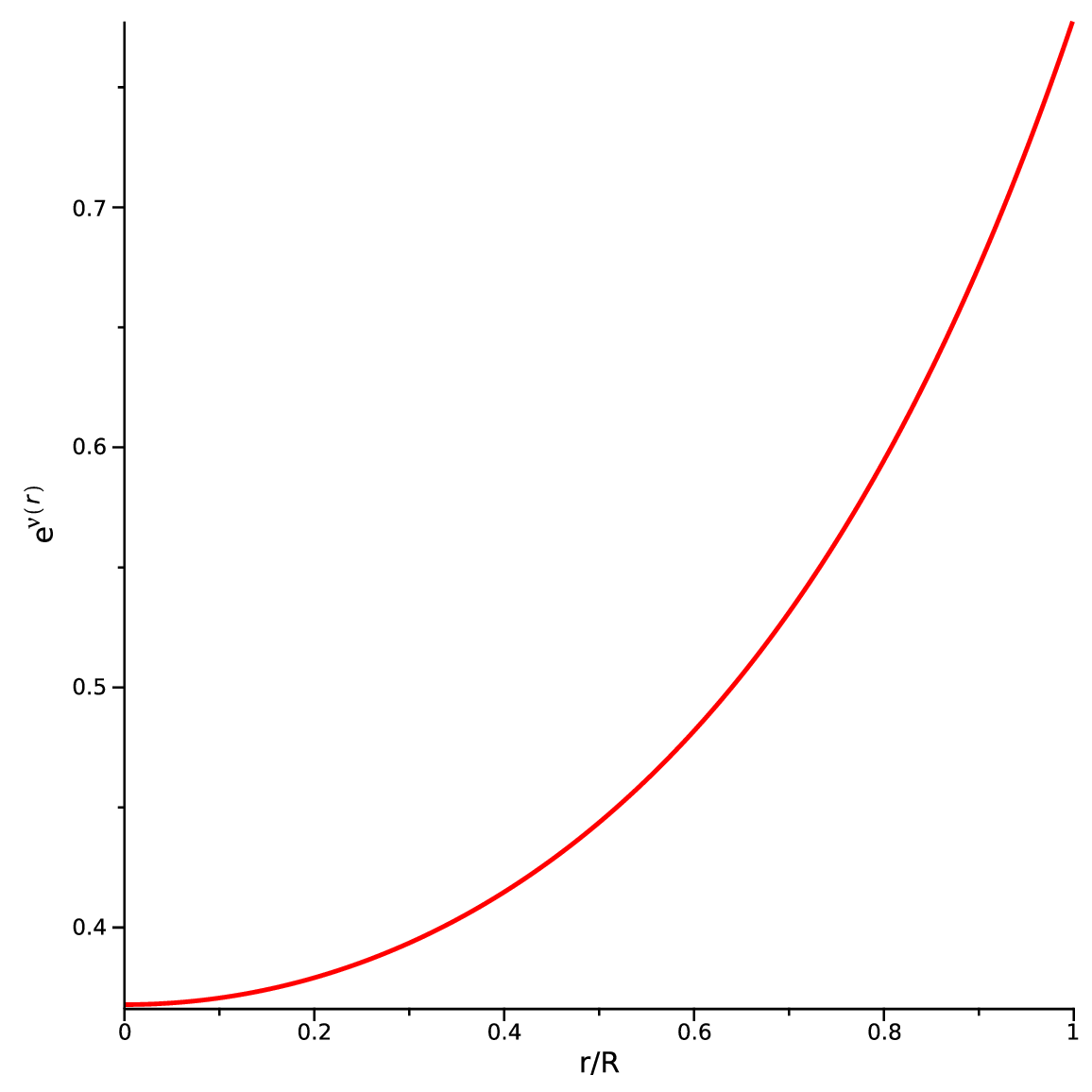}}
		\end{minipage}
		\hfill
		\begin{minipage}[t]{0.31\textwidth}\subfloat[]	
			{\gfx{IVMp2.eps}}
		\end{minipage}
		\hfill
		\begin{minipage}[t]{0.31\textwidth}\subfloat[]	
			{\gfx{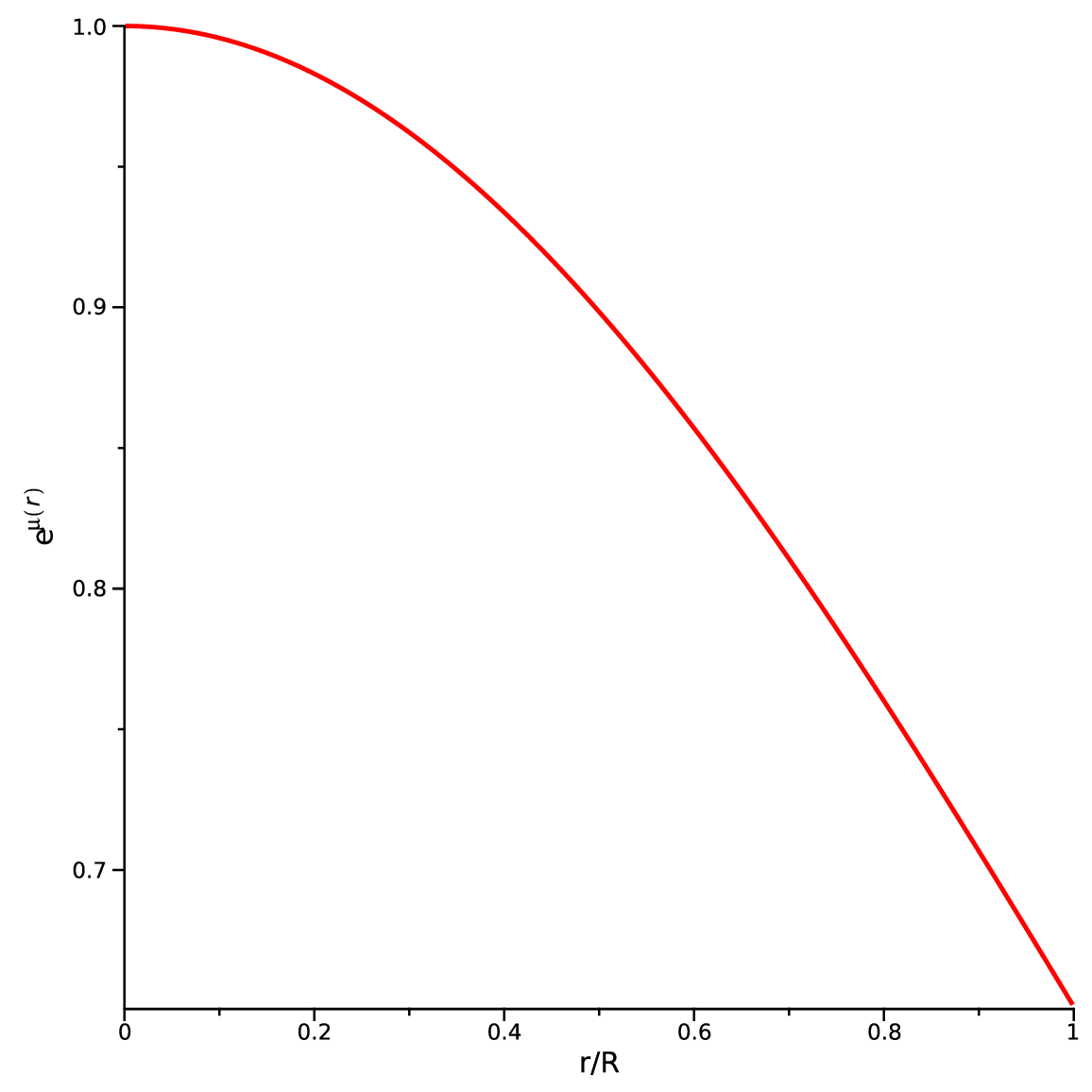}}
		\end{minipage}
		
		\caption{Graph of metric potentials for $a_4=0.75$, $b_4=-1$, $c_4=1$ and $k_4=-0.4285714286$}
		\label{fig.IV1}
	\end{figure}\noindent Because of their physical viability and well-behaved character, the Krori-Barua metric potentials have been frequently employed by several researchers \cite{n46, n47,n48, n49} to describe compact stellar formations. Substituting the metric potentials in the field equations \eqref{III6}-\eqref{III9}, we get the expression of density and pressures in radial, azimuthal and axial directions as
	\begin{align}
		\rho&=e^{-c_4r^2}\left(c_4-2k_4-k_4^2r^2+c_4k_4r^2\right) \label{IV4},
		\\
		P_r&=e^{-c_4r^2}\left(a_4+k_4+a_4k_4r^2\right) \label{IV5},
		\\
		P_\theta&=e^{-c_4r^2}\left(a_4+k_4+r^2\left(a_4^2+(-c_4+k_4)a_4-k_4(c_4-k_4)\right)\right) \label{IV6},
		\\
		P_z&=e^{-c_4r^2}\left(2a_4-c_4+r^2(a_4^2-a_4c_4)\right) \label{IV7}.
	\end{align}

	\begin{figure}
		\centering
		\begin{minipage}[t]{0.46\textwidth}
			{\gfx{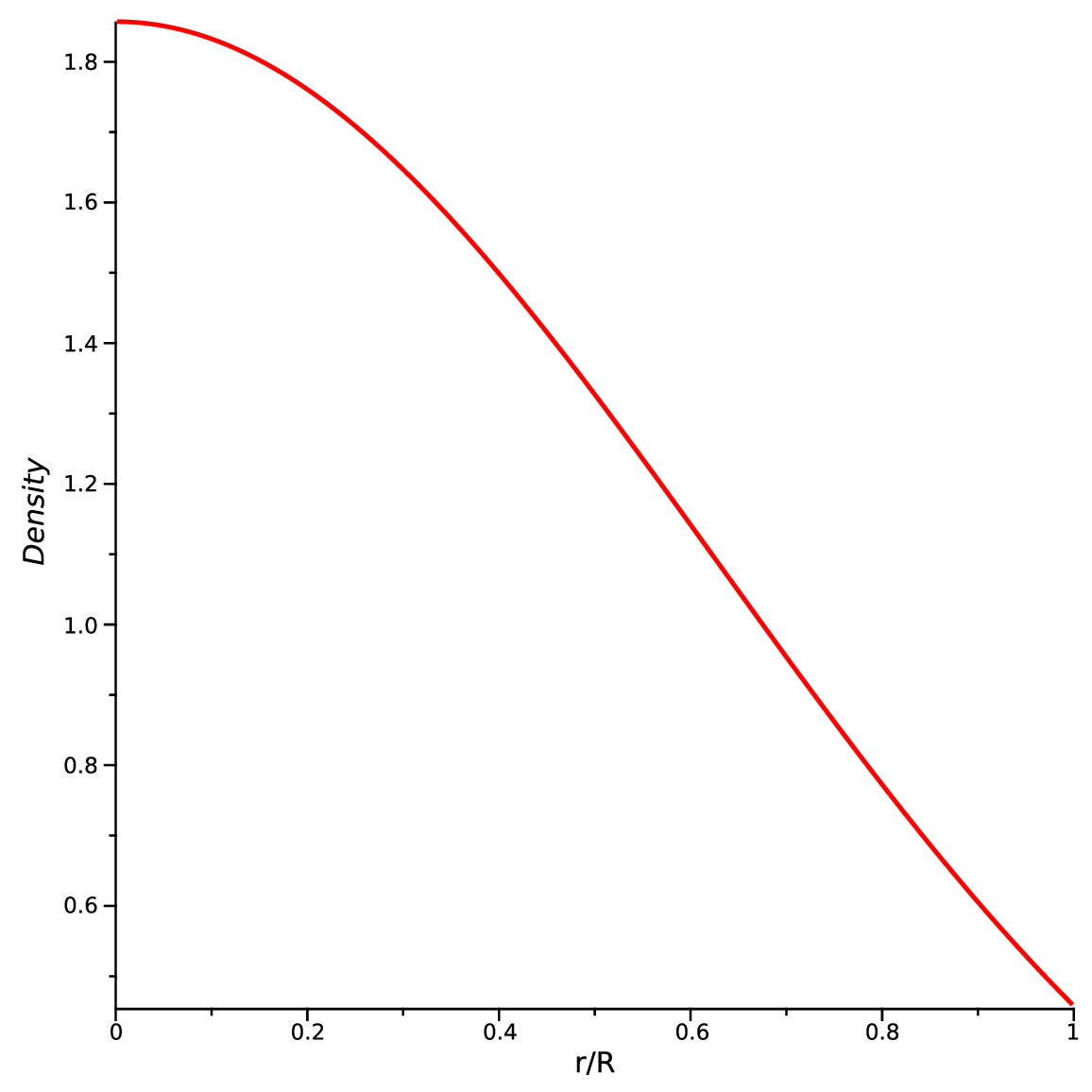}}
			\caption{Graph of energy density for $a_4=0.75$, $c_4=1$ and $k_4=-0.4285714286$}
			\label{fig.IV2}
		\end{minipage}
		\hfill
		\begin{minipage}[t]{0.46\textwidth}
			{\gfx{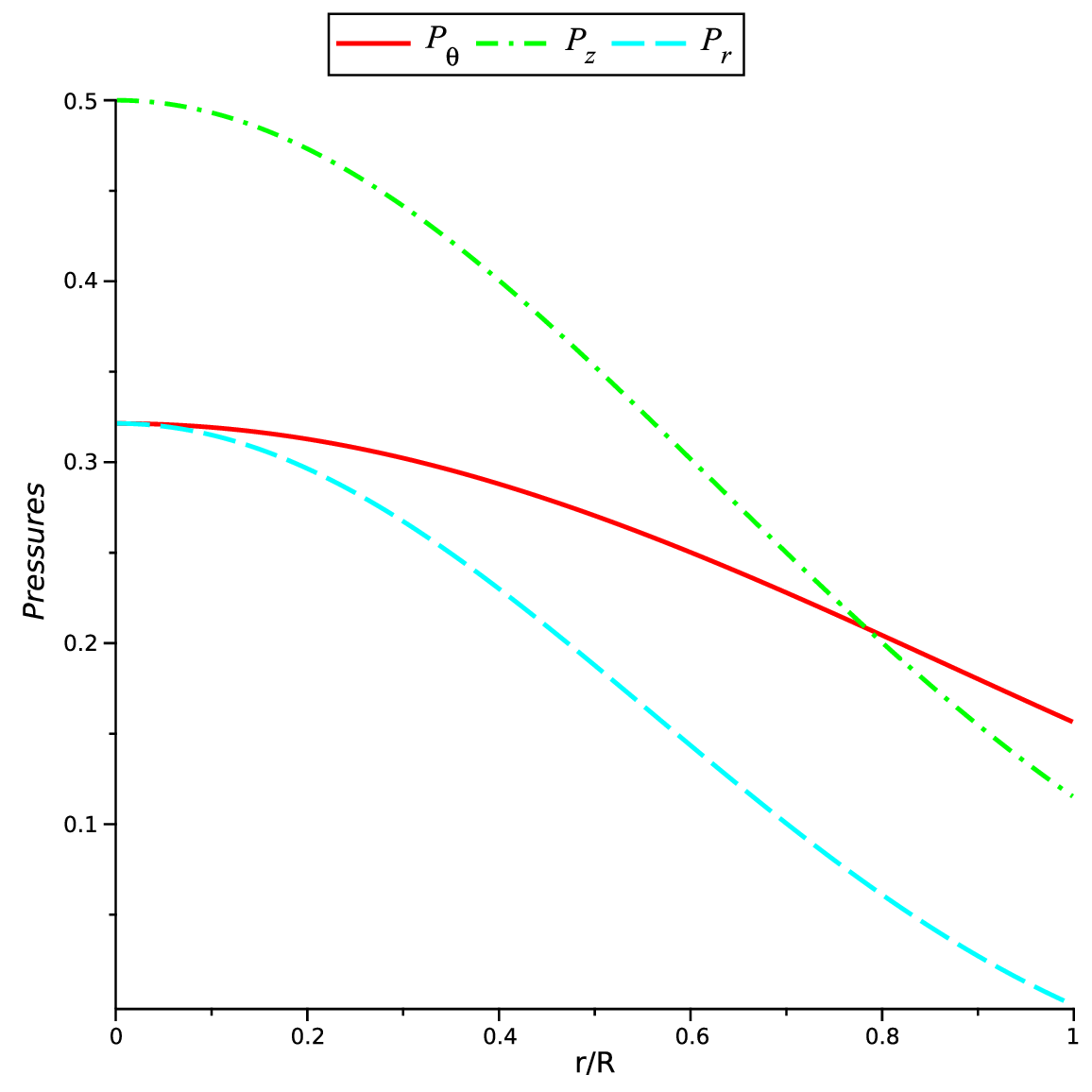}}
			\caption{Graph of pressures for $a_4=0.75$, $c_4=1$ and $k_4=-0.4285714286$}
			\label{fig.IV3}
		\end{minipage}
		
	\end{figure}
	\noindent Since radial pressure must vanish at the boundary i.e. $P_r|_{(r=R)}=0$, it gives the expression of constant
	\begin{equation}
		k_4=\frac{-a_4}{1+a_4}\label{IV8}.
	\end{equation}
	\noindent Consider $a_4=0.75$ then equation \eqref{IV8} gives $k_4=-0.4285714286$, the profiles of metric potentials for these constants are shown in Fig. \ref{fig.IV1} depicting the singularity free nature. Moreover, $(e^\lambda)|_{r=0}$ is $1$ and $e^\nu$ equals the positive constant $0.3678794412$ at the center.
	
	\noindent Fig.s \ref{fig.IV2} and \ref{fig.IV3} illustrate the behavior of density and pressures in all directions, all four are positive and decrease monotonically, also the radial and azimuthal pressures are same at the center $P_\theta=P_r|_{r=0}=a_4+k_4$. This can be verified by expressions of density and pressure gradients, given as 
	\begin{figure}
		\centering
		\begin{minipage}[t]{0.46\textwidth}
			{\gfx{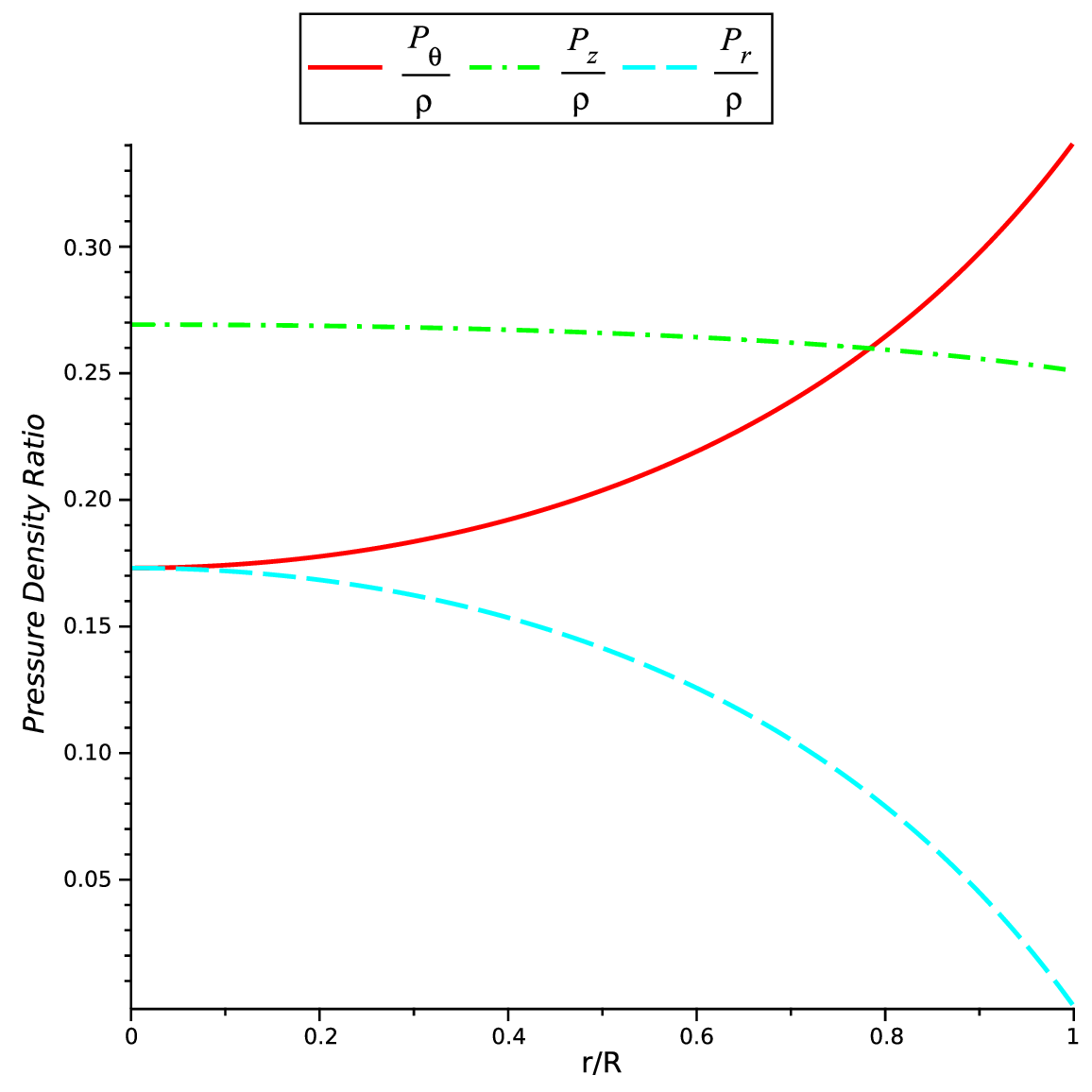}}
			\caption{Graphs of pressure density ratio for $a_4=0.75$,$c_4=1$ and $k_4=-0.4285714286$}
			\label{fig.IV5}
		\end{minipage}
		\hfill
		\begin{minipage}[t]{0.46\textwidth}
			{\gfx{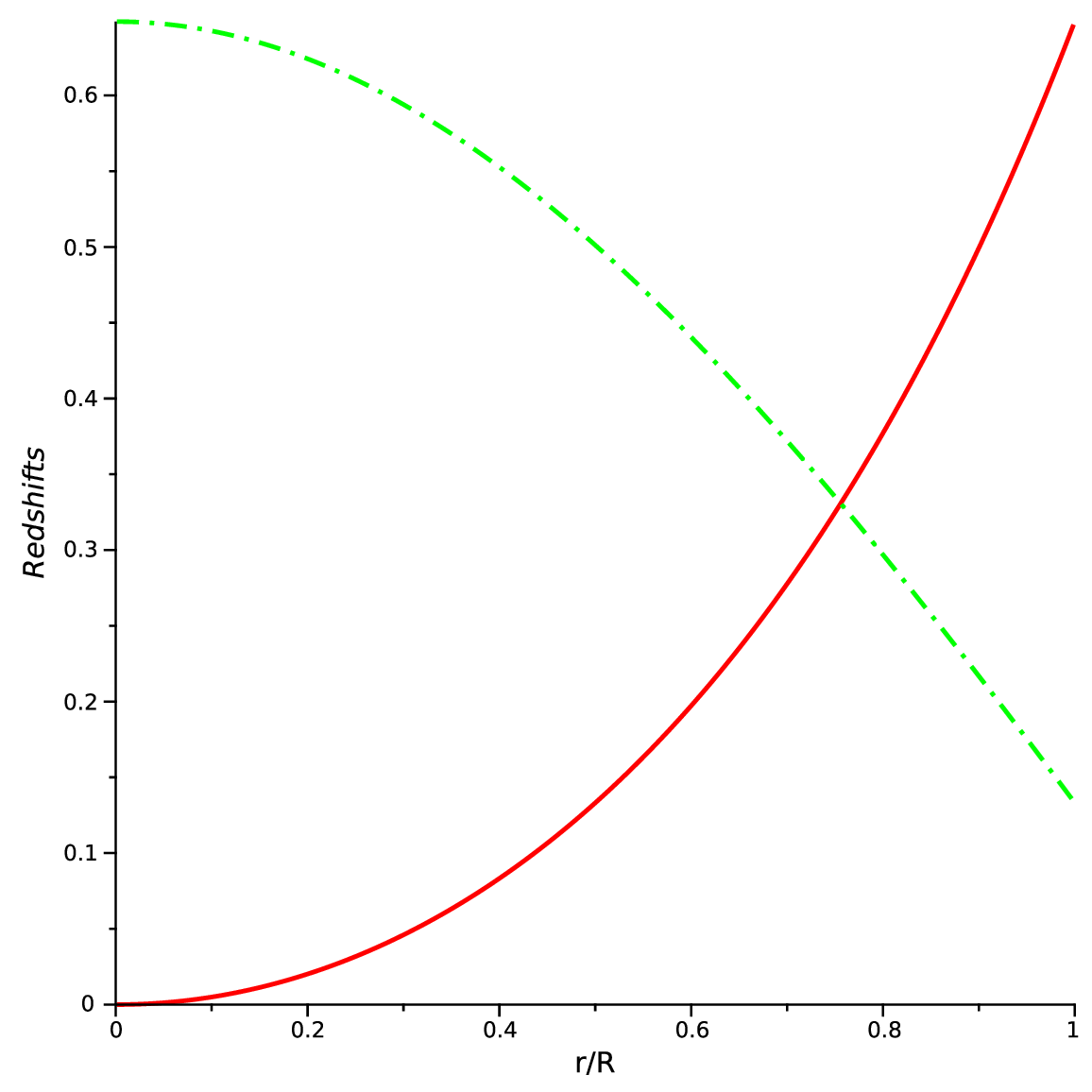}}
			\caption{Graph of redshift for $a_4=0.75$,$c_4=1$ and $k_4=-0.4285714286$}
			\label{fig.IV6}
		\end{minipage}
	\end{figure}
	\begin{align}
		\frac{d\rho}{dr}&=-2re^{-c_4r^2}\left(c_4k_4r^2(c_4-k_4)+c_4^2-3c_4k_4+k_4^2\right),\label{IV9}
		\\
		\frac{dP_r}{dr}&=-2re^{-c_4r^2}\left(a_4(k_4c_4r^2-k_4+c_4)+k_4c_4\right),
		\\
		\frac{dP_\theta}{dr}&=-2re^{-c_4r^2}\left(-c_4^2r^2(k_4+a_4)+c_4\left(a_4^2r^2+\left(a_4(k_4r^2+2)\right.\right.\right.\nonumber
		\\
		&\left.\left.\left.\hspace{4cm} +k_4^2r^2+2k_4\right) \right) -a_4^2-a_4k_4-k_4^2\right),
		\\
		\frac{dP_z}{dr}&=-2re^{-c_4r^2}\left(a_4c_4r^2(a_4-c_4)-a_4^2+3a_4c_4-c_4^2\right).\label{IV12}
	\end{align}
	\noindent The gradients in equations \eqref{IV9}-\eqref{IV12} are negative across the whole interior for these constants, verifying the condition. Moreover, the Zeldovich criterion for this type is satisfied as the ratios of radial, azimuthal, and axial pressures with density are less than 1 as illustrated in Fig. \ref{fig.IV5}.
	
	\noindent Redshift being another important phenomenon, the expressions for gravitational and surface redshifts in this case become
	\begin{align}
		z&=e^{\frac{-(a_4r^2+b_4)}{2}}-1,
		\\
		z_s&=e^{\frac{c_4r^2}{2}}-1.
	\end{align} 
	\noindent The graph of these having increasing and decreasing nature is demonstrated in Fig. \ref{fig.IV6}. For stability we initially check  the speed of sound in radial, azimuthal and axial direction, given in this case as
	\begin{align}
		\frac{dP_r}{d\rho}&=\frac{c_4(a_4+k_4+a_4k_4r^2)-a_4k_4}{c_4^2(1+k_4r^2)+c_4(-3k_4-k_4^2r^2)+k_4^2},
		\\
		\frac{dP_\theta}{d\rho}&=\frac{-c_4^2r^2(a_4+k_4)+c_4(k_4^2r^2+k_4(2+a_4r^2)+a_4^2r^2+2a_4)-a_4^2-a_4k_4-k_4^2}{c_4^2(1+k_4r^2)+c_4(-3k_4-k_4^2r^2)+k_4^2},
		\\
		\frac{dP_z}{d\rho}&=\frac{a_4^2c_4r^2-a_4c_4^2r^2-a_4^2+3a_4c_4-c_4^2}{c_4^2k_4r^2-c_4k_4^2r^2+c_4^2-3c_4k_4+k_4^2}.
	\end{align}

	\begin{figure}
		\centering
		\begin{minipage}[t]{0.46\textwidth}
			{\gfx{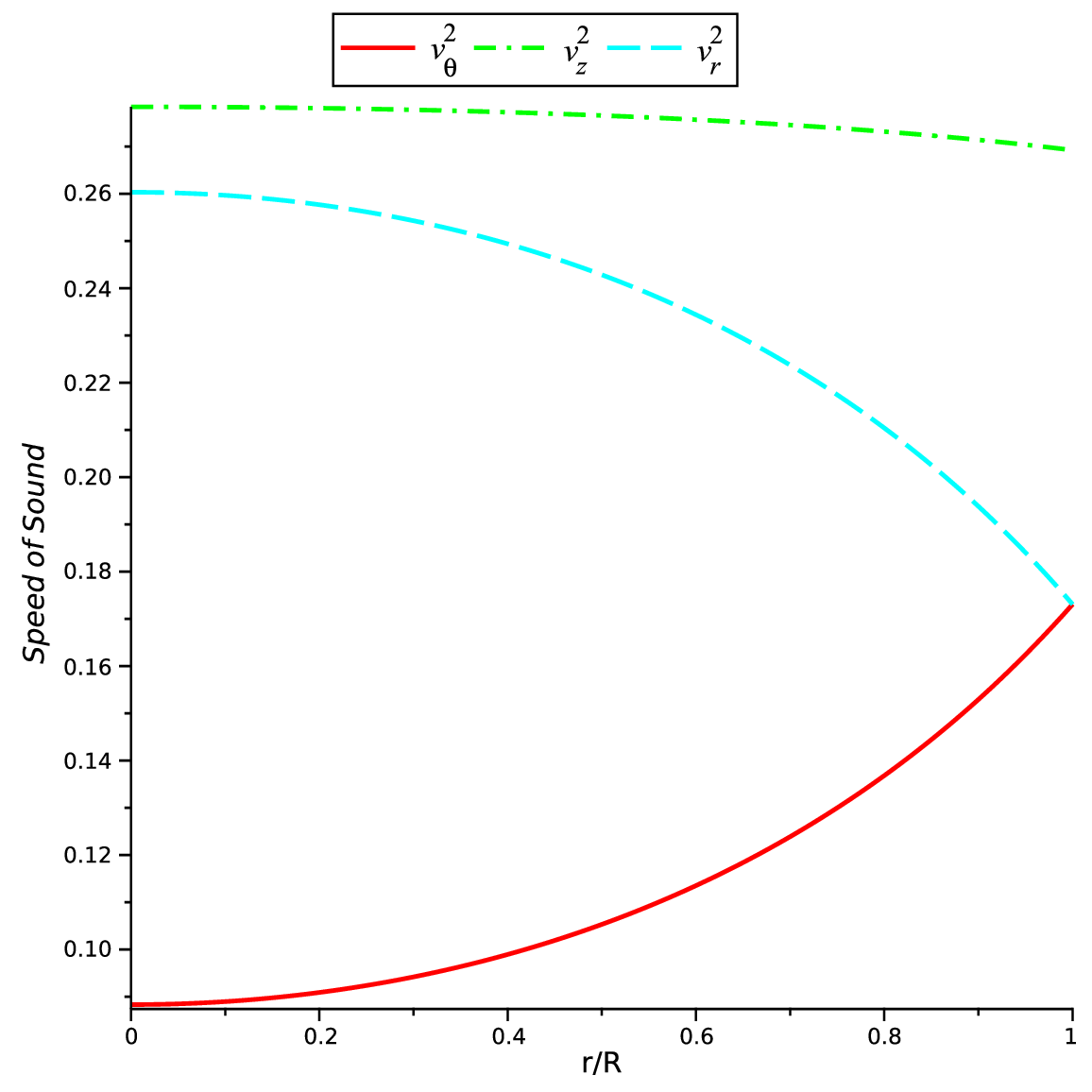}}
			\caption{Graph of causality condition for $a_4=0.75$,$c_4=1$ and $k_4=-0.4285714286$}
			\label{fig.IV7}
		\end{minipage}
		\hfill
		\begin{minipage}[t]{0.46\textwidth}
			{\gfx{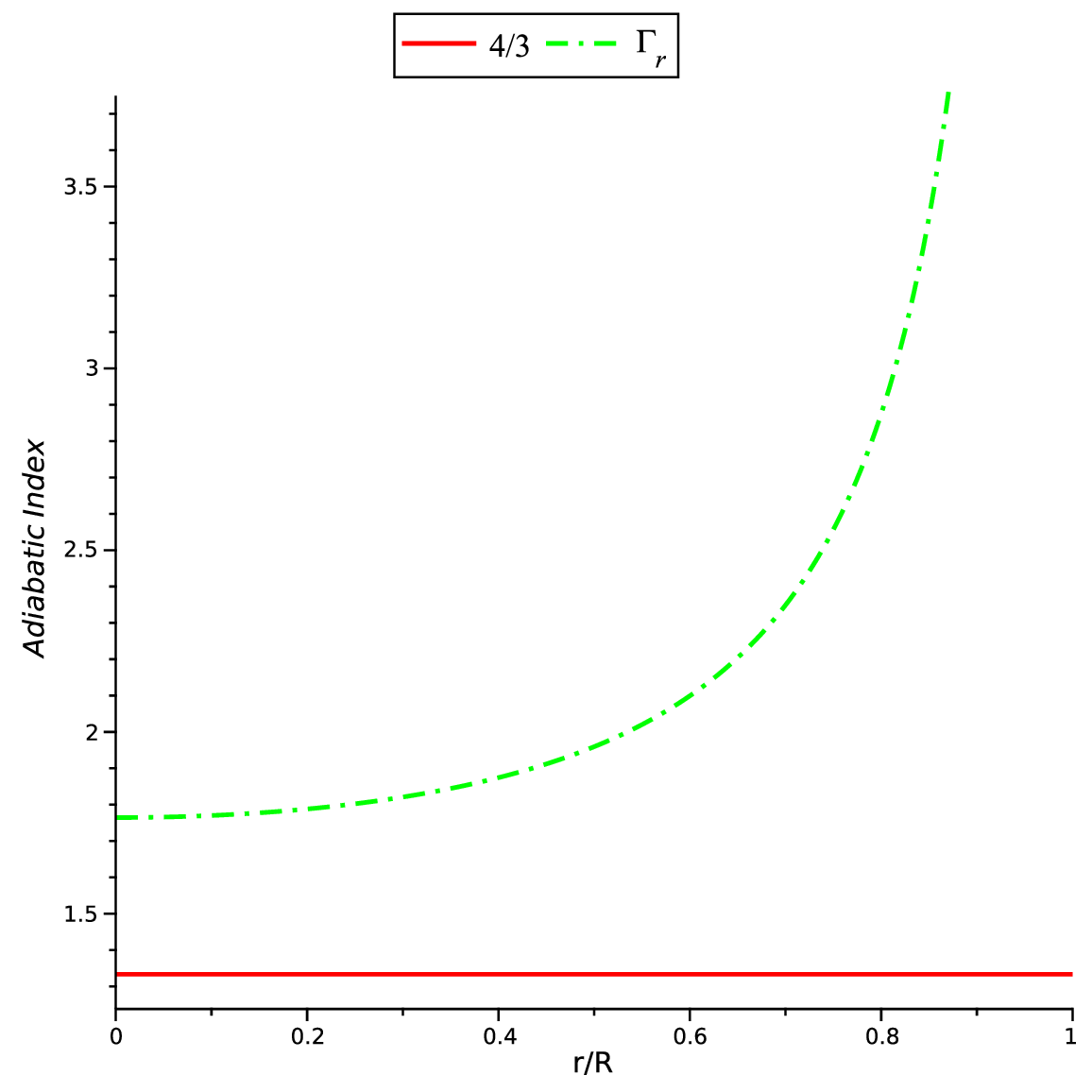}}
			\caption{Graph of adiabatic index for $a_4=0.75$,$c_4=1$ and $k_4=-0.4285714286$}
			\label{fig.IV9}
		\end{minipage}
	\end{figure}
	
	\noindent The behavior of sound in relation to light is depicted in Fig. \ref{fig.IV7}, which guarantees the solution's viability and satisfies the causality criteria. 
	
	\noindent The stellar configuration must adhere to the energy criteria for a physically plausible solution. The energy criteria described in type III are validated in this situation since the fluid distribution in this Segre type is taken similarly to that. The density and all three pressures remain positive (Fig.s \ref{fig.IV2} and \ref{fig.IV3}), thereby satisfying the NEC, WEC, and SEC. The DEC is also satisfied, as the pressure-to-density ratios remain below $1$ throughout (Fig. \ref{fig.IV5}). If $\Gamma_r$, the adiabatic index, is larger than 4/3, the matter composition is stable; the equation for this is as follows. 
	\begin{equation}
		\Gamma_r=\frac{(a_4+c_4-k_4)(1+k_4r^2)(a_4c_4+k_4(a_4c_4r^2-a_4+c_4))}{(a_4+k_4(1+a_4r^2))(k_4^2(1-c_4r^2)+k_4(c_4^2r^2-3c_4)+c_4^2)}.
	\end{equation}
	Fig. \ref{fig.IV9} depicts the stability of the solution satisfying the criteria. The acquired solution in this Segre type $[1,111]$ satisfies all the stability and physical requirements, hence, the proper solution in cylindrical geometry is achieved where the eigenvalue degeneracy relates to all different values. In previous studies, the Krori-Barua functions given in Eqs.~\eqref{IV1} and \eqref{IV2} have been used to describe spherically symmetric compact objects \cite{n46,n47,n48,n49}. Here, we extend their application to cylindrical geometry by combining them with Eq.~\eqref{IV3}. The resulting solution is non-conformally flat.

	\section*{V:\ \ Segre Type $[(1,11)1]$}
	\noindent The Segre type $[(1,11)1]$ represents the presence of tachyonic fluid \cite{41,41a} which results in the existence of an exotic form of matter. Although tachyonic fluids behave hypothetically and speculatively, contradicting physical intuition, mathematical solutions are acceptable in GR, while the required physical criteria are yet unknown. 
	{
		
		\noindent Here the timelike, radial and axial directions share a single eigenvalue and only the azimuthal one differs. We keep the EMT of type III, equations \eqref{III4} and \eqref{III5}, so the field equations are again \eqref{III6}-\eqref{III9}. The non-zero NP scalars are
		\begin{equation}\label{VI1}
			\phi_{02}=\phi_{20}=2\phi_{11}\ne0,
		\end{equation}
		with $\phi_{00}=\phi_{22}=0$. The constraints are equations \eqref{I8} and \eqref{III2}, equation \eqref{III3} coming from $\phi_{00}=\phi_{22}=0$, and
		\begin{equation}\label{VI3}
			2\mu''+\mu'^2-4\nu''-2\nu'^2+\mu'(\nu'-\lambda')+2\nu'\lambda'+\frac{2}{r}\left(\lambda'-\nu'+2\mu'\right)=0
		\end{equation}
		from $\phi_{02}-2\phi_{11}=0$. As in type III, adding equations \eqref{III6} and \eqref{III7} and using equation \eqref{III3} gives $\rho=-P_r$, and equation \eqref{VI3} turns out to be equivalent to $P_z=-\rho$. So the timelike, radial and axial directions share the eigenvalue $-\rho$ and only the azimuthal pressure stands apart. A known source of this form is the Goldstone field outside the core of a global cosmic string, with $P_\theta=\rho$ \cite{CK}.
		
		\noindent We have seven unknowns $(\nu,\lambda,\mu,\rho,P_r,P_\theta,P_z)$ but only six equations, namely the four field equations \eqref{III6}-\eqref{III9} together with \eqref{III3} and \eqref{VI3}, which leaves one function free. We choose
		\begin{equation}\label{VI4}
			\mu=k_5r^2.
		\end{equation}
		Equation \eqref{III3} then gives $\nu+\lambda=k_5r^2+c_5$, and equation \eqref{VI3} reduces to $\lambda''+\left(2k_5r+\frac{1}{r}\right)\lambda'-\lambda'^2=0$, whose only solution regular on the axis is $\lambda'=0$. Taking $\left(e^{\lambda}\right)|_{r=0}=1$, the metric potentials become
		\begin{equation}\label{VI5}
			\nu=k_5r^2+c_5,\hspace{1cm}\lambda=0,
		\end{equation}
		and the field equations \eqref{III6}-\eqref{III9} give
		\begin{align}
			\rho&=-P_r=-P_z=-k_5\left(2+k_5r^2\right),\label{VI6}
			\\
			P_\theta&=k_5\left(2+3k_5r^2\right).\label{VI7}
		\end{align}
		Requiring the radial pressure to vanish at the boundary, $P_r|_{(r=R)}=0$, fixes
		\begin{equation}\label{VI8}
			k_5=\frac{-2}{R^2},
		\end{equation}
		i.e. $k_5=-2$, which gives $\rho=4(1-r^2)$ and $P_\theta=-4(1-3r^2)$. So, unlike the other types, this solution leaves no free constant apart from the boundary radius.
	}
	\begin{figure}
		\centering
		\begin{minipage}[t]{0.31\textwidth}
			\centering
			{\gfx{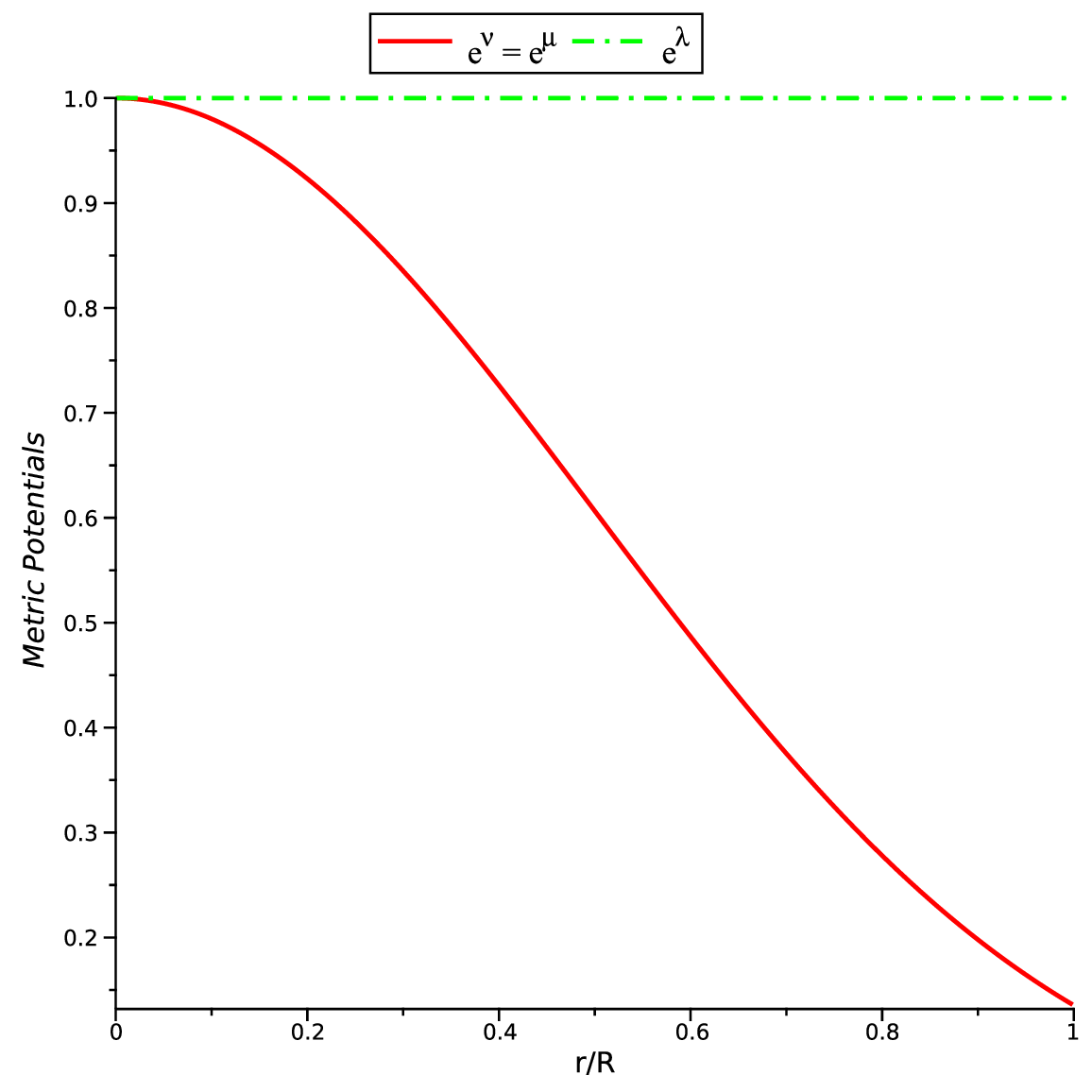}}
			\caption{Graph of metric potentials for $k_5=-2$ and $c_5=0$}
			\label{fig.VI1}
		\end{minipage}
		\hfill
		\begin{minipage}[t]{0.31\textwidth}
			\centering
			{\gfx{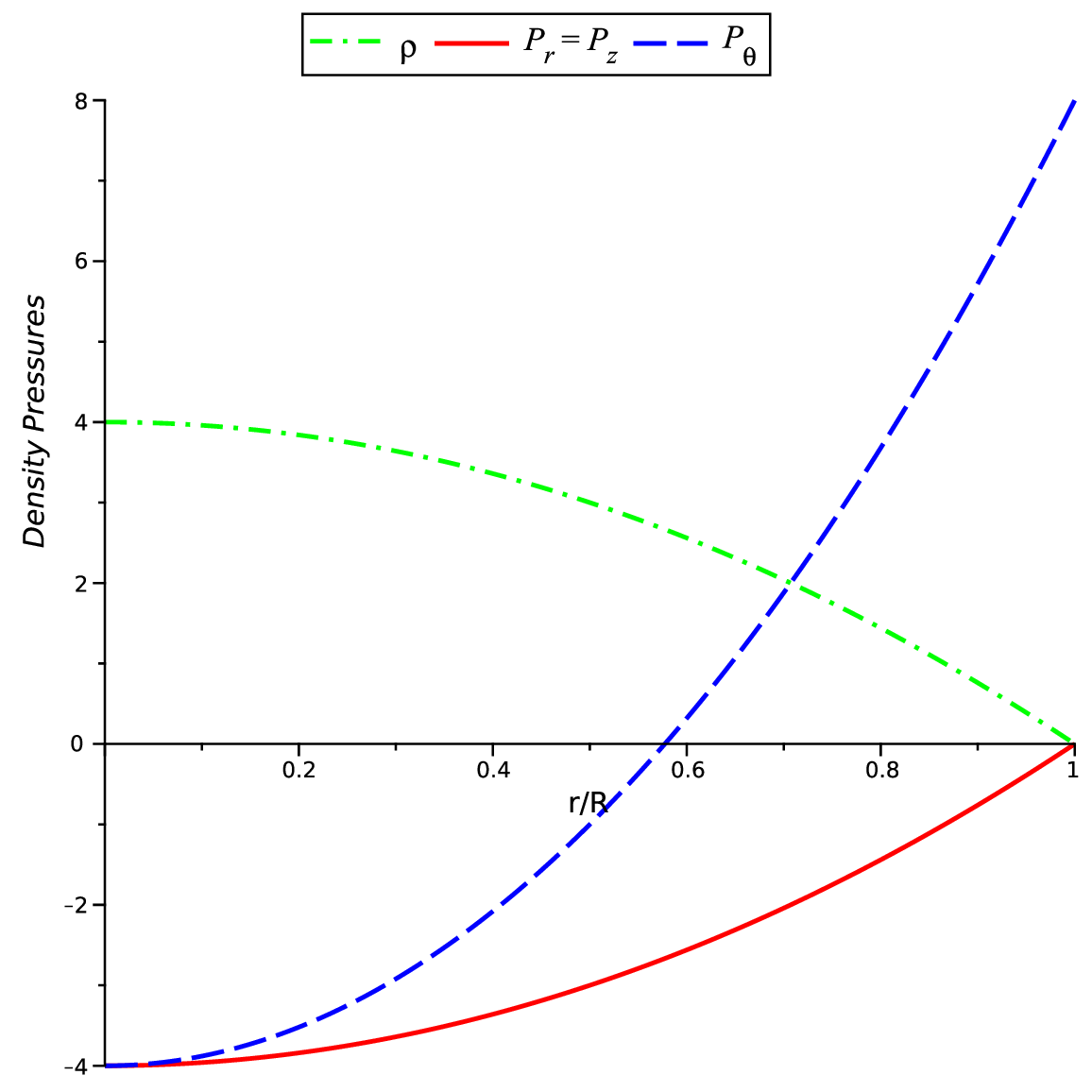}}
			\caption{Graph of energy density and pressures for $k_5=-2$}
			\label{fig.VI2}
		\end{minipage}
		\hfill
		\begin{minipage}[t]{0.31\textwidth}
			\centering
			{\gfx{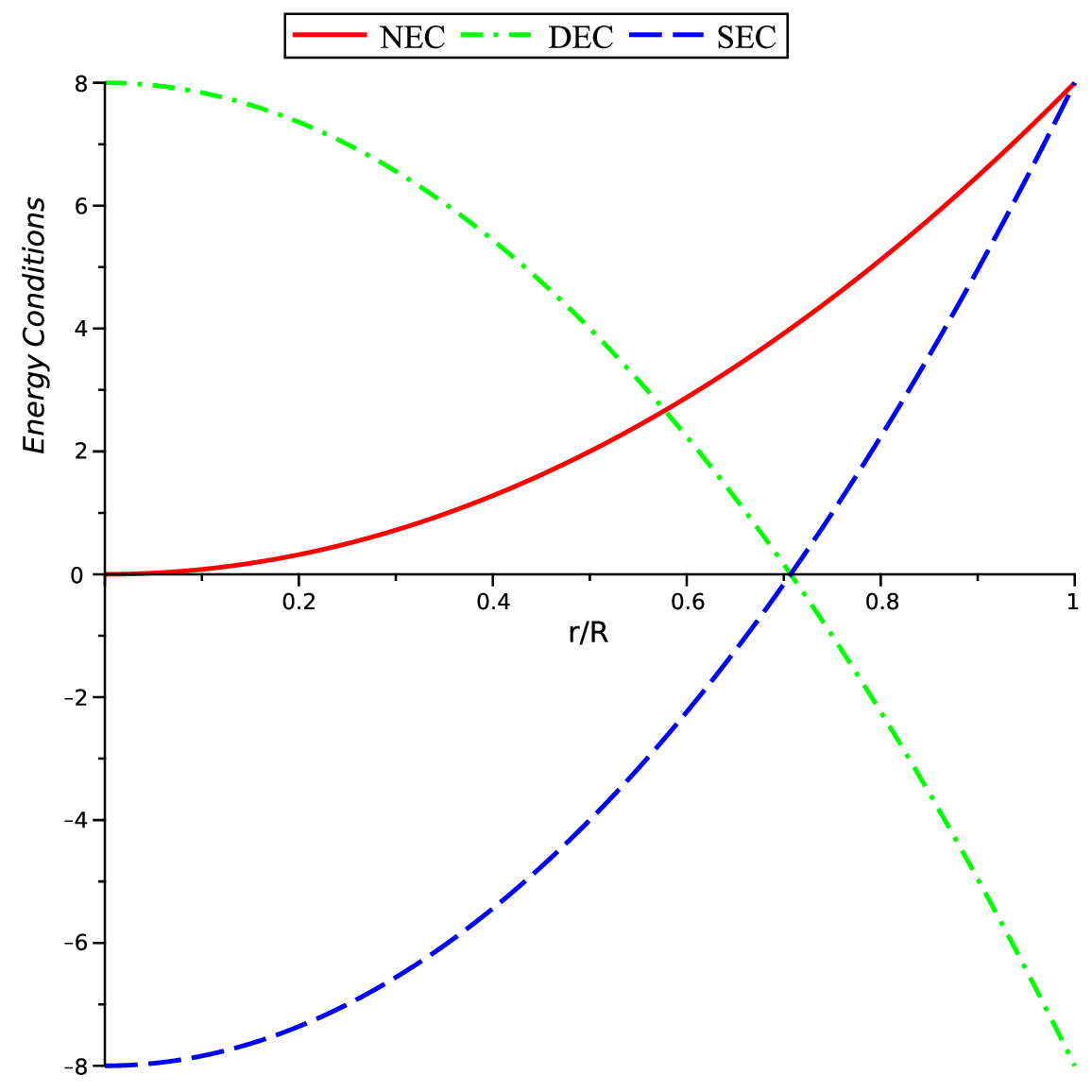}}
			\caption{Graph of energy conditions for $k_5=-2$}
			\label{fig.VI3}
		\end{minipage}
	\end{figure}
	{
		\noindent For $c_5=0$ the metric potentials in Fig. \ref{fig.VI1} are regular everywhere. In Fig. \ref{fig.VI2} the density is positive and decreasing, and it vanishes at the boundary together with the radial and axial pressures. The azimuthal pressure is negative in the core, turns positive towards the boundary, and equals the radial pressure at the center, i.e. $(P_r=P_\theta)|_{r=0}=2k_5$.
		
		\noindent The energy conditions are defined as in type III. Fig. \ref{fig.VI3} shows that NEC and WEC are satisfied, SEC is violated in the core, as expected for dark energy, and DEC fails near the boundary. Because the radial and axial pressures equal $-\rho$, the speed of sound and the adiabatic index do not apply here. The solution for Segre type $[(1,11)1]$ is therefore a regular, dark energy like configuration with the tachyonic fluid form of EMT. Compared with earlier work, the exterior of a global string \cite{CK} has the same form of EMT, whereas the present solution is a regular and bounded interior.}

	\section*{VI:\ \ Segre Type $[(1,1)(11)]$}
	\noindent The type $[(1,1)(11)]$ signifies the presence of a non-null Maxwell field \cite{37}. In cylindrical geometry, this leads to a constraint equation due to $\phi_{11}\ne0$ given by equation \eqref{I8}. In this case, $\phi_{02}=\phi_{20}=0$ and $\phi_{00}=\phi_{22}=0$ lead to the two equations given by \eqref{I9} and \eqref{III3}. These equations result in the expression $\lambda'=\frac{-r}{2}\mu'\nu'$  and $\nu'=\frac{-2\mu''-\mu'^2}{r/2\mu'^2-2/r}$. A suitable form of function satisfying the physical criteria can be worked out here meeting the necessary requisites for solution. {
		
		\noindent For a pure Maxwell field the EMT takes the form
		\begin{equation}\label{VII1}
			T^{\ \ (em)}_{ab}=2\left(F_a^{\ d}F_{bd}-\frac{1}{4}g_{ab}F_{cd}F^{cd}\right),
		\end{equation}
		where the factors of $\pi$ are absorbed in the normalization of $F_{ab}$, so that the same $\kappa=1$ applies. A static radial electric field has only $F_{tr}$ non-zero, and with $E^2=-F_{tr}F^{tr}$ this gives
		\begin{equation}\label{VII2}
			T^{a\ (em)}_{b}=\mathrm{diag}(-E^2,-E^2,E^2,E^2).
		\end{equation}
		This has the eigenvalue structure $[(1,1)(11)]$ and is trace-free, so on top of the two equations above the Ricci scalar \eqref{8} must vanish as well. The source-free Maxwell equations $\nabla_bF^{ab}=0$ then give
		\begin{equation}\label{VII3}
			E=\frac{q\,e^{-\mu/2}}{r},
		\end{equation}
		where the constant $q$ is proportional to the charge per unit length. This makes three equations for three metric functions, so no free function remains and the electrovac solution is fixed up to constants. Its simplest member, in the form of metric \eqref{2}, is
		\begin{equation}\label{VII4}
			e^{\nu}=e^{-\lambda}=\frac{k_0}{r}+\frac{q^2}{r^2},\hspace{1cm}e^{\mu}=r^2,
		\end{equation}
		for which $E^2=q^2/r^4$. This is the analogue of the Reissner-Nordstr\"om solution in this geometry (compare the charged black strings and planes with $\Lambda<0$ in \cite{Lemos,CaiZhang}). For $q=0$ it reduces to the Levi-Civita metric with $\sigma=-1/2$, and the general family is given by equation \eqref{J14} \cite{Bonnor53,37}. Since $\rho=-P_r=P_\theta=P_z=E^2$, every energy condition holds, but the electric field diverges on the axis. It can therefore describe only the field outside a charged source, and a regular interior is still needed.
		
		\noindent Following \cite{35}, a regular source of the same Segre type can be had by combining an anisotropic fluid with the electromagnetic field, $T_{ab}=T^{\ \ (apf)}_{ab}+T^{\ \ (em)}_{ab}$, so that
		\begin{equation}\label{VII5a}
			T^a_b=\mathrm{diag}(-\rho-E^2,P_r-E^2,P_\theta+E^2,P_z+E^2).
		\end{equation}
		Imposing $\phi_{00}=\phi_{22}=0$ and $\phi_{02}=\phi_{20}=0$ gives
		\begin{equation}\label{VII5b}
			\rho=-P_r,\hspace{1cm}P_\theta=P_z.
		\end{equation}
		The fluid is therefore vacuum-like in the radial direction \cite{Dym}, and equation \eqref{VII3} still holds with $q$ replaced by $q(r)$, the charge per unit length enclosed within radius $r$. Since the Segre type fixes only the eigenvalue structure of the total EMT, the unknowns are $(\nu,\lambda,\mu,\rho,P_\theta,E)$, and as in \cite{35} the metric function $\mu$ and the electric field can be chosen freely. With $\mu=k_6r^2$ the two equations above integrate to
		\begin{equation}\label{VII5}
			e^{\nu}=\frac{c_6}{1-k_6r^2},\hspace{1cm} e^{\lambda}=\left(1-k_6r^2\right)e^{k_6r^2},
		\end{equation}
		and the field equations \eqref{9}-\eqref{12} give
		\begin{equation}\label{VII6}
			\rho+E^2=\frac{-2k_6e^{-k_6r^2}}{\left(1-k_6r^2\right)^2},\hspace{1cm} P_\theta+E^2=\frac{2k_6\left(1+k_6r^2\right)e^{-k_6r^2}}{\left(1-k_6r^2\right)^3}.
		\end{equation}
		For the second choice, the electric field must vanish on the axis, so we take
		\begin{equation}\label{VII7}
			E^2=\frac{-2k_6r^2e^{-k_6r^2}}{R^2\left(1-k_6r^2\right)^2},\hspace{1cm}q(r)=\frac{\sqrt{-2k_6}\,r^2}{R\left(1-k_6r^2\right)},
		\end{equation}
		which needs $k_6<0$. The density and pressures then become
		\begin{align}
			\rho&=-P_r=\frac{-2k_6e^{-k_6r^2}}{\left(1-k_6r^2\right)^2}\left(1-\frac{r^2}{R^2}\right),\label{VII8}
			\\
			P_\theta&=P_z=\frac{2k_6e^{-k_6r^2}}{\left(1-k_6r^2\right)^3}\left(1+k_6r^2+\frac{r^2}{R^2}\left(1-k_6r^2\right)\right).\label{VII9}
		\end{align}
	}
	\begin{figure}
		\centering
		\begin{minipage}[t]{0.31\textwidth}\subfloat[]	
			{\gfx{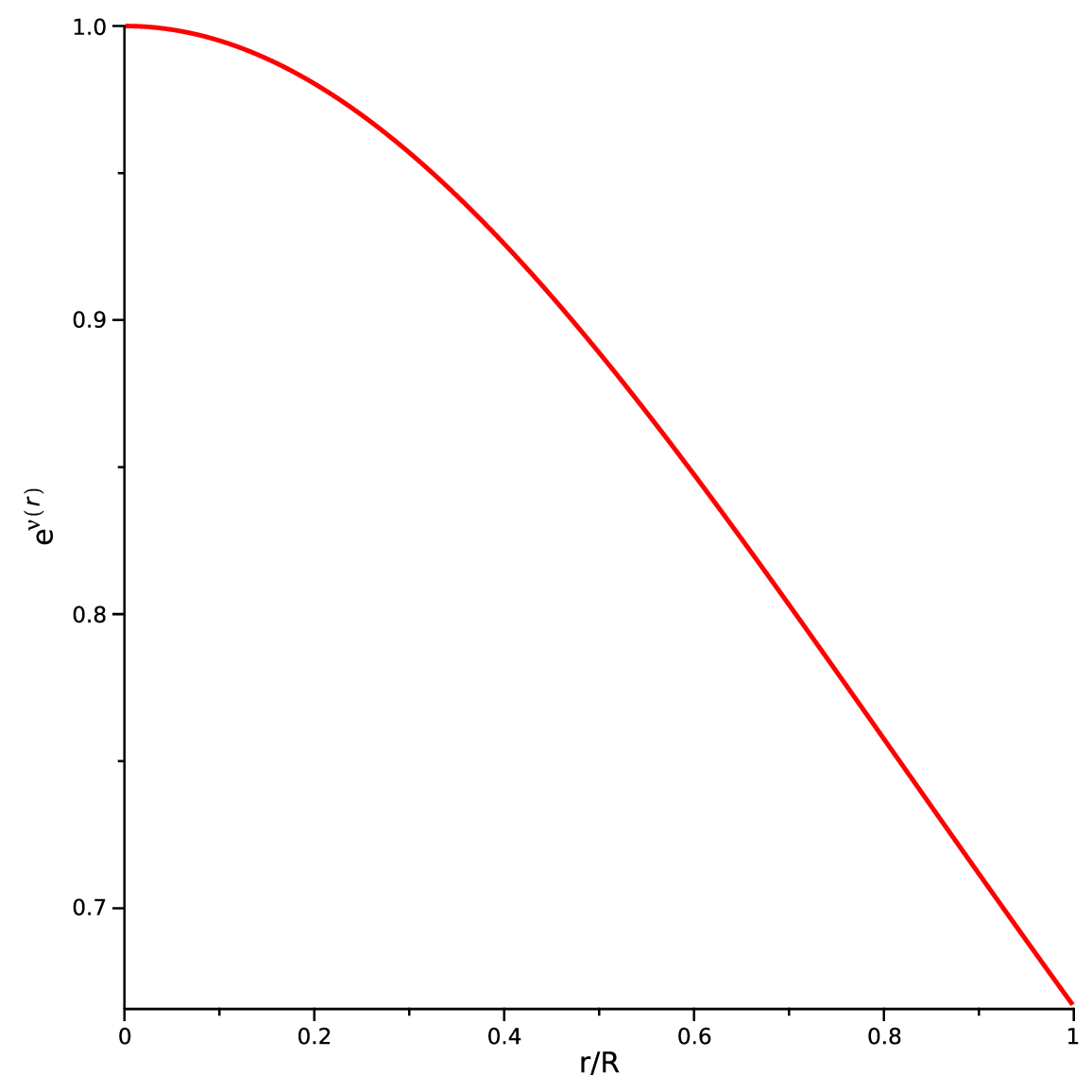}}
		\end{minipage}
		\hfill
		\begin{minipage}[t]{0.31\textwidth}\subfloat[]	
			{\gfx{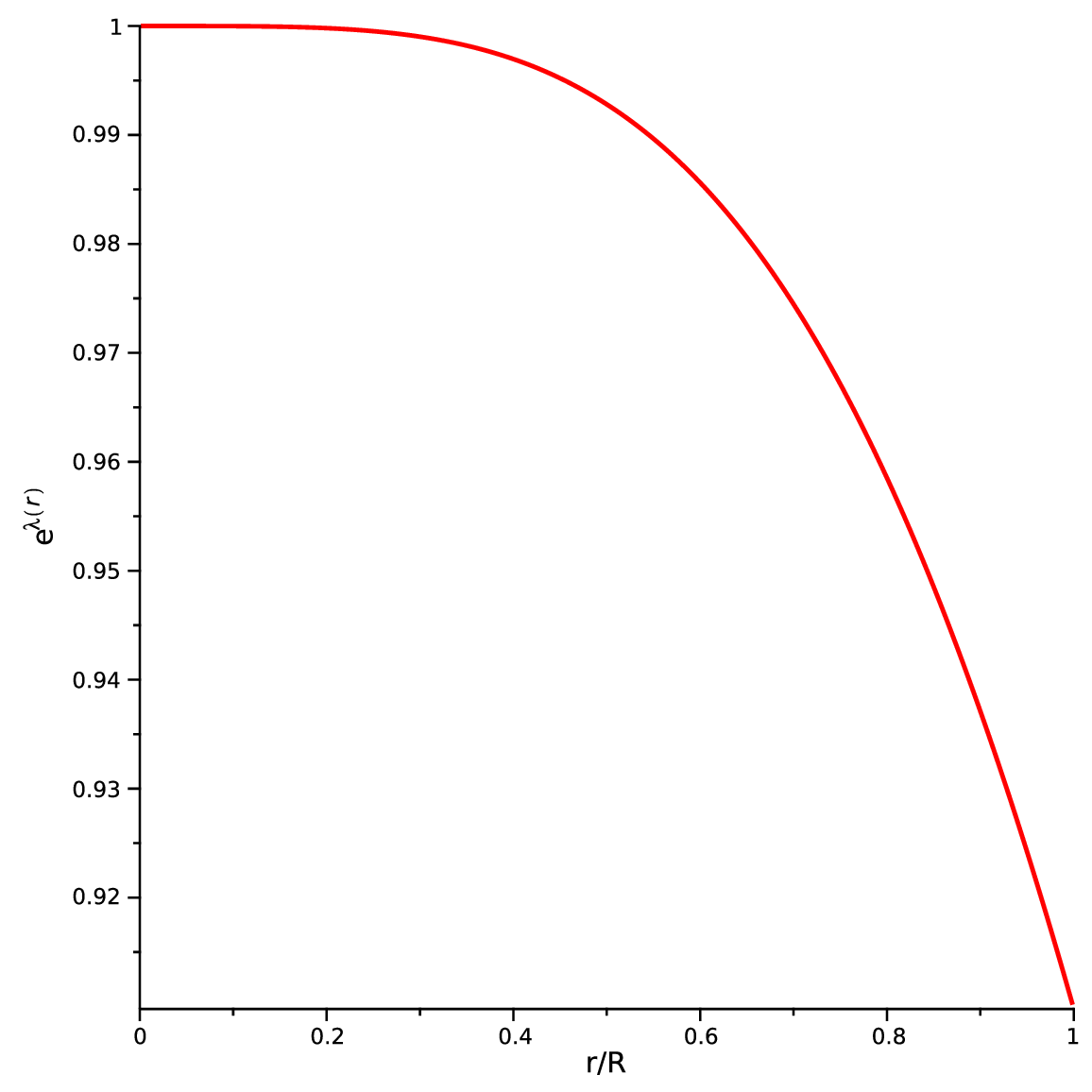}}
		\end{minipage}
		\hfill
		\begin{minipage}[t]{0.31\textwidth}\subfloat[]	
			{\gfx{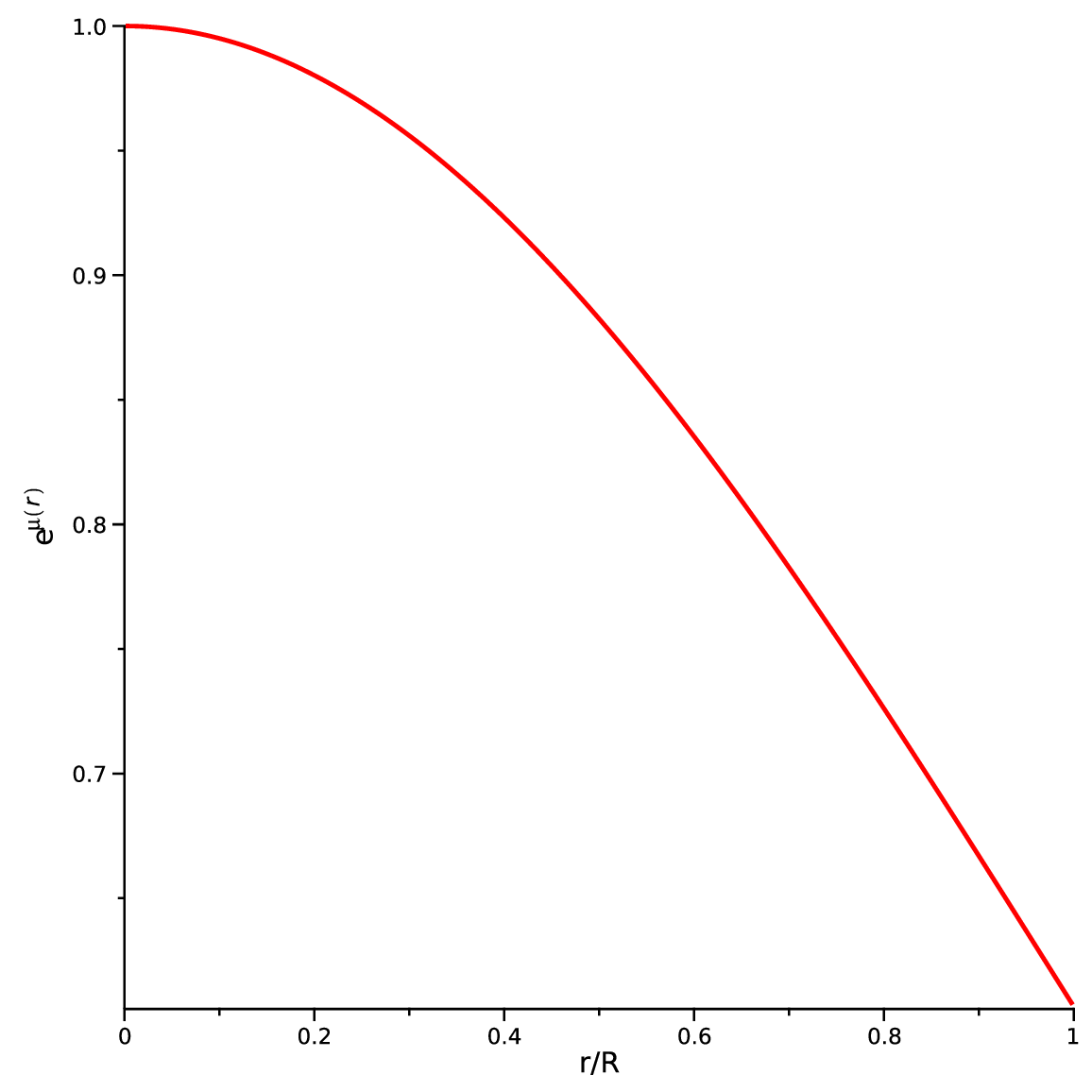}}
		\end{minipage}
		
		\caption{Graph of metric potentials for $k_6=-0.5$ and $c_6=1$}
		\label{fig.VII1}
	\end{figure}
	
	\begin{figure}
		\centering
		\begin{minipage}[t]{0.31\textwidth}
			\centering
			{\gfx{VIIpP.eps}}
			\caption{Graph of energy density and pressures for $k_6=-0.5$}
			\label{fig.VII2}
		\end{minipage}
		\hfill
		\begin{minipage}[t]{0.31\textwidth}
			\centering
			{\gfx{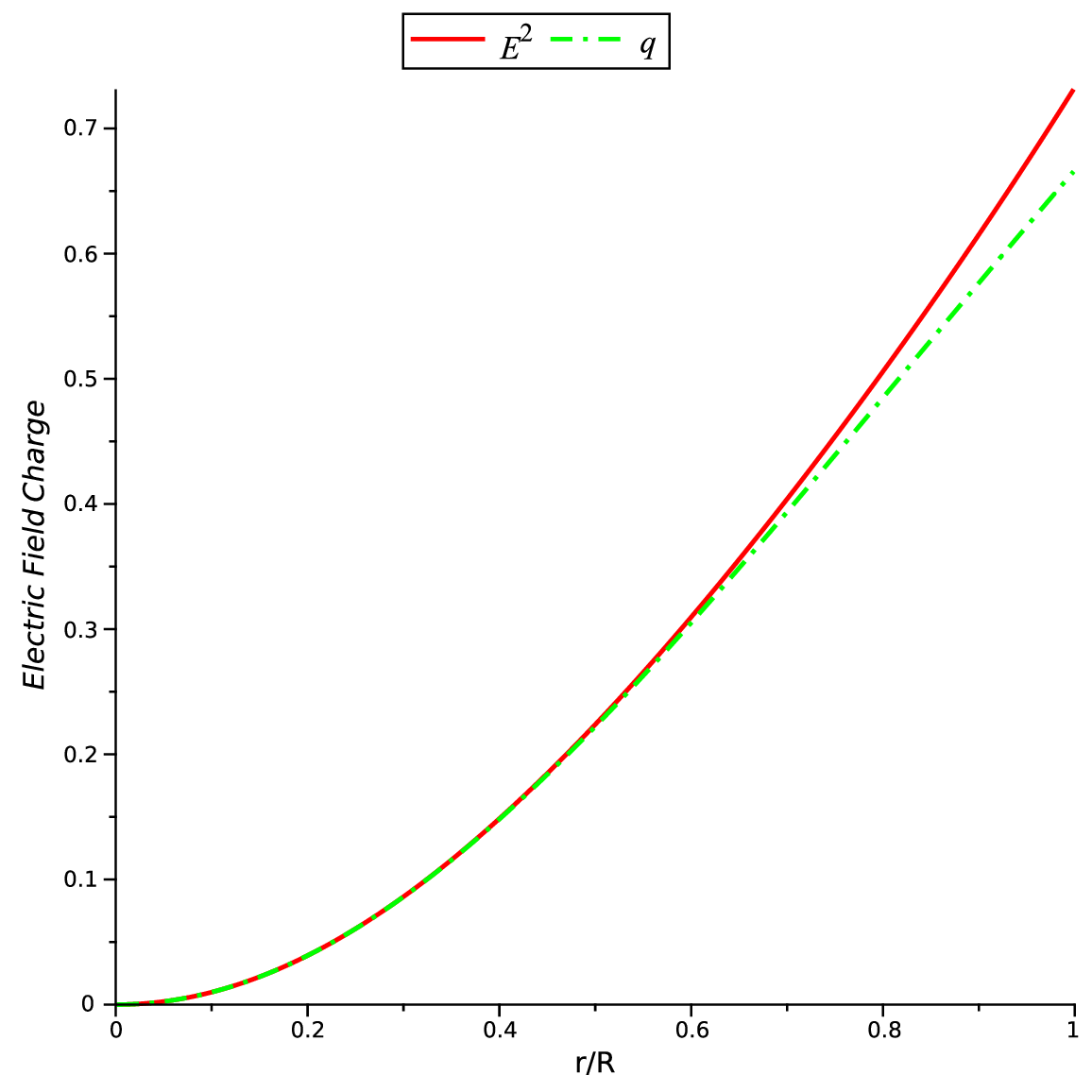}}
			\caption{Graph of electric field and charge for $k_6=-0.5$}
			\label{fig.VII3}
		\end{minipage}
		\hfill
		\begin{minipage}[t]{0.31\textwidth}
			\centering
			{\gfx{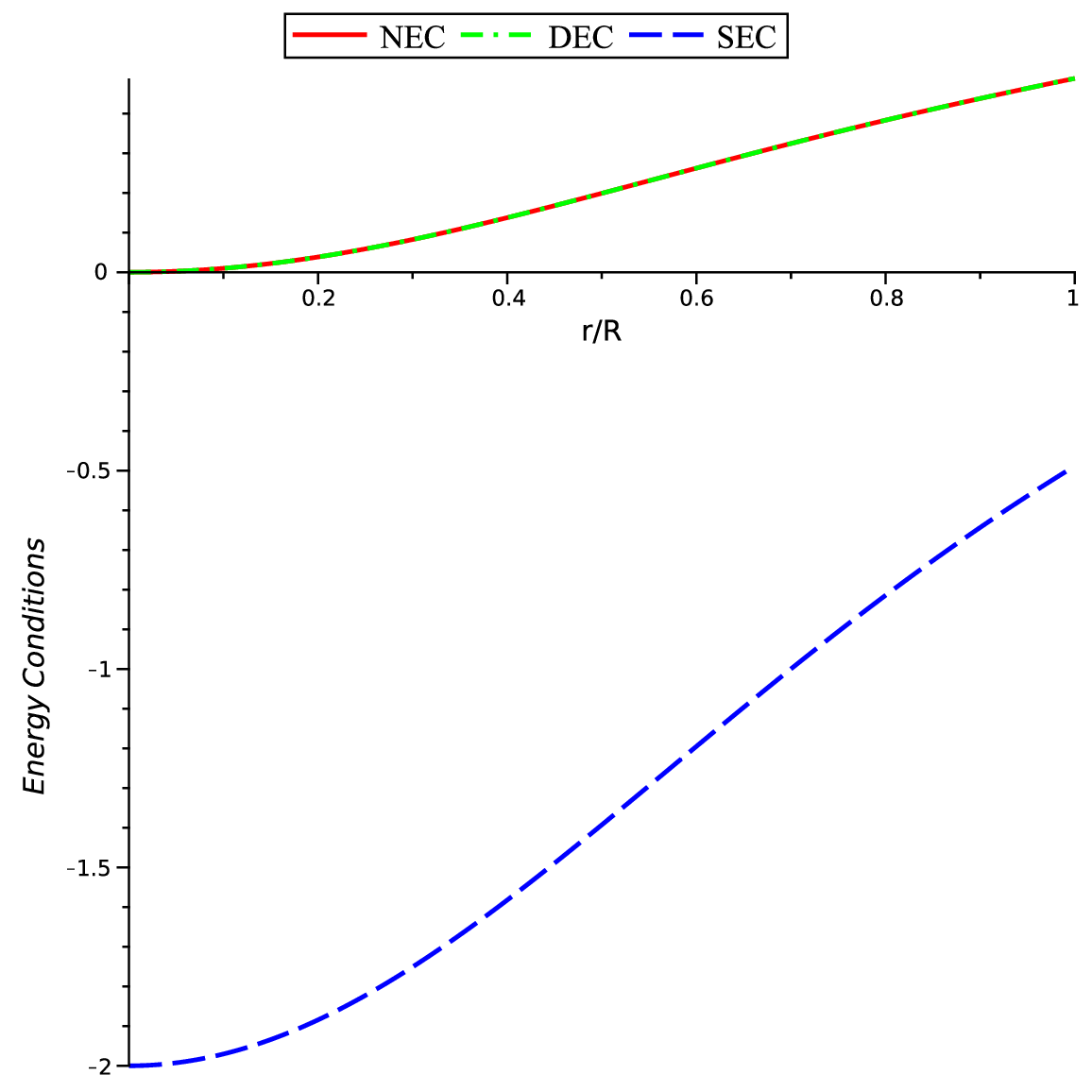}}
			\caption{Graph of energy conditions for $k_6=-0.5$}
			\label{fig.VII4}
		\end{minipage}
	\end{figure}
	{
		\noindent For $k_6=-0.5$ and $c_6=1$ the metric potentials in Fig. \ref{fig.VII1} are regular, with $\left(e^{\lambda}\right)|_{r=0}=1$ and $e^{\nu}$ equal to the positive constant $c_6$ at the center. The decrease of $e^{\lambda}$ from $1$ towards the boundary follows from the field equations of this Segre type and not from the chosen function. In Fig. \ref{fig.VII2} the density is positive, decreasing and vanishes at the boundary together with the radial pressure, while the pressures stay negative, so this type is again dark energy like as in type III. Fig. \ref{fig.VII3} gives the electric field and the charge, both zero on the axis.
		
		\noindent For the energy conditions the total EMT is used. Fig. \ref{fig.VII4} shows that NEC, WEC and DEC are satisfied, whereas SEC is violated. Since $\rho=-P_r$, the causality condition and the adiabatic index do not apply here, as in \cite{35}. The solution for Segre type $[(1,1)(11)]$ is therefore a regular charged anisotropic interior in the presence of the non-null electromagnetic field. Compared with earlier work, the electrovac member is the known solution \cite{Bonnor53,37}, and the construction follows the spherical case \cite{35}.}
	
	\section*{VII:\ \ Segre Type $[(1,111)]$}
	\noindent In the Segre type $[(1,111)]$, the eigenvalue degeneracy refers to the spacelike and timelike eigenvectors having similar eigenvalues. This particular type corresponds to the vacuum cases when finding the solution i.e. $R_{ab}=0$ \cite{37}. The vacuum solutions in cylindrical geometry \eqref{2} for the Einstein field equations are discussed in detail by \cite{24}. It is worth noting that a vanishing trace-free Ricci tensor describes Einstein spaces in general, so this type also includes the static cylindrically symmetric solutions with a cosmological constant \cite{Linet,Tian}.
	
	\section{Junction Conditions, C-energy and Comparison of the Solutions}\label{S8}
	{
		\noindent In every Segre type above the radial pressure was taken to vanish at the boundary $r=R$. In this section we check that this condition gives a smooth junction with the static solution outside the source, calculate the C-energy, which plays the role of the mass function in cylindrical geometry, and compare the solved types.}
	\subsection{Matching with the Levi-Civita exterior}
	\noindent{The most general static cylindrically symmetric vacuum solution is the Levi-Civita metric \cite{15a,23}, which can be written as
		\begin{equation}\label{J1}
			ds^2=-A^2\varrho^{4\sigma}dt^2+B^2\varrho^{4\sigma(2\sigma-1)}d\varrho^2+C^2\varrho^{2(1-2\sigma)}d\theta^2+D^2\varrho^{4\sigma(2\sigma-1)}dz^2,
		\end{equation}
		where $A$, $B$, $C$ and $D$ are constants and $\sigma$ is the Levi-Civita parameter. For small $\sigma$ it can be taken as the Newtonian mass per unit length of the source, and this holds for $0<\sigma<1/4$, the range where timelike circular geodesics exist \cite{23}. We join the interior region $r\le R$, described by metric \eqref{2}, to \eqref{J1} across the hypersurface $\Sigma$: $r=R$ $(\varrho=\varrho_\Sigma)$ using the Darmois conditions \cite{Darmois,Israel}, i.e. continuity of the first and second fundamental forms. Both metrics are static and diagonal, so with the unit normal $n^a=e^{-\lambda/2}\delta^a_r$ the extrinsic curvature is $K_{ij}=\frac{1}{2}e^{-\lambda/2}\partial_rg_{ij}$ for $i,j=t,\theta,z$. Continuity of the metric components gives
		\begin{equation}\label{J2}
			e^{\nu(R)}=A^2\varrho_\Sigma^{4\sigma},\hspace{0.7cm} R^2=C^2\varrho_\Sigma^{2(1-2\sigma)},\hspace{0.7cm} e^{\mu(R)}=D^2\varrho_\Sigma^{4\sigma(2\sigma-1)},
		\end{equation}
		and continuity of $K_{ij}$ gives
		\begin{equation}\label{J3}
			e^{-\frac{\lambda(R)}{2}}\nu'(R)=\frac{4\sigma\epsilon}{B\varrho_\Sigma^{N}},\hspace{0.25cm} e^{-\frac{\lambda(R)}{2}}\frac{2}{R}=\frac{2(1-2\sigma)\epsilon}{B\varrho_\Sigma^{N}},\hspace{0.25cm} e^{-\frac{\lambda(R)}{2}}\mu'(R)=\frac{4\sigma(2\sigma-1)\epsilon}{B\varrho_\Sigma^{N}},
		\end{equation}
		where $N=4\sigma^2-2\sigma+1$ and $\epsilon=\mathrm{sign}(1-2\sigma)$, which is needed because for $\sigma>1/2$ the circumferential radius of \eqref{J1} grows as $\varrho$ decreases. Dividing the first and the third relations by the second, the constants $B$ and $\varrho_\Sigma$ cancel and
		\begin{equation}\label{J4}
			\frac{R\nu'(R)}{2}=\frac{2\sigma}{1-2\sigma},\hspace{1cm}\frac{R\mu'(R)}{2}=-2\sigma.
		\end{equation}
		The second relation gives the Levi-Civita parameter directly from the interior solution,
		\begin{equation}\label{J5}
			\sigma=-\frac{R\mu'(R)}{4},
		\end{equation}
		and putting it in the first relation leaves $\mu'\nu'+\frac{2}{R}(\mu'+\nu')=0$ at $r=R$, which by equation \eqref{10} is exactly $P_r|_{(r=R)}=0$. Thus the vanishing of the radial pressure at the boundary, which we imposed in every Segre type, is both necessary and sufficient for the matching with the Levi-Civita exterior.
		
		\noindent With the function $\mu$ of each type, equation \eqref{J5} gives
		\begin{equation}\label{J7}
			\begin{aligned}
				\sigma_{I}&=\frac{-a_1R^2}{2(1+a_1R^2)},\hspace{0.5cm}\sigma_{II}=\frac{-b_2R^2}{2(1+b_2R^2)},\\
				\sigma_{III}&=\frac{-b_3R^2}{2},\hspace{0.5cm}\sigma_{IV}=\frac{-k_4R^2}{2},\hspace{0.5cm}\sigma_{V}=\frac{-k_5R^2}{2},
			\end{aligned}
		\end{equation}
		For the constants used in the graphical analysis these become $\sigma_{I}=1/6$, $\sigma_{II}=1/8$, $\sigma_{III}=0.995$, $\sigma_{IV}=3/14$ and $\sigma_{V}=1$. Types I, II and IV therefore lie in the range $0<\sigma<1/4$, which means positive mass per unit length and an attractive field outside. For types III and V we have $\sigma>1/2$ and $\nu'(R)<0$, so $g_{tt}$ decreases outward at the boundary and the gravitational field there is repulsive.}
	
	\subsection{Charged exterior for type $[(1,1)(11)]$}
	\noindent{For type VI the electric field extends outside the boundary, so the exterior is the static cylindrically symmetric electrovac solution \cite{Bonnor53,37,23}
		\begin{equation}\label{J14}
			ds^2=-\frac{A^2}{G^2}dt^2+B^2\varrho^{2m^2}G^2d\varrho^2+C^2\varrho^2G^2d\theta^2+D^2\varrho^{2m^2}G^2dz^2,
		\end{equation}
		where $G=C_1\varrho^m+C_2\varrho^{-m}$. It is of Segre type $[(1,1)(11)]$ and goes back to the Levi-Civita metric with $\sigma=-m/2$ for $C_2=0$. Applying the Darmois conditions in the same way, with $\varrho_\Sigma=1$ and $g=(\varrho G'/G)|_\Sigma$, we get
		\begin{equation}\label{J16}
			\frac{R\nu'(R)}{2}=\frac{-g}{1+g},\hspace{1cm}\frac{R\mu'(R)}{2}=\frac{g+m^2}{1+g}.
		\end{equation}
		For the interior solution \eqref{VII5} these give
		\begin{equation}\label{J17}
			g=-k_6R^2,\hspace{1cm}m^2=k_6R^2\left(2-k_6R^2\right)<0,
		\end{equation}
		so $m=ip$ is imaginary and $G=\cos\left(p\ln\varrho+\psi\right)/\cos\psi$ with $p^2=-m^2$ and $\tan\psi=-g/p$. Since $\rho(R)=0$, the electric field of the exterior at $\Sigma$ equals $E^2(R)=\rho_T(R)$ of the interior, so it is continuous across the boundary, the exterior charge is $q(R)$ and no surface charge is needed. For $k_6=-0.5$ and $R=1$ this gives $g=0.5$, $p=1.1180$ and $\psi=-0.4205$.}
	
	\subsection{C-energy}
\begin{figure}
		\centering
		\begin{minipage}[t]{0.46\textwidth}
			{\gfx{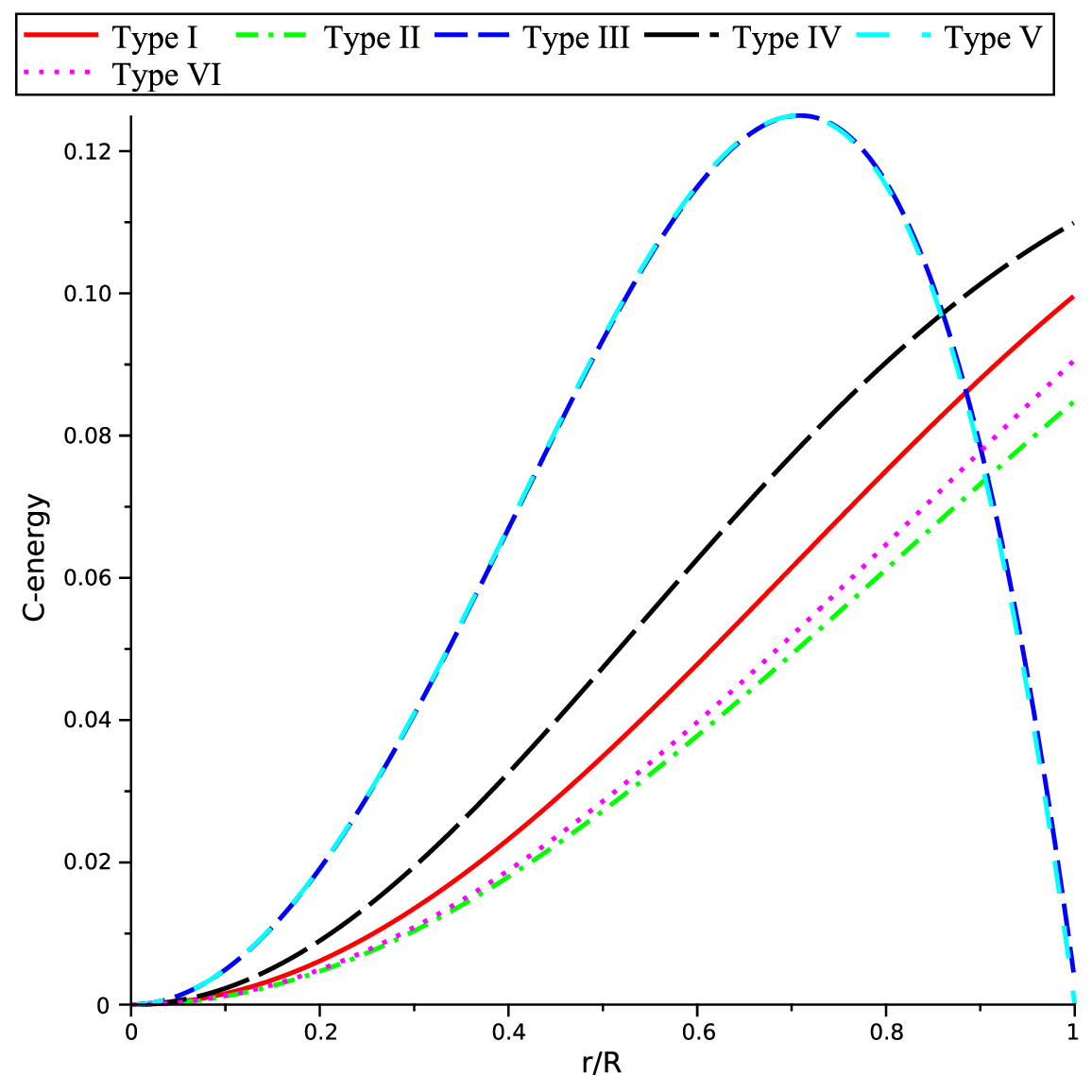}}
			\caption{Graph of C-energy for types I, II, III, IV, V and VI with the constants used in each type}
			\label{fig.CE}
		\end{minipage}
	\end{figure}

	\noindent{In cylindrical geometry the counterpart of the mass function of spherical models is the C-energy introduced by Thorne \cite{Thorne}, and here we use the version proposed in \cite{Hayward}. With the circumferential radius $\sqrt{g_{\theta\theta}}$, the specific length $\ell=\sqrt{g_{zz}}$ and the areal radius $\tilde{r}=\ell\sqrt{g_{\theta\theta}}$ it reads $E=\frac{1}{8}\left(1-\ell^{-2}g^{ab}\partial_a\tilde{r}\,\partial_b\tilde{r}\right)$, and for metric \eqref{2} we have $\tilde{r}=re^{\mu/2}$, so that
		\begin{equation}\label{J8}
			E(r)=\frac{1}{8}\left(1-e^{-\lambda}\left(1+\frac{r\mu'}{2}\right)^2\right),
		\end{equation}
		which vanishes on the axis because $\left(e^{\lambda}\right)|_{r=0}=1$. Putting the metric functions of each solved type in it, we get
		\begin{align}
			E_{I}&=\frac{1}{8}\left(1-\frac{e^{-\frac{b_1r^2}{2}}\left(1+2a_1r^2\right)^2}{1+a_1r^2}\right),\label{J9}
			\\
			E_{II}&=\frac{1}{8}\left(1-e^{-a_2r^2\cosh(b_2r^2)}\left(\frac{1+2b_2r^2}{1+b_2r^2}\right)^2\right),
			\\
			E_{III}&=\frac{1}{8}\left(1-e^{-a_3r^2\cosh(b_3r^2)}\left(1+b_3r^2\right)^2\right),
			\\
			E_{IV}&=\frac{1}{8}\left(1-e^{-c_4r^2}\left(1+k_4r^2\right)^2\right),\label{J12}
			\\
			E_{V}&=\frac{1}{8}\left(1-\left(1+k_5r^2\right)^2\right),
			\\
			E_{VI}&=\frac{1}{8}\left(1-\frac{e^{-k_6r^2}\left(1+k_6r^2\right)^2}{1-k_6r^2}\right).
		\end{align}
		For types I, II and IV the C-energy increases monotonically from the axis to the boundary and reaches $0.0997$, $0.0849$ and $0.1100$ respectively, and type VI behaves in the same way with the boundary value $0.0907$. Type III is different. The areal radius $re^{b_3r^2/2}$ is not monotonic, it is stationary at $r=1/\sqrt{|b_3|}=0.709R$, where $E_{III}$ reaches $1/8$, and then the C-energy falls to $0.0033$ at the boundary. Type V shows the same behavior, $E_{V}$ reaches $1/8$ at $r=R/\sqrt{2}$, where SEC is restored, and vanishes at the boundary. The profiles for all types are shown in Fig. \ref{fig.CE}.}
	{
		
		\subsection{Comparison of the solved types and limiting cases}
		\noindent Table \ref{T2} collects the main characteristics of the solved types, using the constants from the graphical analysis.
		\begin{table}[ht]
			\centering
			\setlength{\tabcolsep}{5pt}
			\renewcommand{\arraystretch}{1.6}
			\resizebox{\textwidth}{!}{%
				\begin{tabular}{|c|c|c|c|c|c|c|c|}
					\hline
					Type & Fluid & $\rho(0)$ & $P_r(0)$ & $P_z(0)$ & $\sigma$ & $E(R)$ & SEC
					\\ \hline
					I & $P_r=P_\theta=P_z$ & $1.2500$ & $0.2500$ & $0.2500$ & $0.1667$ & $0.0997$ & satisfied
					\\ \hline
					II & $P_\theta=P_z$ & $0.9501$ & $0.1501$ & $0.1501$ & $0.1250$ & $0.0849$ & satisfied
					\\ \hline
					III & $\rho=-P_r$ & $3.9818$ & $-3.9818$ & $-3.9855$ & $0.9950$ & $0.0033$ & violated, $r<0.709R$
					\\ \hline
					IV & $P_r\ne P_\theta\ne P_z$ & $1.8571$ & $0.3214$ & $0.5000$ & $0.2143$ & $0.1100$ & satisfied
					\\ \hline
					V & $\rho=-P_r=-P_z$ & $4.0000$ & $-4.0000$ & $-4.0000$ & $1.0000$ & $0.0000$ & violated, $r<R/\sqrt{2}$
					\\ \hline
					VI & $\rho=-P_r$, $P_\theta=P_z$, charged & $1.0000$ & $-1.0000$ & $-1.0000$ & $-$ & $0.0907$ & violated
					\\ \hline
			\end{tabular}}
			\caption{Comparison of the solved Segre types (in units $\kappa=1$, $R=1$). The violation of SEC in types III, V and VI is the expected behaviour of dark energy like sources. In all types $P_\theta(0)=P_r(0)$, and for type VI the exterior is the electrovac solution \eqref{J14}}\label{T2}
		\end{table}
		
		\noindent As a check, each solution reduces to flat spacetime when its free parameter is switched off, by equations \eqref{I16}, \eqref{II13} and \eqref{III15} for types I, II and III. In type IV, $k_4\to0$ forces $a_4\to0$ through equation \eqref{IV8} and the solution becomes the string-like source $\rho=-P_z=c_4e^{-c_4r^2}$, $P_r=P_\theta=0$ of Segre type $[(1,1)(11)]$, whose exterior is the Levi-Civita metric with $\sigma=0$, a locally flat conical spacetime with deficit angle $2\pi\left(1-e^{-c_4R^2/2}\right)$, as for the interior solutions of cosmic strings \cite{Gott,His}.
		
		\noindent The Kretschmann scalar $K=R_{abcd}R^{abcd}$ is finite throughout the interior of every solution, and on the axis it takes the values $4.75$, $2.51$, $63.42$, $9.97$, $64$ and $4$ for types I, II, III, IV, V and VI respectively, in the units used here. The Weyl invariant $C_{abcd}C^{abcd}$ is non-zero throughout the interior of every solution, with the values $1.33$, $0.65$, $21.12$, $2.33$, $21.33$ and $1.33$ on the axis, so none of the solutions is conformally flat, which is consistent with \cite{HLMS}, where it is shown that a conformally flat static cylindrical source can not be matched to the Levi-Civita exterior.
	}
	
	\FloatBarrier
	\section{Discussion} \label{S7}
	
	\noindent In this article, the primary goal was to find the Segre classification of the cylindrically symmetric static metric, and for the corresponding types find the solutions of the Einstein field equations. Since Segre classification provides the algebraic structure of the tensor field, it is an important aspect for determining the distribution of matter and energy in spacetime. Each Segre type is related to a set of canonical forms of Ricci spinors represented by NP scalars. For metric \eqref{2} the non-zero NP scalars found are $\phi_{00}=\phi_{22}$, $ \phi_{02}=\phi_{20}$ and $\phi_{11}$ which lead to the Segre types $[1,(111)]$, $[1,1(11)]$, $[(1,1)11]$, $[1,111]$, $[(1,11)1]$, $[(1,1)(11)]$ and $[(1,111)]$. Here, the first integer before comma represents the timelike eigenvalue whereas the others are spacelike. The other motive was exploring solutions of corresponding Segre types by figuring out the suitable EMT. A detailed analysis of all the Segre characteristics obtained for cylindrical geometry is done, like
	\begin{itemize}
		\item $[1,(111)]$: The symmetry in the spatial directions gives an isotropic distribution of matter and energy. The solution is regular, the density and pressure are positive and decreasing, and all the stability conditions are fulfilled, so this type gives a reliable solution in cylindrical geometry.
		
		\item $[1,1(11)]$: Two spatial eigenvectors share an eigenvalue and the third is different, so the fluid is anisotropic. A new function $\lambda=a_2{r^2}\cosh(b_2r^2)$ is introduced, and the resulting solution is regular and satisfies the physical and stability criteria, the energy conditions and the causality condition included.
		
		\item $[(1,1)11]$: The similar timelike and spacelike eigenvalue gives $\rho=-P_r$, which indicates the presence of dark energy. SEC is violated in the core, while NEC and WEC are violated only marginally through the axial stress, as shown in Fig.s \ref{fig.III4} and \ref{fig.III5}. DEC is also marginally violated through $P_z$ and, for $r>0.709R$, through $P_\theta$.
		
		\item $[1,111]$: All three pressures are different, and the solution is found by taking the Krori and Barua functions. The profiles are shown in Fig.s \ref{fig.IV1}-\ref{fig.IV9}, and all the physical and stability criteria are satisfied.
		
		\item $[(1,11)1]$: One eigenvalue is shared by the timelike, radial and axial directions, $\rho=-P_r=-P_z$, which is the tachyonic fluid form of the EMT. The solution is regular and has no free constant, SEC is violated in the core and DEC beyond $r=R/\sqrt{2}$.
		
		\item $[(1,1)(11)]$: This type corresponds to a non-null Maxwell field. The electrovac solution is singular on the axis, so a regular charged interior with $\rho=-P_r$ and $P_\theta=P_z$ is constructed. NEC, WEC and DEC hold for the total EMT while SEC is violated, and the interior joins the electrovac exterior without a surface charge.
		
		\item $[(1,111)]$: This is the vacuum type, studied in detail in \cite{24}. Its solution is the Levi-Civita metric, which is used here as the exterior of the interior solutions.
	\end{itemize}
	In cylindrical geometry, metric \eqref{2} gives seven Segre types, and a solution is presented for each of them: six interior solutions are obtained here and the seventh is the known vacuum one. The condition $P_r|_{(r=R)}=0$ is equivalent to the Darmois matching with the Levi-Civita exterior and fixes the Levi-Civita parameter as $\sigma=-R\mu'(R)/4$, giving $0<\sigma<1/4$ for types $[1,(111)]$, $[1,1(11)]$ and $[1,111]$ and $\sigma>1/2$, with a repulsive field at the boundary, for types $[(1,1)11]$ and $[(1,11)1]$. The C-energy is obtained for every solution, and none of them is conformally flat. Other choices of the free functions and the extension to modified theories of gravity can be investigated further.
	~\\
	~\\
	\noindent \textbf{Data Availability Statement}\ \ This manuscript has no associated data.
	
\end{document}